\documentclass{elsarticle}

\usepackage[british]{babel}
\usepackage{graphicx}
\usepackage{amsmath}
\usepackage{amssymb}
\usepackage{amsfonts}

\begin{document}

\begin{frontmatter}

\title{Semiclassical theory of potential scattering for massless Dirac fermions}

\author[RUN]{K J A Reijnders}
\ead{K.Reijnders@science.ru.nl}
\author[RUN]{T Tudorovskiy}
\author[RUN]{M I Katsnelson}
\address[RUN]{Radboud University Nijmegen, Institute for Molecules and Materials, \\Heyendaalseweg 135, 6525 AJ Nijmegen, The Netherlands}

\begin{abstract}
In this paper we study scattering of two-dimensional massless Dirac fermions by a potential that depends on a single Cartesian variable. Depending on the energy of the incoming particle and its angle of incidence, there are three different regimes of scattering. To find the reflection and transmission coefficients in these regimes, we apply the Wentzel-Kramers-Brillouin (WKB), also called semiclassical, approximation. We use the method of comparison equations to extend our prediction to nearly normal incidence, where the conventional WKB method should be modified due to the degeneracy of turning points. We compare our results to numerical calculations and find good agreement.
\end{abstract}

\begin{keyword}
Massless Dirac fermions, Semiclassical approximation, Scattering, Graphene, Topological insulators
\PACS{81.05.ue, 03.65.Vf, 03.65.Sq, 03.65.Nk, 73.40.Gk}

\end{keyword}

\end{frontmatter}

In this paper we present a systematic theory of potential scattering for massless Dirac fermions. Being the effective charge carriers in graphene~\cite{r1,Katsnelson12,r3,r7}, and topological insulators~\cite{Moore09,Hasan10,Qi11}, these particles attracted a keen interest. The discovery of massless Dirac fermions in condensed matter systems stimulated the fabrication of `artificial graphene', a material with a hexagonal lattice, where quantum dots~\cite{Singha11}, or molecules~\cite{Gomes12}, play the role of carbon atoms. The electron excitations in these materials give rise to massless Dirac fermions. The main feature of massless Dirac fermions is chirality (as it is called for graphene) or helicity (for topological insulators), i.e. an additional degree of freedom that relates to two kinds of particles (electrons and holes) simultaneously present in the system. Chirality makes the behavior of massless Dirac fermions dramatically different from that of Schr\"odinger particles. One of the most prominent examples is Klein tunneling~\cite{Katsnelson12,KatsnelsonNovGeim2006,Cheianov06,Silvestrov07,Shytov08,
Tudorovskiy12,YoungKim2009,StanderHuardGoldhaberGordon2009}. Due to this effect, a massless Dirac fermion normally incident on an electrostatic potential will be transmitted with unit probability.

\begin{figure}[tb]
  \begin{center}
    \includegraphics[width=0.7\textwidth]{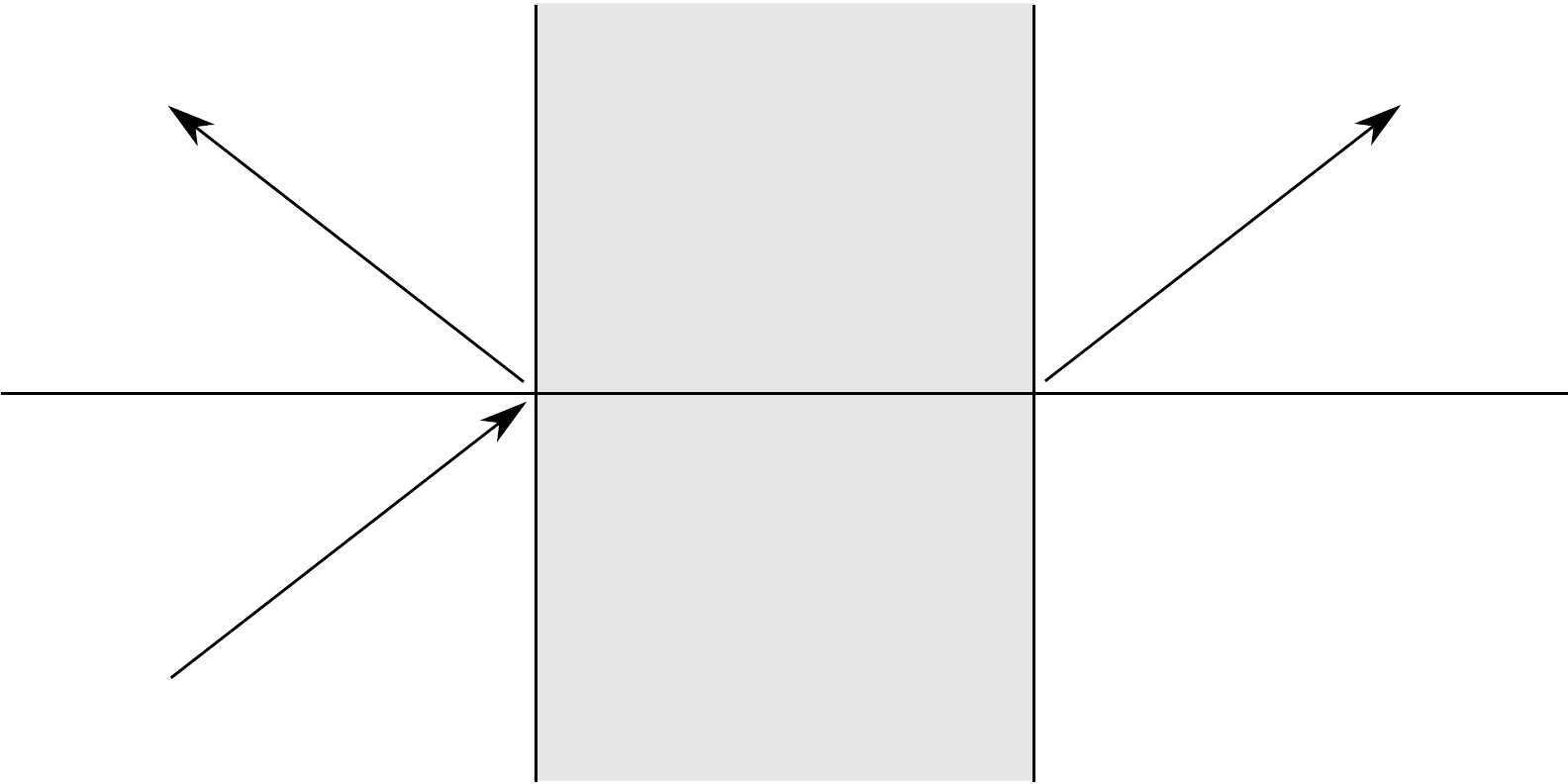}
    \caption{Illustration of angular scattering by a quasi one-dimensional potential barrier (gray area). The incoming and reflected waves are shown by arrows on the left of the potential and the transmitted wave is shown on the right.}
    \label{fig:one-d}
  \end{center}
\end{figure}

In this paper we consider scattering of massless Dirac fermions by quasi-one dimensional potential barriers (the corresponding potentials depend on a single Cartesian variable, see figure~\ref{fig:one-d}). Such barriers occur for instance in graphene heterostructures that were fabricated in~\cite{YoungKim2009,StanderHuardGoldhaberGordon2009}. They can also be intrinsic, as in the case of puddles in graphene~\cite{Mar08,KatsnelsonNovGeim2006}. We always assume that the potential profile is smooth enough, so that the Wentzel-Kramers-Brillouin (WKB) or semiclassical approximation~\cite{Heading62,Froeman65,Fedoruk66,Berry72,Landau77,Froeman02} can be used. The latter allows us to obtain generic formulas valid for arbitrary potentials. Another method that can be used to study generic potentials numerically was suggested in~\cite{Stone12}.

We distinguish three different regimes of scattering and show that the massless Dirac equation is equivalent to a pair of effective Schr\"odinger equations with complex potentials.
We then solve the scattering problem for each of these three regimes with the help of the WKB approximation. The specific formulation we use is the one pioneered by Zwaan~\cite{Zwaan29}, and further developed in~\cite{Kemble35,Furry47,Heading62,Froeman65,Fedoruk66,Froeman02}. Since we do not expect that all readers are familiar with this technique, it is summarized in~\ref{app:WKB}.

We start from preliminary considerations based on classical mechanics, section~\ref{sec:regimes}. In section~\ref{sec:scatstates} we introduce the semiclassical scattering states. Then, in section \ref{sec:WKB}, we formulate a set of simple rules that form the basis of the WKB  method and are sufficient to solve the scattering problem for angular scattering. Here and further on we use the term ``angular scattering'' for incidence far from both normal and tangential (see figure \ref{fig:one-d}). In section~\ref{sec:overdense}, we consider tunneling through a barrier supporting hole states, or an $n$-$p$-$n$ junction. Due to the classically allowed hole states within the barrier, one finds Fabry-P\'erot oscillations in the transmission coefficient~\cite{KatsnelsonNovGeim2006,Shytov08,Tudorovskiy12}, which were used to experimentally verify Klein tunneling~\cite{YoungKim2009}. We show that the WKB approximation does not accurately describe near-normal incidence on this barrier, since the classical turning points are nearly degenerate in this case. To circumvent this obstacle and to obtain a solution that is uniformly valid in the entire range of incidence angles, we use the technique of comparison equations, developed in~\cite{Fock34,Langer49,Cherry50,Lynn70,Zauderer72} and summarized in~\ref{app:comp-eq}. The results presented in this section were already published by the authors in~\cite{Tudorovskiy12}, but no proofs were given there. Here we present a complete and systematic treatment, including the detailed derivation.

In section~\ref{sec:above-barr}, we apply the WKB approximation to the case of above-barrier scattering, and in section~\ref{sec:barr-noholes} we consider tunneling through the barrier without hole states. 
In section~\ref{sec:tanh}, we consider an exact solution, first constructed in~\cite{Miserev12}, which is used to analyze the case of a single monotonous $n$-$n$ junction. Surprisingly, when applying the WKB approximation to this case, we find that along a certain path in the complex plane this situation can be reduced to the case of Klein tunneling. To stress the interconnection, this case is referred to as ``virtual Klein tunneling''. Finally, in section~\ref{sec:numerics}, we compare our predictions with numerical calculations.

Our main results are presented in the form of easy-to-use analytic expressions for reflection and transmission coefficients.

\section{Preliminary considerations: three regimes of scattering} \label{sec:regimes}

The wave function $\Psi$ of a massless Dirac fermion obeys the effective Dirac equation
\begin{equation}
  \left[ v \boldsymbol\sigma \cdot \hat{\boldsymbol p} + u(x/l,y/l) \right] \Psi(x,y) = E \Psi(x,y) ,
\end{equation}
where $v$ is the Fermi velocity, $\boldsymbol\sigma =(\sigma_x,\sigma_y)$ is the two-dimensional vector of Pauli matrices, $\hat{\boldsymbol p}=-i\hbar\nabla$ is the momentum operator and $l$ is the characteristic scale of change of the potential. In this paper we consider a potential $u$ that depends on $x$ only. Then the separation of variables gives $\Psi(x,y)=\Psi(x) \exp(i p_y y/\hbar)$, and we obtain
\begin{equation}
 \left[v \left(\begin{array}{cc}0 & \hat p_x-ip_y\\ \hat p_x+ip_y & 0 \end{array}\right)+u(x/l)\right]\Psi=E\Psi.
\label{eq:single-dim}
\end{equation}
Denoting the characteristic value of $|u-E|$ as $v p_0$ and introducing the dimensionless variables $\tilde x=x/l$, $\tilde p_x=-ihd/d \tilde x$, $\tilde p_y=p_y/p_0$, $h=\hbar/p_0l$, $\tilde u=u/v p_0$ and $\widetilde E=E/v p_0$, we can write (\ref{eq:single-dim}) in the form
\begin{equation}
 \left[\left(\begin{array}{cc}0 & \tilde p_x-i \tilde p_y\\
\tilde p_x+i \tilde p_y & 0 \end{array}\right)+\tilde u(\tilde x) \right]\Psi=\widetilde E\Psi,
\label{eq:single-dimless}
\end{equation}
or, equivalently,
\begin{equation}
 \bigl(\boldsymbol{\sigma} \cdot \boldsymbol{p}+u(x)\bigr)\Psi=E\Psi.
\label{eq:sigma-single}
\end{equation}
Here and further on we omit the tildes.

Let us consider the classically different scattering regimes comprised in equation~(\ref{eq:sigma-single}). It is well known \cite{Maslov81, Tudorovskiy12} that the classical Hamiltonian functions corresponding to the matrix quantum Hamiltonian are given by the eigenvalues of this matrix, where momentum operators are replaced by $c$-numbers and corrections of the order $h$ are neglected. Applying this prescription to (\ref{eq:sigma-single}) we obtain two Hamiltonian functions
\begin{equation}
 L_0^\pm(p_x,x)=\pm |\boldsymbol{p}|+u(x),
\label{eq:Ham-func}
\end{equation}
where $L_0^+$ and $L_0^-$ give rise to the electron and hole dynamical systems respectively. These Hamiltonian functions coincide at the point $x_0$ if for a certain energy $u(x_0)=E$. This implies $p_x=p_y=0$, i.e. the electron and hole systems merge for the case of normal incidence for sub-barrier scattering. The intersection of classical Hamiltonian functions is the origin of the Klein paradox \cite{KatsnelsonNovGeim2006,Tudorovskiy12}. It implies that the electron and hole systems cannot be treated separately for near-normal incidence. This forces us to change the representation to the one where electrons and holes are treated together. Such a representation can be easily found from (\ref{eq:Ham-func}). Indeed, for a given energy $E$ we have
\begin{equation}
 E=\pm |\boldsymbol{p}|+u(x),
\label{eq:Ham-func1}
\end{equation}
or $\mp |\boldsymbol{p}|=v(x)$, where we introduced the short-hand notation $v(x) = u(x)-E$. Squaring the last equality, we find
\begin{equation}
 \mathcal{L}(p_x,x)=p_x^2-v^2(x)=-p_y^2.
\label{eq:Ham-func2}
\end{equation}
where $\mathcal{L}(p_x,x)$ can be treated as the new Hamiltonian function and the parameter $\epsilon=-p_y^2$ plays the role of energy,
so that the level lines of $\mathcal{L}(p_x,x)$ corresponding to $\epsilon$ coincide with level lines of $L_0^\pm(p_x,x)$ corresponding to the energy $E$.
In the representation given by $\mathcal{L}$, electrons and holes are treated together.

\begin{figure}[tb]
  \begin{center}
    \includegraphics[width=1.0\textwidth]{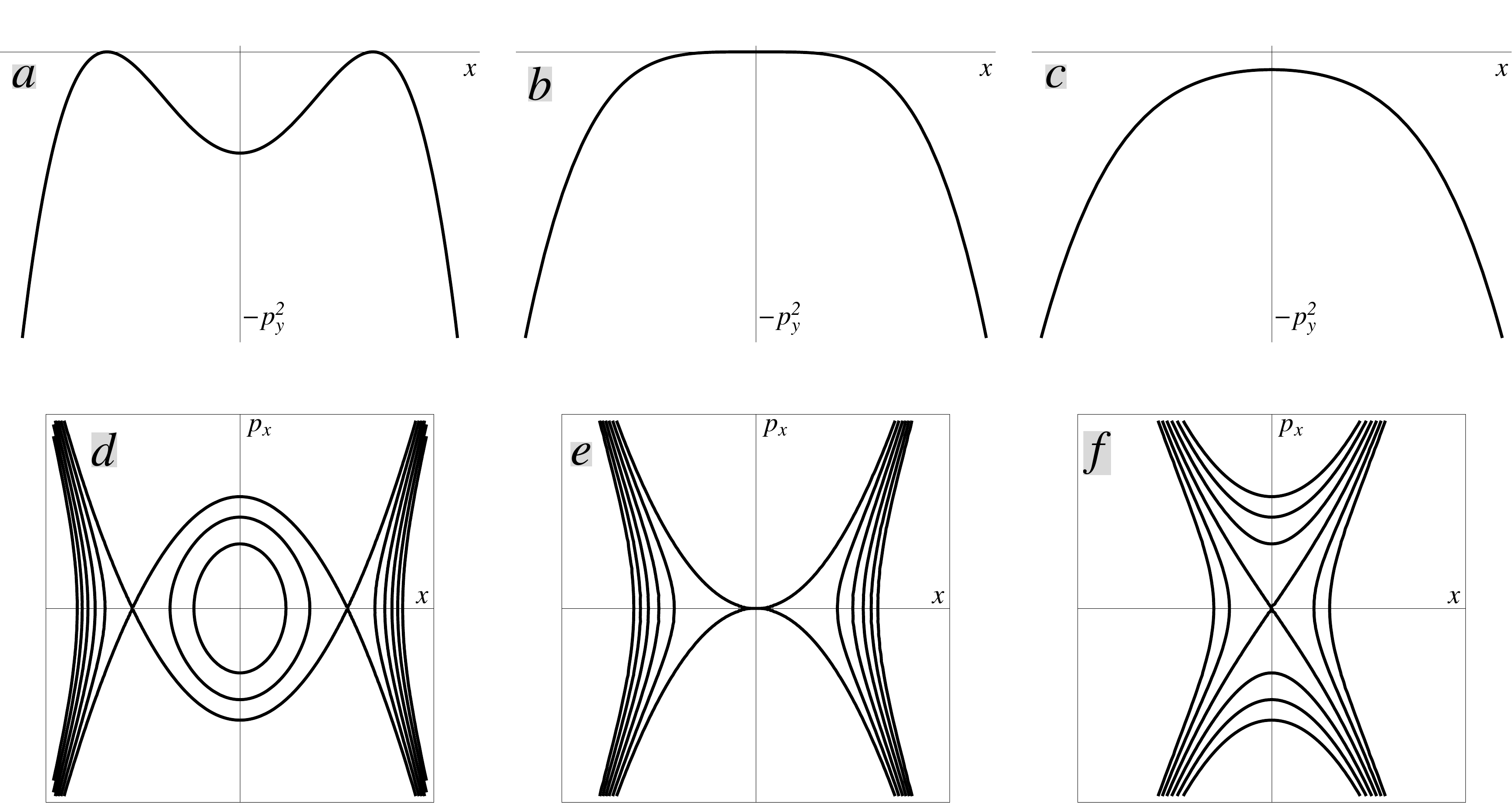}
    \caption{Effective potentials and phase portraits in the combined representation for the cases
    $E<0$: \textit{a}, \textit{d}; $E=0$: \textit{b}, \textit{e} and $E>0$: \textit{c}, \textit{f}. In the figure $u(x)=-x^2$.}
    \label{fig:regimes}
  \end{center}
\end{figure}

The phase portraits of the Hamiltonian systems that originate from $\mathcal{L}$ are cuts of the original four-dimensional phase space $\{p_x,p_y,x,y\}$ by the hyperplane $E=\textrm{const}$. Every individual trajectory in this cut is defined by a certain value~$-p_y^2$. In figure~\ref{fig:regimes} one sees the effective potentials $-v^2(x)$ for different values of $E$ and the corresponding phase portraits. These pictures describe all qualitatively different regimes for a Dirac particle scattered by a single hump potential.

When the energy $E$ does not exceed $u_0$, the maximal value of $u(x)$ (figure~\ref{fig:regimes}\,a,\,d), there exist either four (for small $|p_y|$) or two real turning points (for larger $|p_y|$). In the opposite case (figure~\ref{fig:regimes}\,c,\,f), when $E$ is larger than $u_0$, real turning points are absent for small $|p_y|$ and there appear two of them for larger values of $|p_y|$. Thus we differentiate three different scattering regimes:
\begin{enumerate}
  \item $E < u_0$, $|p_y| < u_0 - E$: Klein tunneling regime, or tunneling through a barrier supporting hole states
  \item $E > u_0$, $|p_y| < E - u_0$: above-barrier scattering
  \item $E < u_0$ and $|p_y| > u_0 - E$, or $E > u_0$, $|p_y| > E - u_0$: conventional tunneling regime, tunneling through a barrier without hole states.
\end{enumerate}
For each of these scattering regimes we will construct a separate description.

Representation (\ref{eq:Ham-func2}) allowed us to combine electrons and holes within a single dynamical system. This transformation has a straightforward quantum analogue.
Indeed, let us rewrite (\ref{eq:sigma-single}) as $(\boldsymbol\sigma\cdot\boldsymbol p+v(x))\Psi=0$. We can now act on this equation from the left with the operator $(\boldsymbol\sigma\cdot\boldsymbol p -v(x))$ to obtain~\cite{Tudorovskiy12},
\begin{equation}
\bigl(\boldsymbol{\sigma} \cdot \boldsymbol{p}-v(x)\bigr)\bigl(\boldsymbol{\sigma} \cdot \boldsymbol{p}+v(x)\bigr)\Psi=
  \bigl(\hat{p}_x^2 +p_y^2 -v^2(x) -i h \sigma_x v'(x) \bigr) \Psi  = 0.
\end{equation}
Since the last equation contains only a single Pauli matrix, it can be diagonalized by writing
\begin{equation}
  \Psi = \left( \begin{array}{c} 1 \\ 1 \end{array} \right) \eta_1 + \left( \begin{array}{c} 1 \\ -1 \end{array} \right) \eta_2 ,
  \label{eq:psiexpeta}
\end{equation}
and one obtains
\begin{equation}
\left(h^2\frac{d^2}{dx^2}+v^2(x)-p_y^2 \pm ihv'(x)\right)\eta_{1,2}=0.
\label{eq:reduction}
\end{equation}
The functions $\eta_{1}$ and $\eta_2$ are not independent, they are related by:
\begin{equation}
 \eta_{2,1}=\frac{1}{p_y}\left(h\frac{d}{dx}\pm iv(x)\right)\eta_{1,2},
\label{eq:connection}
\end{equation}
as can be found from equation~(\ref{eq:sigma-single}). The real part of equation~(\ref{eq:reduction}) corresponds to (\ref{eq:Ham-func2}), and the imaginary part gives a quantum correction to the classical transformation.

\section{Semiclassical scattering states} \label{sec:scatstates}

Before we can solve the scattering problem for the different regimes outlined in the previous section, we first have to define the asymptotic scattering states. From now on we assume that $v(z)$ is an analytic function in the complex plane. Therefore we can consider equations~(\ref{eq:reduction}) and~(\ref{eq:connection}) for $\eta_1$ in the complex plane:
\begin{equation}
 \left(h^2\frac{d^2}{dz^2}+v^2(z)-p_y^2 + ihv'(z)\right)\eta_{1}(z)=0, \label{eq:reduction-complex}
\end{equation}
and
\begin{equation}
 \eta_2=\frac{1}{p_y}\left(h\frac{d}{dz}+iv(z)\right)\eta_1(z). \label{eq:connection-complex}
\end{equation}

The semiclassical solution for equation (\ref{eq:reduction-complex}) has the form
\begin{equation}
 \eta_1(z)=A(z,h) e^{i s(z)/h} ,
\end{equation}
where $A(z,h)=A_0(z)+hA_1(z)+\ldots$ is a power series in $h$. Substituting this function into equation~(\ref{eq:reduction-complex}), we find
\begin{equation}
 \left[\left(h\frac{d}{dz}+i s'(z)\right)^2+v^2(z)-p_y^2 + ihv'(z)\right]A(z,h)=0.
\end{equation}
Equating the terms on the left hand side to zero for all powers of $h$, we obtain equations that determine $s(z)$ and $A(z,h)$. The terms of order $h^0$ give
\begin{equation}
 (s'(z))^2=v^2(z)-p_y^2,
 \label{eq:HJ}
\end{equation}
whilst collecting the terms of order $h^1$, we find
\begin{equation}
 2s'(z)A_0'(z)+s''(z)A_0(z)+v'(z)A_0(z)=0.
\end{equation}
Multiplying the latter equation by $A_0(z)$, we obtain
\begin{equation}
 \frac{d}{dz}[s'(z)A_0^2(z)]+v'(z)A_0^2(z)=0.
\end{equation}
Assuming that $s'(z)$ does not vanish we get
\begin{equation}
 A_0^2(z)=\frac{B}{s'(z)}\exp\left(-\int_{z_0}^z d\zeta \frac{v'(\zeta)}{s'(\zeta)}\right), \label{eq:A0}
\end{equation}
where $B$ is a constant, $z_0$ is an (up to now) arbitrary point and the integration should be performed along a suitable path in the complex plane. Equation (\ref{eq:HJ}) has two solutions, namely $\pm s(z_0,z)$, where
\begin{equation}
 s(z_0,z)=\pm \int_{z_0}^z p_x(\zeta) d\zeta, \quad p_x(z) = \left(v^2(z)-p_y^2\right)^{1/2}. \label{eq:action}
\end{equation}
Note that the square root is not a single-valued function in the complex plane, which means that we have to insert branch cuts emanating from every point where its argument vanishes. To distinguish the square root as an analytic function defined as discussed above from the positive square root of a positive number, we will denote the former by $z^{1/2}$, and the latter by $\sqrt{x}$.
Combining equations~(\ref{eq:A0}) and~(\ref{eq:action}) and choosing the constants $B_\pm$ in an appropriate way, we obtain two asymptotic solutions,
\begin{equation}
  \widetilde \eta_1^{\,\pm}(z)=\frac{1}{p_x^{1/2}(z)}\exp\left(\mp\frac{1}{2}\int_{z_0}^z d\zeta \frac{v'(\zeta)}{p_x(\zeta)}\right) \exp\left(\pm\frac{i}{h} s(z_0,z) \right) .
  \label{eq:eta1-sol-cmplx}
\end{equation}
The integral in the exponent can also be computed explicitly,
\begin{equation}
 \int^z_{z_0} d\zeta \frac{v'(\zeta)}{p_x(\zeta)}=\int_{v(z_0)}^{v(z)}\frac{dv}{(v^2-p_y^2)^{1/2}}=
 \ln\left[\frac{v(z)+(v^2(z)-p_y^2)^{1/2}}{|p_y|}\right]+\textrm{const} ,
\end{equation}
which gives rise to the representation
\begin{equation}
 \eta_1^{\pm}(z)=\frac{g^{\mp 1/2}(z)}{p_x^{1/2}(z)}e^{\pm i s(z_0,z)/h}, \quad g(z)=\frac{v(z)+p_x(z)}{|p_y|}.
 \label{eq:eta1-sol-cmplx-g}
\end{equation}
It is important to note that $g(z)$ does not vanish at any point $z$ if $p_y$ does not vanish and $|z|<\infty$.

At this point, let us come back to the real axis and introduce refection and transmission coefficients.
First we establish the current conservation condition. It is convenient to start from equation (\ref{eq:sigma-single}), and to multiply it from the left by $\Psi^\dag$. This gives
\begin{equation}
 \Psi^\dag\bigl(\boldsymbol{\sigma} \cdot \boldsymbol{p}+u(x)\bigr)\Psi(x)=E\Psi^\dag(x)\Psi(x).
 \label{eq:psi-h-psi}
\end{equation}
Subtracting from equation~(\ref{eq:psi-h-psi}) its complex conjugate, we obtain
\begin{equation}
 -ih\Psi^\dag(x)\sigma_x\Psi'(x) - ih[\Psi^\dag(x)]'\sigma_x\Psi(x)=-ih[\Psi^\dag(x)\sigma_x\Psi(x)]'=0.
\end{equation}
Hence the conserved current is given by $j_x(x)=\Psi^\dag(x)\sigma_x\Psi(x)$, which can also be written as $\textrm{Re}\,[\psi_1^*(x)\psi_2(x)]=\textrm{const}$. Using relation~(\ref{eq:psiexpeta}) between $\Psi$ and $\eta_{1,2}$, we find
\begin{equation}
|\eta_1(x)|^2-|\eta_2(x)|^2=\textrm{const}.
\label{eq:conserve}
\end{equation}
It is useful to understand the conservation equation (\ref{eq:conserve}) from the point of view of the effective equation~(\ref{eq:reduction}), and equation~(\ref{eq:connection}). For the equation
\begin{equation}
\left(h^2\frac{d^2}{dx^2}+v^2(x)-p_y^2+ihv'(x)\right)\eta_1(x)=0 ,
\label{eq:reduction2}
\end{equation}
the conserved quantity is given by the Wronskian,
\begin{equation}
 W=\left|\begin{array}{cc}\eta_1(x) & \eta(x) \\ \eta'_1(x) & \eta'(x) \end{array}\right|,
\label{eq:W}
\end{equation}
where $\eta_1(x)$ and $\eta(x)$ are linear independent solutions of (\ref{eq:reduction2}). Since, after complex conjugation, equation~(\ref{eq:reduction}) for $\eta_2$ coincides with equation~(\ref{eq:reduction2}), one can choose $\eta(x)=\eta_2^*(x)$, where the star denotes the complex conjugation. From (\ref{eq:W}) we then find
\begin{equation}
 \eta_1(x) (\eta_2^*)'(x)-\eta_2^*(x)\eta_1'(x)=\textrm{const}.
\end{equation}
Using equation~(\ref{eq:connection}) to eliminate the derivatives, we arrive at (\ref{eq:conserve}).

Now we introduce the scattering solutions in a given classically allowed region,
\begin{align}
  \eta_{1}(x) = \frac{a_1}{\sqrt{p_x(x)}\sqrt{G(x)}} e^{i S(x_0,x)/h} + a_2 \frac{\sqrt{G(x)}}{\sqrt{p_x(x)}} e^{-i S(x_0,x)/h} . \label{eq:scat-sol}
\end{align}
Note that we now use the symbol $\sqrt{x}$, i.e. we assume that we are on the real axis. We denoted
\begin{equation}
 G(x)=\left(\frac{|v(x)|+p_x(x)}{|p_y|}\right)^{\nu}, \quad \nu=\textrm{sgn}[v(x_0)] \label{eq:def-G}
\end{equation}
and the momentum $p_x(x)$ and the action $S(x_0,x)$ are defined through
\begin{equation}
  p_x(x) = \sqrt{v^2(x)-p_y^2}, \quad S(x_0,x) = \int_{x_0}^x p_x(\zeta)d\zeta. \label{eq:action-realaxis}
\end{equation}
In the electron region one has $E>u(x)$, hence $v(x)<0$.
Applying equation~(\ref{eq:connection}) to the expansion~(\ref{eq:scat-sol}), we can find $\eta_2(x)$ as
\begin{equation}
  \eta_2(x) = i\nu \frac{|p_y|}{p_y} \left(
    a_1 \frac{\sqrt{G(x)}}{\sqrt{p_x(x)}} e^{i S(x_0,x)/h} +
    a_2 \frac{1}{\sqrt{p_x(x)}\sqrt{G(x)}} e^{-i S(x_0,x)/h}
    \right). \label{eq:eta2-el}
\end{equation}
To obtain (\ref{eq:eta2-el}), we used the equalities
\begin{equation}
 \frac{v(x)+p_x(x)}{|p_y|}=\nu G, \quad \frac{v(x)-p_x(x)}{|p_y|}=\frac{\nu}{G}. \label{eq:equal-G}
\end{equation}
Inserting $\eta_1(x)$ and $\eta_2(x)$ into the current conservation law~(\ref{eq:conserve}), we find
\begin{eqnarray}
  |\eta_1(x)|^2-|\eta_2(x)|^2&=&\frac{1}{p_x(x)}\left(\frac{1}{G(x)}-G(x)\right)(|a_1|^2-|a_2|^2),\nonumber\\
  &=&-\frac{2\nu}{|p_y|}(|a_1|^2-|a_2|^2) \label{eq:el-current},
\end{eqnarray}
where we assumed that the action $S(x_0,x)$ is purely real.

When we consider scattering from an electron region on the left to an electron region on the right, we can introduce the coefficients $a_1 = 1$ and $a_2=r$ on the left, and $a_1 = t$ and $a_2=0$ on the right. Since $\nu=-1$ on both sides, equation~(\ref{eq:el-current}) tells us that
\begin{equation}
 |r|^2+|t|^2=1.
\label{eq:conservation}
\end{equation}
Therefore $r$ and $t$ can be treated as the reflection and transmission coefficient respectively.

Now let us turn to scattering from an electron region on the left to a hole region on the right. From equation~(\ref{eq:Ham-func}) one infers that holes with positive velocity $v_x=\partial L_0^-(p_x,x)/\partial p_x$ have negative momentum, see also~\cite{Tudorovskiy12}. Therefore the coefficient of the right-moving hole state is $a_2$, and we set $a_2 = t$ and $a_1=0$. Since $v(x)>0$, we also have $\nu=+1$. Inserting this into the current conservation law~(\ref{eq:conserve}), we  again obtain equation~(\ref{eq:conservation}).

In a previous paper~\cite{Tudorovskiy12}, the authors constructed the asymptotic scattering states with the help of a different method, and defined expansion coefficients that slightly differ from those given above. The relation between these two expansions is given in~\ref{app:scatstates}.

\section{Stokes diagrams and the WKB approximation in the complex plane} \label{sec:WKB}

In the previous section we established that the functions $\eta_1^\pm(z)$, given by~(\ref{eq:eta1-sol-cmplx}), solve the reduced equation~(\ref{eq:reduction}) for $\eta_1$ up to order $h$. Let us now choose the constant $z_0$ to be a turning point, i.e. a point where $p_x(z)$ vanishes; $p_x(z_0)=0$. There are lines, called \emph{anti-Stokes} lines~\cite{Heading62,Froeman65,Froeman02}, emanating from the point $z_0$ along which the imaginary part of the function $s(z_0,z)$, given by equation~(\ref{eq:action}), vanishes, that is $\textrm{Im}[s(z_0,z)]=0$.\footnote{We remark that what we call an anti-Stokes line is sometimes denoted as Stokes line, see e.g. \cite{Fedoruk66}.} Along an anti-Stokes line both asymptotic solutions are of order one with respect to the small parameter $h$. On the anti-Stokes line $\gamma$ the exact solution $\psi(z)$ can be represented as
\begin{equation}
 \psi(z)= C^\gamma_+\eta_1^+(z)+C^\gamma_-\eta_1^-(z),  \label{eq:eta_asymptotic}
\end{equation}
where the superscript `$\gamma$' relates to the anti-Stokes line. Equality~(\ref{eq:eta_asymptotic}) only holds up to order $h$.

\begin{figure}
\begin{center}$
\begin{array}{ccc}
\includegraphics[height=3.42cm]{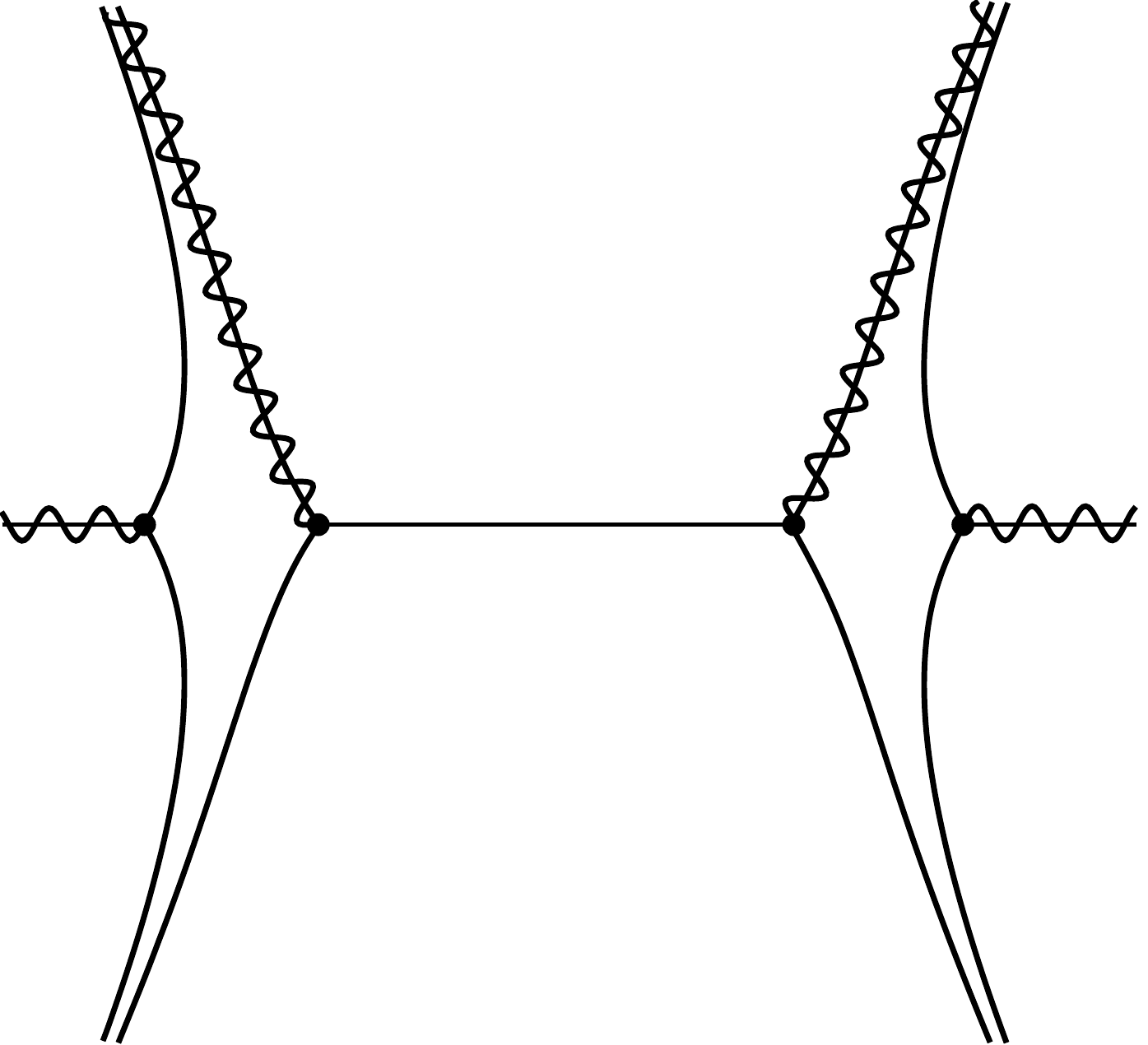} &
\includegraphics[height=3.42cm]{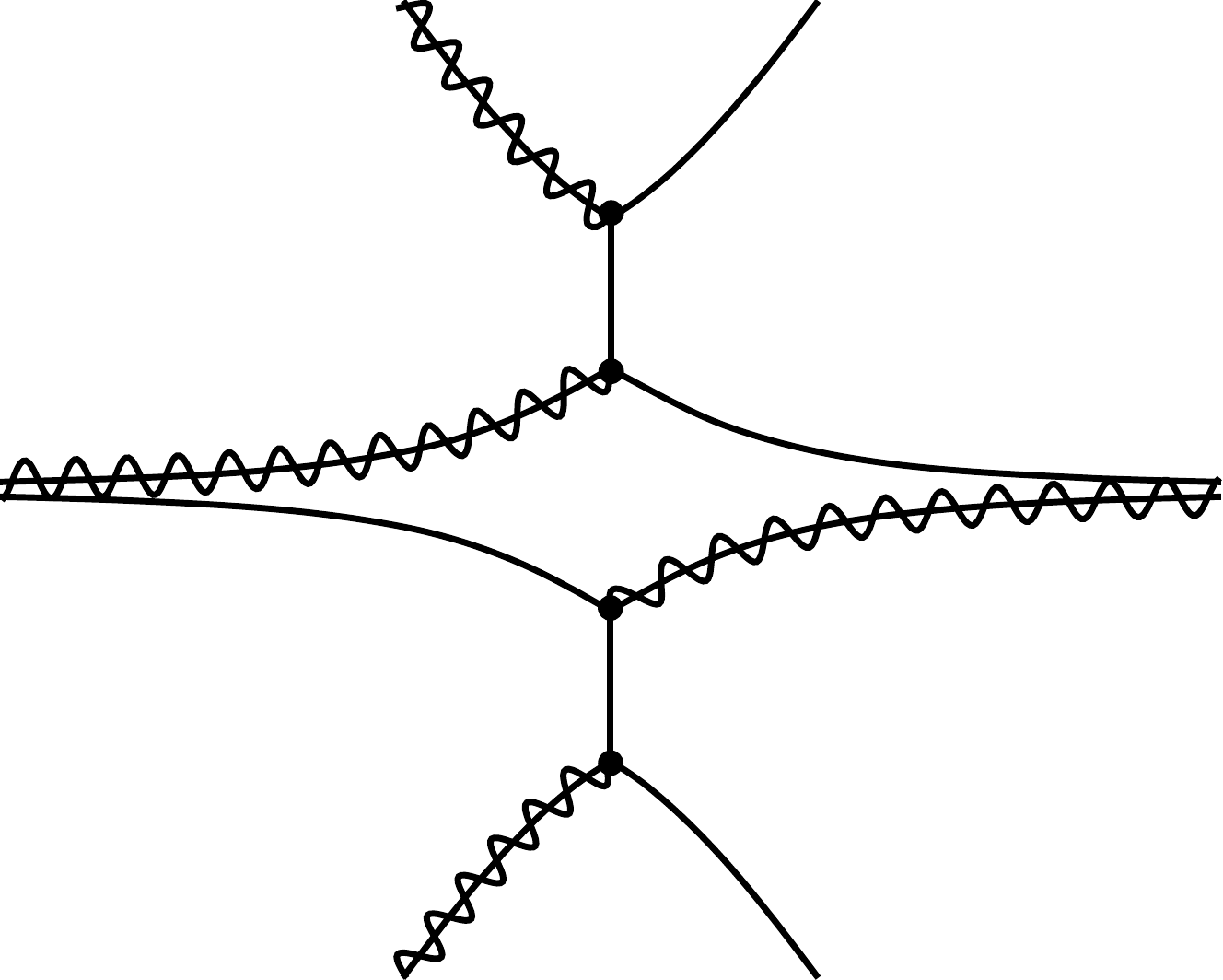} &
\includegraphics[height=3.42cm]{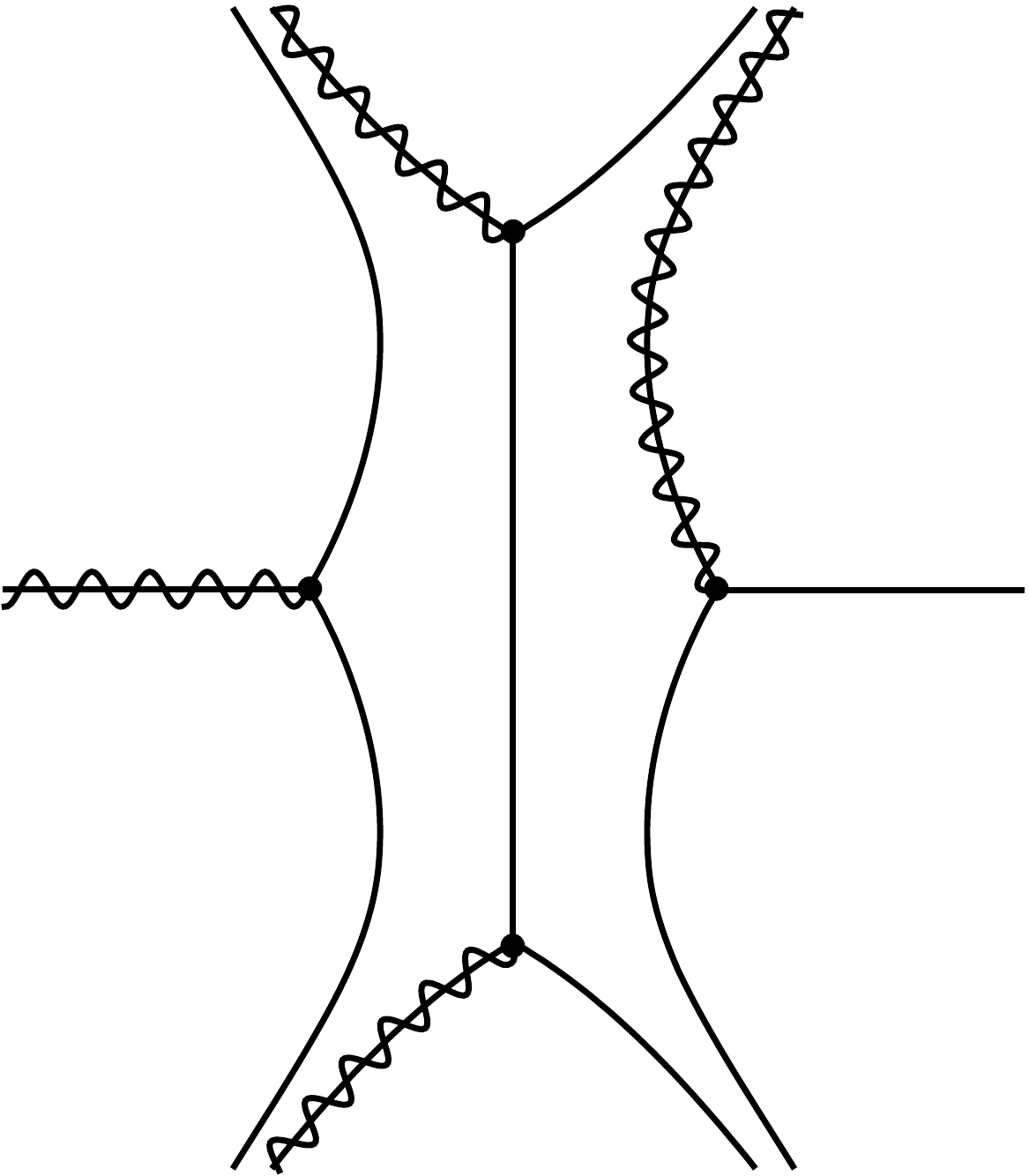} \\
a) & b) & c)
\end{array}$
\end{center}
\caption{Stokes diagrams for the three different regimes outlined in section~\ref{sec:regimes}: a) Klein tunneling, b) above-barrier scattering and c) conventional tunneling. Bold points show the turning points, the solid lines correspond to anti-Stokes lines and the wavy lines designate branch cuts of the function $(z-z_0)^{1/2}$. In the figure $u(z)=-z^2$.}
\label{fig:Stokes_diagrams}
\end{figure}

Let us introduce a so-called \emph{Stokes diagram}, in which the anti-Stokes lines are drawn in the complex plane, together with the choice of the branch cuts of the square root in the definition of $p_x(z)$. In figure~\ref{fig:Stokes_diagrams}, we show the Stokes diagrams for the regimes outlined in section~\ref{sec:regimes}. The bold circles depict the turning points, the solid lines depict the anti-Stokes lines, and the wavy lines represent the branch cuts. Near a simple turning point, which means that $p_x(z)$ has a simple root $z=z_0$, we can approximate $p_x(z)$ by $\alpha(z-z_0)^{1/2}$, where $\alpha$ is some constant. It is then easy to show~\cite{Heading62,Froeman65,Fedoruk66,Froeman02} that three anti-Stokes lines emanate from it.

In figure \ref{fig:Stokes_diagrams}\,a), corresponding to the regime of Klein tunneling, one sees four real turning points. There are classically forbidden regions between the left two and between the right two turning points, while the anti-Stokes line connecting the middle two turning points represents the classically allowed hole region. The position of these turning points depends on the transversal momentum $|p_y|$. When it goes to zero, the two turning points on the left (and on the right) come close together and eventually merge, that is, the classically forbidden region disappears, as discussed in~\cite{Tudorovskiy12}. When $|p_y|$ becomes larger, the turning points on the left (and on the right) move further apart, and whenever $u_0-E=|p_y|$, the middle two turning points merge. If $|p_y|$ increases further, they disappear off the real axis, and there are only two real turning points left. However, in this case $p_x(z)$ acquires two complex roots, and hence we have two complex turning points. This is the situation in the conventional tunneling regime, the corresponding Stokes diagram is shown in figure~\ref{fig:Stokes_diagrams}\,c). We use this term because the situation is similar to that of a Schr\"odinger particle that tunnels through a potential hump. Obviously, the analogy can be used only if the complex turning points are sufficiently far from the real axis.

In the regime of above-barrier scattering, figure~\ref{fig:Stokes_diagrams}\,b), all four turning points are complex. Since the potential $u(x)$ is real on the real axis, the turning points come in complex conjugate pairs. Each of the two turning points closest to the real axis gives rise to one finite anti-Stokes line, and to two infinite anti-Stokes lines. If the potential $u(x)$ vanishes along the real axis at $|x|\to\infty$, the infinite anti-Stokes lines approach horizontal asymptotes. If $u(x)$ is unbounded at $|x|\to\infty$, as in the figure \ref{fig:Stokes_diagrams}\,b), the infinite anti-Stokes lines approach the real axis. When $|p_y|$ becomes smaller, the upper (and lower) two turning points come close together and eventually merge whenever $|p_y|$ vanishes. When $|p_y|$ increases, the distance between points closest to the real axis becomes smaller, and when $u_0-E=|p_y|$ they merge. When $|p_y|$ grows further, we once again end up in the conventional tunneling regime, figure \ref{fig:Stokes_diagrams}\,c).

Now we can reformulate the scattering problem in terms of the Stokes diagram. Every scattering problem can be reduced to the determination of the coefficients $C_+^{\gamma_1}$, $C_-^{\gamma_1}$ along the anti-Stokes line $\gamma_1$ emanating from the turning point $z_1$, under the assumption that the expansion coefficients $C_+^{\gamma}$, $C_-^{\gamma}$ along the anti-Stokes line $\gamma$ emanating from the turning point $z_0$ are known. Generally speaking $z_0$ and $z_1$ do not coincide. The problem of establishing the connection between the expansion coefficients at different anti-Stokes lines is known as the \emph{connection problem}. It was first found by Stokes~\cite{Stokes57}, and was further elaborated by many others, see e.g.~\cite{Froeman65,Froeman02}. There are various ways to solve it. The approach that was used to produce the first connection formulas~\cite{Rayleigh12,Gans15,Jeffreys25}, and is usually taken in textbooks on quantum mechanics, e.g.~\cite{Griffiths05}, is to approximate the potential near the turning point, and to solve the resulting equation exactly. In its most rigorous form, this method is known as the method of comparison equations~\cite{Fock34, Langer49,Cherry50,Heading62, Lynn70, Zauderer72}, and is summarized and applied in~\ref{app:comp-eq}. In the remainder of this section we introduce a different approach, that was pioneered by Zwaan~\cite{Zwaan29}, and further developed in~\cite{Kemble35,Furry47,Heading62,Fedoruk66,Froeman65,Froeman02}. In this method one passes from one anti-Stokes line to another along a suitable path in the complex plane avoiding the vicinities of turning points. In the rest of this section we give an introduction to this method, more details can be found in~\ref{app:WKB}.

\begin{figure}[tb]
  \begin{center}
    \includegraphics[width=0.5\textwidth]{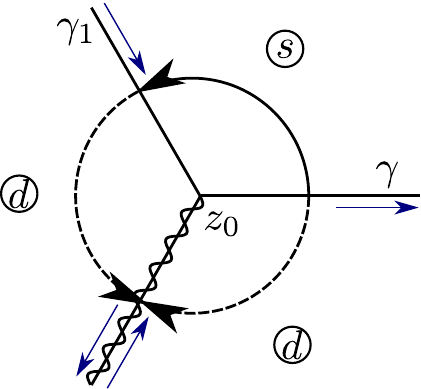}
    \caption{The Stokes diagram for a simple turning point $z_0$. The wavy line depicts the cut. The blue arrows show the direction of the growth of the action $s(z_0,z)$. The letters `s' and `d' indicate the sectors where $\eta_1^+$ is subdominant and dominant respectively}
    \label{fig:Airy}
  \end{center}
\end{figure}

Following the idea first set forth by Furry~\cite{Furry47}, see also~\cite{Heading62,Fedoruk66,Froeman65,Froeman02}, we first consider the transition between two anti-Stokes lines emanating from the same turning point $z_0$. We notice that though $\eta_1^+(z)$ and $\eta_1^-(z)$ are asymptotic solutions for (\ref{eq:reduction2}), every single of them \textit{does not} provide an asymptotic solution of any exact solution within an area around $z_0$. This already becomes clear if we take into account that $\eta_1^{\pm}$ have cuts emanating from $z_0$, while the exact solution does not have such cuts. These cuts correspond to the definition of $p_x(z)$ in the complex plane. To resolve this apparent contradiction, let us first consider an exact solution $\eta_1(z)$ which has an asymptotic expansion $\eta_1^+(z)$ along an anti-Stokes line $\gamma$. We assume that the action $s(z_0,z)$ grows along $\gamma$ (as indicated by a blue arrow in figure \ref{fig:Airy}). For convenience let us choose the cut in the definition of $p_x(z)$ along the anti-Stokes line next to $\gamma$ in the clockwise direction. If we now leave $\gamma$ in the counterclockwise direction, the action $s(z_0,z)$ acquires a positive imaginary part and $\eta_1^+(z)$ gets exponentially small. Along the anti-Stokes line $\gamma_1$ we thus recover an asymptotic expansion in the form of a single ``incoming'' wave, meaning that $s(z_0,z)$ decays along this line. The $2\pi/3$-sector between $\gamma$ and $\gamma_1$ is designated by `\textit{s}' (subdominant) in figure \ref{fig:Airy}, meaning that within it $\eta_1^+(z)$ gets exponentially small values. In the next $2\pi/3$-sector in the counterclockwise direction, designated by `\textit{d}' (dominant),  $\eta_1^+(z)$ becomes exponentially large. It still gives the asymptotic expansion of $\eta_1(z)$ everywhere apart from some vicinity of the anti-Stokes line that coincides with the cut. In this vicinity the asymptotic representation fails since in the considered sector we can not avoid the appearance of the additional term given by $\eta_1^-(z)$ in the asymptotic of $\eta_1(z)$ against the background of exponentially large values of $\eta_1^+(z)$. Thus $\eta_1^+(z)$ gives only one term of the asymptotic expansion of $\eta_1(z)$ along the right lip of the cut. Left and right lips of the cut are defined with respect to an observer standing on the cut with the turning point behind him.
A similar consideration shows that the other term of the asymptotic expansion of $\eta_1(z)$ is given by $\eta_1^+(z)$ on the left lip of the cut.

For any point $z$ on the cut we can define a nearby point on the right lip $z_r$ and a nearby point on the left lip $z_l$.
Then one has $p_x(z_l)=e^{-i\pi}p_x(z_r)$, \mbox{$g^{1/2}(z_l)=\nu g^{-1/2}(z_r)$}, see also~\ref{app:WKB}. Therefore we obtain the following asymptotic representation of the exact solution $\eta_1(z_r)$, i.e. on the right lip of the cut,
\begin{equation}
\eta_1(z_r)=\frac{g^{1/2}(z_r)}{p_x^{1/2}(z_r)}e^{i s(z_0,z_r)/h}+i\nu\,
\frac{g^{-1/2}(z_r)}{p_x^{1/2}(z_r)}e^{-i s(z_0,z_r)/h}.
\end{equation}
The latter can be written as
\begin{equation}
\eta_1(z_r)=\eta_1^+(z_r)+i\nu \eta_1^-(z_r).
\label{eq:eta1Stokes}
\end{equation}

The fact that $\eta_1(z)$ can not be approximated by $\eta_1^+(z)$ along every anti-Stokes line is associated with the \emph{Stokes phenomenon}~\cite{Stokes57}. The constant $i$ in front of $\eta_1^-(z)$ in (\ref{eq:eta1Stokes}) is called the \textit{Stokes constant}. In contrast to the Schr\"odinger equation~\cite{Heading62,Froeman65,Fedoruk66,Froeman02} we found an extra factor $\nu$ in front $\eta_1^-(z_r)$, which relates to the additional amplitude factor $g(z)$, see \ref{app:WKB}.

Now we turn to the case where the anti-Stokes lines emanate from different turning points. When we are dealing with a finite anti-Stokes line that connects the two turning points, the initial expansion is valid along the entire line. However, we do have to change the reference point of the action,  that is, write $s(z_0,z)=s(z_0,z_1)+s(z_1,z)$. This introduces the additional phase factor $e^{is(z_0,z_1)}$ by which the coefficients have to be multiplied.

\begin{figure}[tb]
  \begin{center}
    $\begin{array}{ccc}
    \includegraphics[width=.4\textwidth]{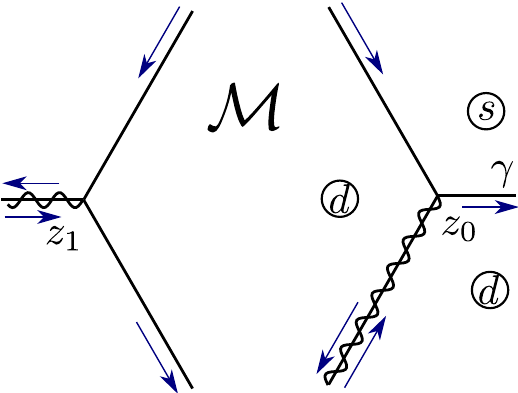} & \quad &
    \includegraphics[width=.4\textwidth]{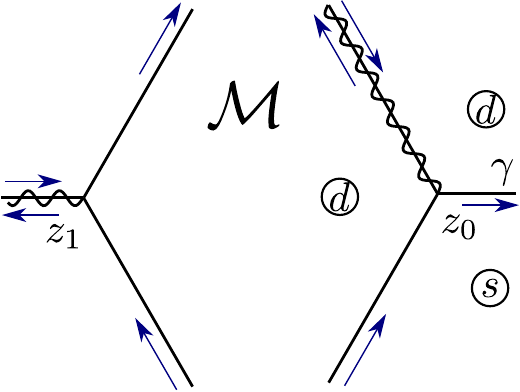} \\
    a) & \quad & b)
    \end{array}$
    \caption{The Stokes diagram for two simple turning points $z_0$ and $z_1$. The wavy lines depict cuts. Blue arrows show the direction of the growth of the action $s(z_0,z)$. The letters in circles are given with respect to $z_0$. Diagram \textit{a)} corresponds to $\eta_1(z)=\eta_1^+(z)$ along $\gamma$ and diagram b) to $\eta_1(z)=\eta_1^-(z)$.}
    \label{fig:reference}
  \end{center}
\end{figure}

When we are dealing with two turning points $z_0$ and $z_1$ that are not connected by an anti-Stokes line, the situation is more complicated. In figure~\ref{fig:reference} one sees two simple turning points and corresponding anti-Stokes lines. Let us consider the same asymptotic solution $\eta_1^+(z)$ as was considered above. In the region $\mathcal{M}$ bounded by four anti-Stokes lines, $\eta_1^+(z)$ gets dominant when viewed from the reference point $z_0$.  Against the background of exponentially large values of $\eta_1^+(z)$,  an exponentially small term in (\ref{eq:eta1Stokes}), given by $\eta_1^-(z)$, should be neglected within the accuracy of the WKB approximation. This means that the connection procedure prescribed by WKB-method is not bijective: functions differing by $\eta_1^-(z)$ on the clockwise lip of the cut will be mapped into the same expansions along the anti-Stokes lines emanating from $z_1$. This leads to the so-called one-directional nature of the connection formulae, see~\ref{app:WKB} and~\cite{Heading62,Froeman65,Froeman02,Berry72}.

Let us summarize this section. We have formulated a set of rules on how to pass between different anti-Stokes lines. Using these rules one can solve certain kinds of scattering problems. However degenerate turning points cannot be treated within this approximation. In the next section we consider necessary generalizations of the WKB-method.

\section{Tunneling through a barrier supporting hole states} \label{sec:overdense}

In this section we solve the scattering problem for the first case of section~\ref{sec:regimes}, that is, tunneling through a barrier supporting hole states. The classically allowed region for this case has been extensively studied by the authors in~\cite{Tudorovskiy12} with the canonical operator method~\cite{Maslov81}, and particular emphasis was placed there on its geometric interpretation. In this section we are mainly interested in the transition through the classically forbidden region.

\begin{figure}[t]
  \begin{center}
    \includegraphics[width=0.8\textwidth]{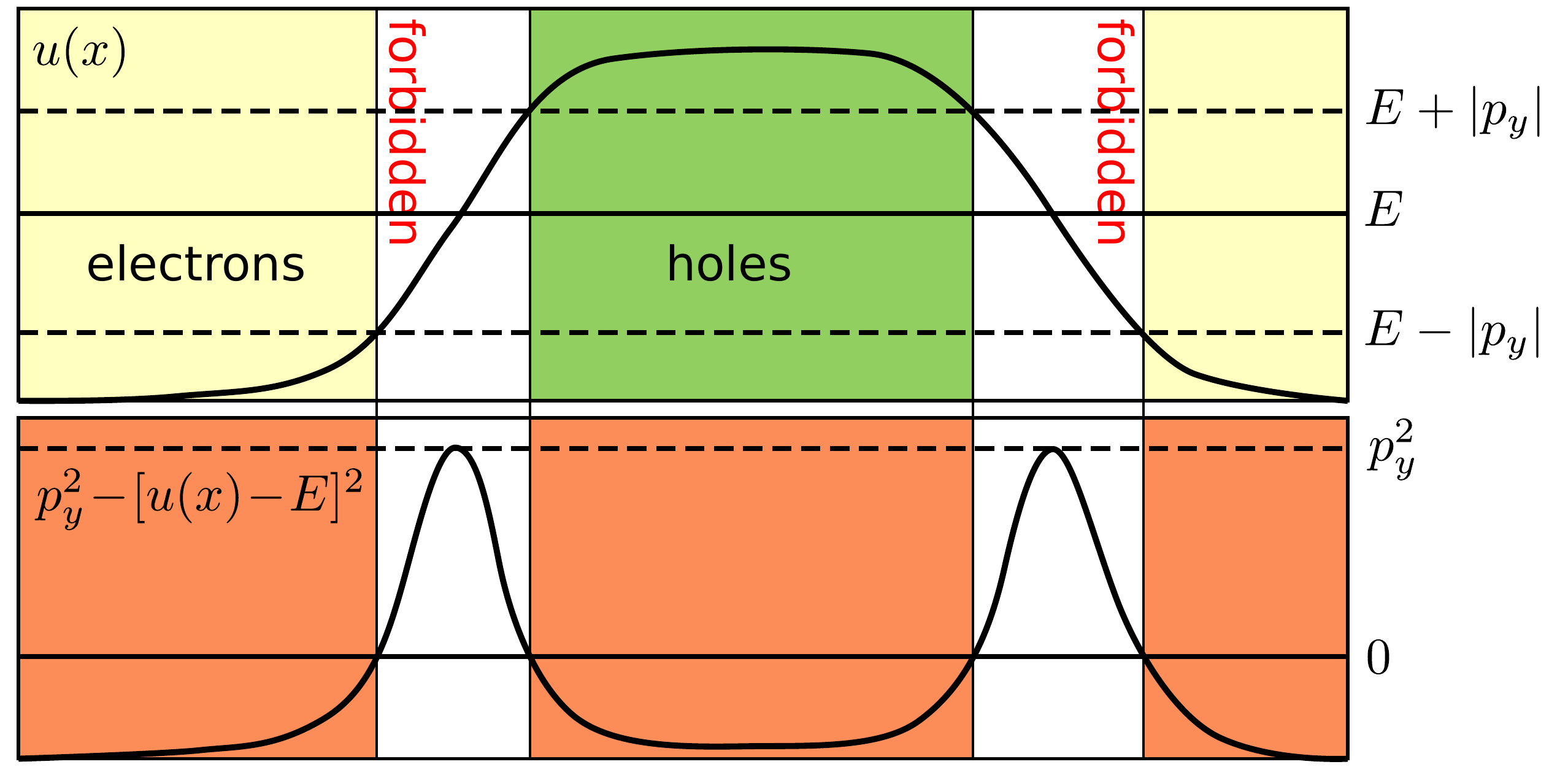}
    \caption{The potential $u(x)$, together with the effective potential $p_y^2-v^2(x)$. There are two classically forbidden regions, separating the classically allowed electron and hole regions.}
    \label{fig:effpot}
  \end{center}
\end{figure}
In figure~\ref{fig:effpot} we show the potential $u(x)$ of the original (Dirac) equation~(\ref{eq:sigma-single}) and the effective (classical) potential $p_y^2-v^2(x)$ in equation~(\ref{eq:reduction}). One sees that there is a classically allowed hole region, separated from the two classically allowed electron regions by two classically forbidden regions. Note that this classically allowed hole region corresponds to the anti-Stokes line in the middle of figure~\ref{fig:Stokes_diagrams}a. We assume that this region is broad enough to use the semiclassical solutions~(\ref{eq:eta1-sol-cmplx}) within it. Therefore we can split the problem of transmission through a barrier supporting hole states into two simpler problems:
\begin{enumerate}
  \item Transmission from the electron region to the hole region; we will call this transmission through an $n$-$p$ junction.
  \item Transmission from the hole region to the electron region; which will be denoted by transmission through a $p$-$n$ junction.
\end{enumerate}

We start our treatment by introducing the transfer matrix, that connects the expansion coefficients $a_{1,2}$ in the electron and hole regions, and relate its elements to the reflection and transmission coefficients for the $n$-$p$ and $p$-$n$ junctions. We proceed by obtaining the reflection and transmission coefficients from the complex WKB method. The formulas we find this way are not valid for near-normal incidence, that is, when the transversal momentum $|p_y|$ is small. Applying the comparison equation technique,~\ref{app:comp-eq}, we finally obtain expressions for the reflection and transmission coefficients that are uniformly valid in the entire range of incidence angles.

\subsection{Transfer matrix}

Let us start by considering an $n$-$p$ junction, with an electron region on the left, and a hole region on the right. In these regions we have the expansions~(\ref{eq:scat-sol}). We introduce coefficients $a^{el}_r$, $a^{el}_l$ and $a^{h}_r$, $a^{h}_l$ corresponding to a right/left moving electron and right/left moving hole respectively.
We then define the transfer matrix $T_{np}$ as the matrix that connects the expansion coefficients $(a_r^{h},a_l^{h})$ and $(a_r^{el},a_l^{el})$,
\begin{equation}
 \left(\begin{array}{c} a_r^{el} \\ a_l^{el}\end{array}\right)=T_{np}
 \left(\begin{array}{c}a_r^{h} \\ a_l^{h}\end{array}\right), \quad T_{np}=\left(\begin{array}{cc}T_{11} & T_{12}\\T_{21} & T_{22}\end{array}\right),
\label{eq:transfer-np-def}
\end{equation}
where the coefficients on the left are defined with respect to the turning point $x_-$, $v(x_-)=-|p_y|$, and the coefficients on the right are defined with reference to the turning point $x_+$, where $v(x_+)=|p_y|$. Now let us express the coefficients of this matrix in terms of the reflection and transmission coefficients.

To determine $T_{np}$ it is enough to know two linear independent solutions. As it was shown in section \ref{sec:scatstates}, one can take $\eta_1$ and $\eta_2^*$. For $\eta_1$ we have
\begin{equation}
 \left(\begin{array}{c}1 \\ r\end{array}\right)=\left(\begin{array}{cc}T_{11} & T_{12}\\T_{21} & T_{22}\end{array}\right)
 \left(\begin{array}{c}t \\ 0\end{array}\right).
\label{eq:transfer-np-eta1}
\end{equation}
Having already constructed $\eta_2$, equation~(\ref{eq:eta2-el}), we obtain
\begin{equation}
 -\left(\begin{array}{c}r^* \\ 1\end{array}\right)=\left(\begin{array}{cc}T_{11} & T_{12}\\T_{21} & T_{22}\end{array}\right)
 \left(\begin{array}{c}0 \\ t^* \end{array}\right),
\label{eq:transfer-np-eta2}
\end{equation}
where the minus sign on the left comes from opposite values of $\nu$ for the electron and hole regions. Since complex conjugation interchanges $e^{iS/h}$ and $e^{-iS/h}$, the coefficients in (\ref{eq:transfer-np-eta2}) are swapped as compared to (\ref{eq:transfer-np-eta1}). 
Solving (\ref{eq:transfer-np-eta1}), (\ref{eq:transfer-np-eta2}), we find that
\begin{equation}
 T_{np}=\left(\begin{array}{cc} 1/t & -r^*/t^* \\ r/t & -1/t^* \end{array}\right) . \label{eq:transfer-np}
\end{equation}

To find transfer matrix $T_{pn}$ for a $p$-$n$ junction,
\begin{equation}
  \left(\begin{array}{c}a_r^{h} \\ a_l^{h}\end{array}\right)=
  T_{pn}\left(\begin{array}{c} a_r^{el} \\ a_l^{el}\end{array}\right), \quad
  T_{pn}=\left(\begin{array}{cc}\tilde T_{11} & \tilde T_{12}\\ \tilde T_{21} & \tilde T_{22}\end{array}\right).
\end{equation}
we exploit the fact that an $n$-$p$ junction for $\eta_1$ is a $p$-$n$ junction for $\eta_2$, a fact that follows from equation~(\ref{eq:reduction}). Therefore two linear independent solutions in this case are $\eta_2$ and $\eta_1^*$.
Starting from the reflection and transmission problem for $\eta_1$, we find the following equation for transfer matrix $T_{pn}$ by considering $\eta_1^*$,
\begin{equation}
 \left(\begin{array}{c}1 \\ r^*\end{array}\right)=\left(\begin{array}{cc}\tilde T_{11} & \tilde T_{12}\\ \tilde T_{21} & \tilde T_{22}\end{array}\right)
 \left(\begin{array}{c}t^* \\ 0 \end{array}\right).
\label{eq:transfer-pn-eta1}
\end{equation}
In obtaining the above result, note that the right-moving hole wave is proportional to $e^{-iS/h}$, and that the electron and hole regions for $\eta_1^*$ are interchanged as compared to $\eta_1$. On the other hand, from the solution given by $\eta_2$ we obtain
\begin{equation}
 -\left(\begin{array}{c}r \\ 1\end{array}\right)=\left(\begin{array}{cc}\tilde T_{11} & \tilde T_{12}\\ \tilde T_{21} & \tilde T_{22}\end{array}\right)
 \left(\begin{array}{c}0 \\ t\end{array}\right).
\label{eq:transfer-pn-eta2}
\end{equation}
From (\ref{eq:transfer-pn-eta1}), (\ref{eq:transfer-pn-eta2}) we obtain the transfer matrix as
\begin{equation}
 T_{pn}=\left(\begin{array}{cc} 1/t^* & -r/t \\ r^*/t^* & -1/t \end{array}\right) .  \label{eq:transfer-pn}
\end{equation}

Formulas~(\ref{eq:transfer-np}) and~(\ref{eq:transfer-pn}) show that once we have the reflection and transmission coefficients for an $n$-$p$ junction, we immediately know the full transfer matrix for both an $n$-$p$ and $p$-$n$ junction. 

Now we come back to our initial problem, transmission through a barrier supporting hole states. The transfer matrix for this problem can be obtained from the two transfer matrices considered before, but one has to take into account that the wave functions in the hole region are defined with respect to different reference points. Therefore we will need the additional matrix
\begin{equation}
 T_{pp}=\left(\begin{array}{cc}e^{i L/h } & 0\\0 & e^{-i L/h}\end{array}\right), \quad L = \int_{x_{1+}}^{x_{2-}} \sqrt{v^2(x)-p_y^2} , \label{eq:def-L}
\end{equation}
where $x_{1+}$ and $x_{2-}$ are the turning points on the left and on the right of the hole region, respectively. To distinguish the two different junctions we also denote
\begin{equation}
T_{pn}=\left(\begin{array}{cc}1/\tilde{t}^* & -\tilde r/\tilde t \\ \tilde r^*/\tilde t^* & -1/\tilde t \end{array}\right) .
\end{equation}
Then we compute the total transfer matrix as
\begin{equation}
 T_{npn}=T_{np}T_{pp}T_{pn} .
\end{equation}

As $x\to\infty$, we have $a_r^{el}=t$, $a_l^{el}=0$ and for $x\to-\infty$ we write  $a_r^{el}=1$, $a_l^{el}=r$. This gives
\begin{equation}
\left(\begin{array}{c}1 \\ r_{npn}\end{array}\right)=
T_{npn}\left(\begin{array}{c}t_{npn} \\ 0\end{array}\right).
\label{eq:tpnp}
\end{equation}
From~(\ref{eq:tpnp}) and the assumption that the transmission coefficient is real, cf.~\cite{Landau77}, we find the total transmission as
\begin{equation}
 t_{npn}=\frac{1}{T_{npn,11}}= \frac{t \tilde t e^{-iL/h} }{1-r^*\tilde r^* e^{-2iL/h}} , \label{eq:FP}
\end{equation}
which is the familiar Fabry-P\'erot formula~\cite{Shytov08,Tudorovskiy12}.

\subsection{Transmission coefficient from the complex WKB method} \label{subsec:contours}

Now let us analyze the problem of reflection and transmission using the theory outlined in section~\ref{sec:WKB}. We consider an $n$-$p$ junction, and assume that the classically forbidden region is broad enough to allow for the use of the semiclassical wave functions between the two turning points, $x_-<x_+$.

On the right-hand side, we start with the transmitted wave,
\begin{equation}
 \eta_1(x)=t \sqrt{\frac{G(x)}{p_x(x)}}\exp\left(-\frac{i}{h}\int_{x_+}^x p_x(x') dx'\right),
 \label{eq:np-eta1-right}
\end{equation}
where $p_x(x) = \sqrt{v^2(x)-p_y^2}$, and
\begin{equation}
 G(x)=\frac{v(x)+\sqrt{v^2(x)-p_y^2}}{|p_y|}.
\end{equation}
We now choose the analytic continuation of the square root such that
\begin{equation}
  (v^2(x)-p_y^2)^{1/2} = \sqrt{v^2(x)-p_y^2}, \quad x>x_+ ,
\end{equation}
so in this region $G(x) = g(x)$, and $\eta_1(x)$ coincides with $\eta_1^-(z)$, as defined in equation~(\ref{eq:eta1-sol-cmplx-g}). Hence the analytic continuation of~(\ref{eq:np-eta1-right}) reads
\begin{equation}
  \eta_1(z)= t \frac{g^{1/2}(z)}{(v^2(z)-p_y^2)^{1/4}}\exp\left(-\frac{i}{h}\int_{x_+}^z (v^2(z')-p_y^2)^{1/2} dz'\right)
\end{equation}

\begin{figure}[tb]
\begin{center}$
\begin{array}{ccc}
\includegraphics[height=3.98cm]{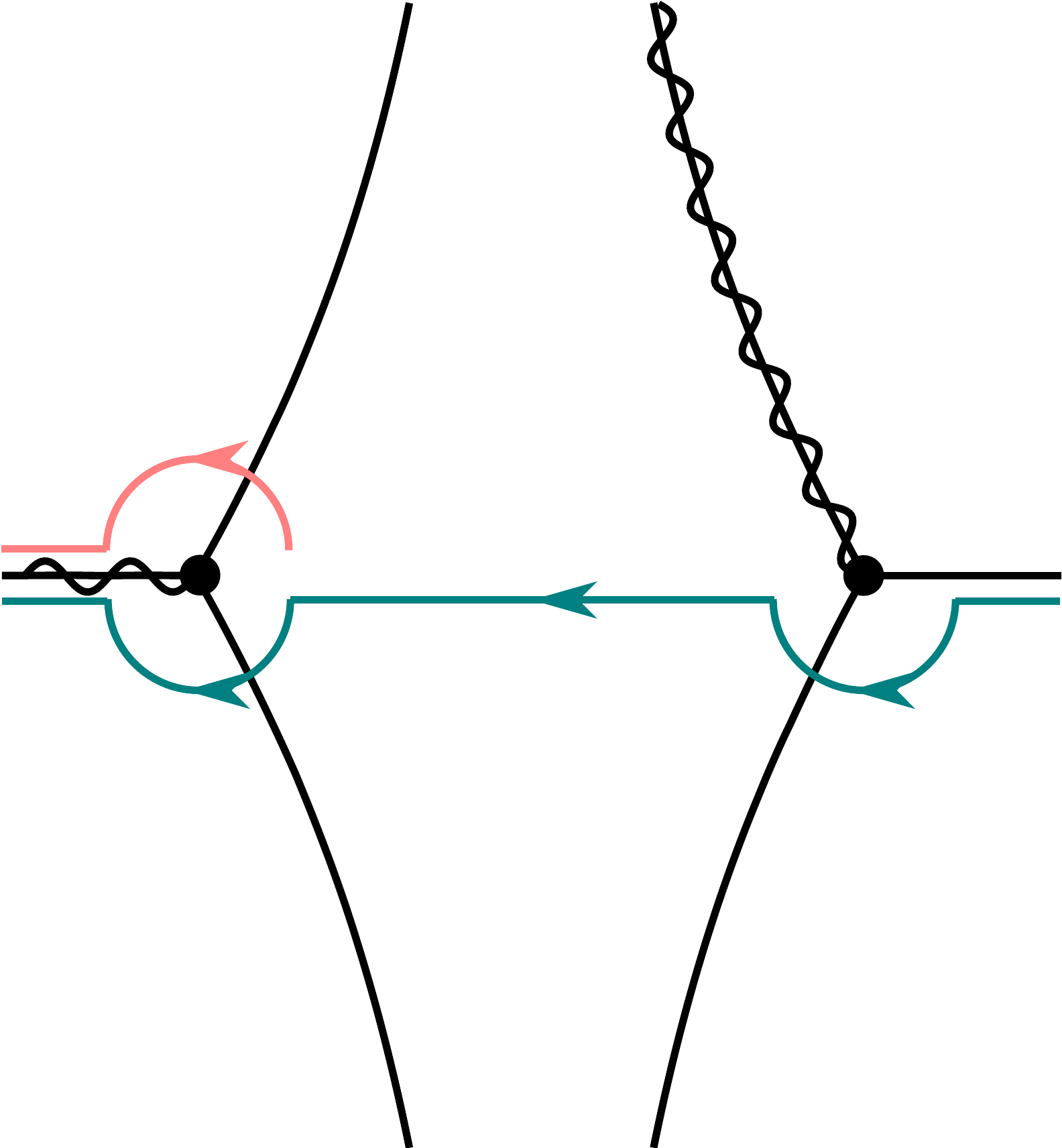} &
\includegraphics[height=3.98cm]{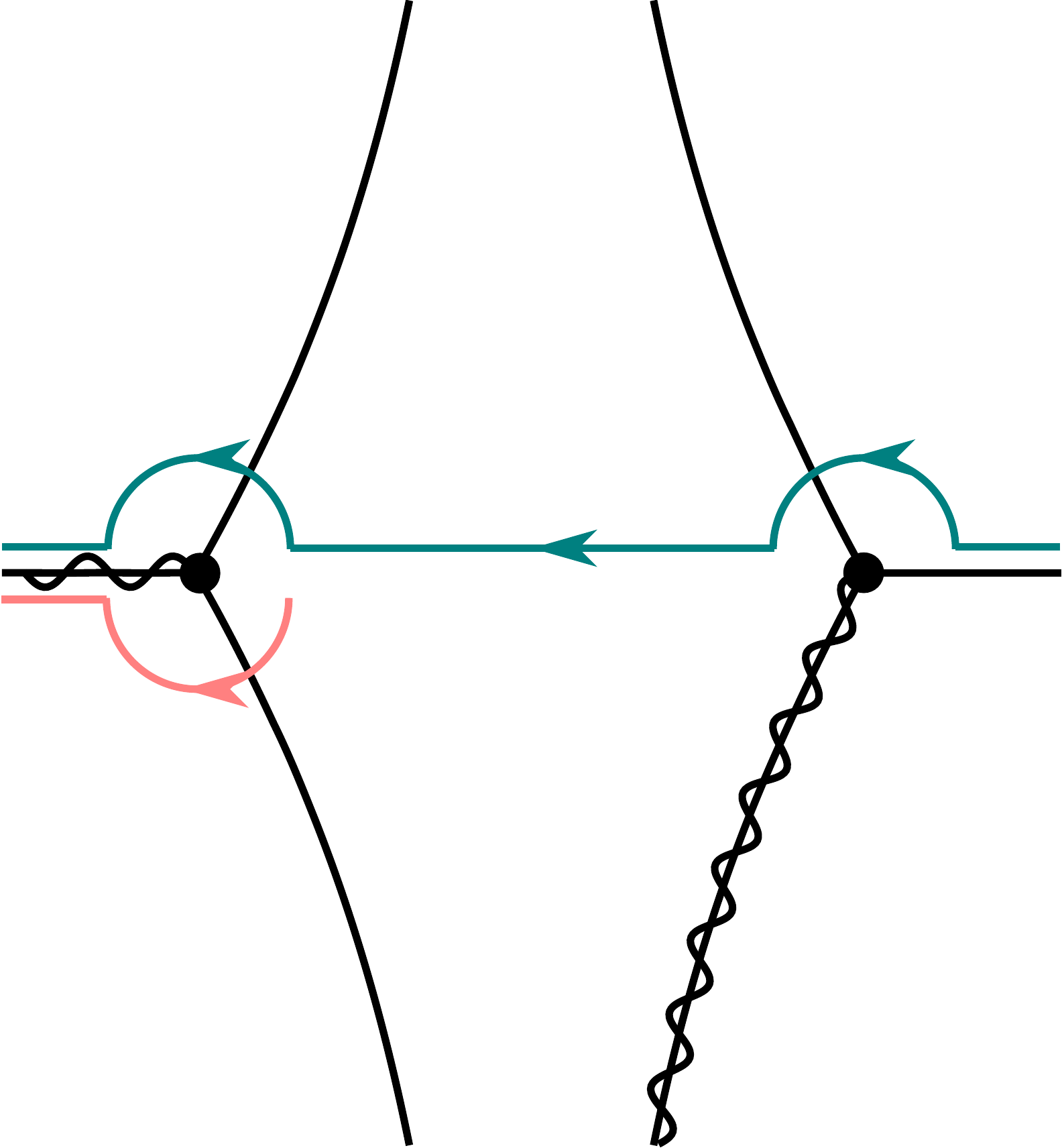} &
\includegraphics[height=3.98cm]{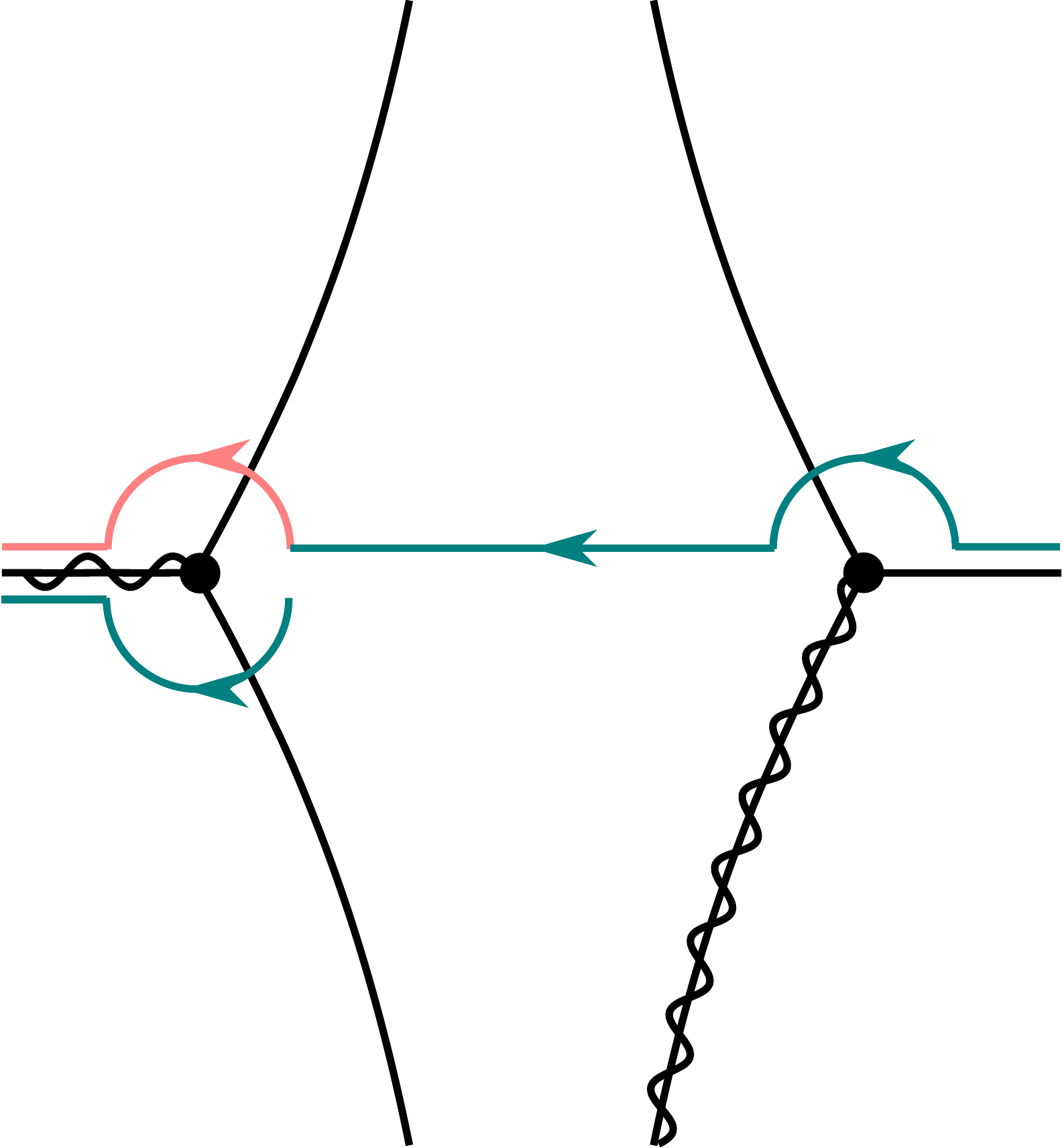} \\
a) & b) & c)
\end{array}$
\end{center}
\caption{Stokes diagrams for a) $n$-$p$ junction, b) $p$-$n$ junction and c) propagation through a single hump potential for a Schr\"odinger particle. Bold points, solid lines and wavy lines depict turning points, anti-Stokes lines and cuts respectively. The colored lines show the paths transferring the transmitted wave on the right of the classically forbidden region to the incoming (blue) and reflected (pink) waves on the left. The qualitative difference between the first two diagrams and the third one, i.e. between Dirac and Schr\"odinger particles, is that to transfer the outgoing wave to the incoming wave, one has to circumvent both turning points in the same direction in the Dirac case, whereas for the Schr\"odinger particle both turning points are circumvented in different directions.}
\label{fig:Landaudiagrams}
\end{figure}

Now we consider the behavior of $\eta_1^+$ along a certain path and compare it to that for the exact solution. The path is chosen by the requirement that the outgoing wave becomes subdominant when we move away from the real axis at positive infinity. Finally the branch cuts are chosen in such a way that the path does not cross them, see figure~\ref{fig:Landaudiagrams}. Following the green path, we first arrive on the anti-Stokes line where the exact solution is still accurately represented by $\eta_1(z)$. According to the arguments from section~\ref{sec:WKB}, this function then becomes dominant between the two turning points, and we can proceed with it. Considering the behavior of $\eta_1(z)$ on both lips of the cut along the negative real axis, we can find the coefficients of the incoming and reflected waves.

For an $n$-$p$ junction, $\eta_1(z)$ becomes subdominant in the lower half-plane, so we choose the contour shown in figure~\ref{fig:Landaudiagrams}\,a). Then we have for $x$ between $x_-$ and $x_+$,
\begin{equation}
  (v^2-p_y^2)^{1/2} = e^{-i\pi/2} \sqrt{p_y^2-v^2(x)} , \quad  x_- < x < x_+ .
\end{equation}
Proceeding along the same contour, we find that on the lower lip of the cut
\begin{equation}
  (v^2-p_y^2)^{1/2} = e^{-i\pi}\sqrt{v^2(x)-p_y^2} , \quad x < x_- .
\end{equation}
Therefore, the process of analytic continuation turns the outgoing hole wave into the incoming electron wave, as can also be seen from figure~\ref{fig:Landaudiagrams}\,a). One sees that both turning points are circumvented in the same direction. This is in stark contrast with the scattering problem for a conventional Schr\"odinger particle: in that case the same path transforms the outgoing wave into the reflected wave, and one has to circumvent both turning points in opposite directions to turn the outgoing wave into the incoming wave, see figure~\ref{fig:Landaudiagrams}\,c).

We just saw that upon analytic continuation along the green contour in figure~\ref{fig:Landaudiagrams}\,a, $p_x(x)$ turns into $-p_x(x)$. However, we see from the definition of $g(z)$ and equation~(\ref{eq:def-G}) that at the same time $G(x)$ turns into $-1/G(x)$, since $v(x)$ changes sign. This means that the ratio $G(x)/p_x(x)$ turns into $1/G(x)p_x(x)$, without an additional sign. To determine its phase on the lower lip of the cut, we can consider the limit $p_y\to 0$. Then in the lower complex half-plane we have \mbox{$(v^2(z)-p_y^2)^{1/2}\to v(z)$} and $g(z)/p_x(z)\to 2/|p_y|$. Hence the process of analytic continuation gives the incoming wave on the left as
\begin{equation}
  \eta_1(x)=\frac{t}{\sqrt{G(x)p_x(x)}} \exp\left( \frac{i}{h}\int_{x_+}^x p_x(x') dx'\right) , \label{eq:eta1in}
\end{equation}
which can be rewritten as
\begin{equation}
 \eta_1(x)=\frac{t\exp\left(K/h\right)}{\sqrt{p_x(x)G(x)}} \exp\left(\frac{i}{h}\int_{x_-}^x p_x(x') dx'\right)  ,
 \label{eq:eta1in-2}
\end{equation}
where
\begin{equation}
 K=\int_{x_-}^{x_+} \sqrt{p_y^2 - v^2(x)} dx. \label{eq:def-K}
\end{equation}
Since the amplitude of the incoming wave should equal one, we find the transmission coefficient as
\begin{equation}
 t=e^{-K/h}.  \label{eq:t}
\end{equation}

\begin{figure}[t]
  \begin{center}
    \includegraphics[height=4cm]{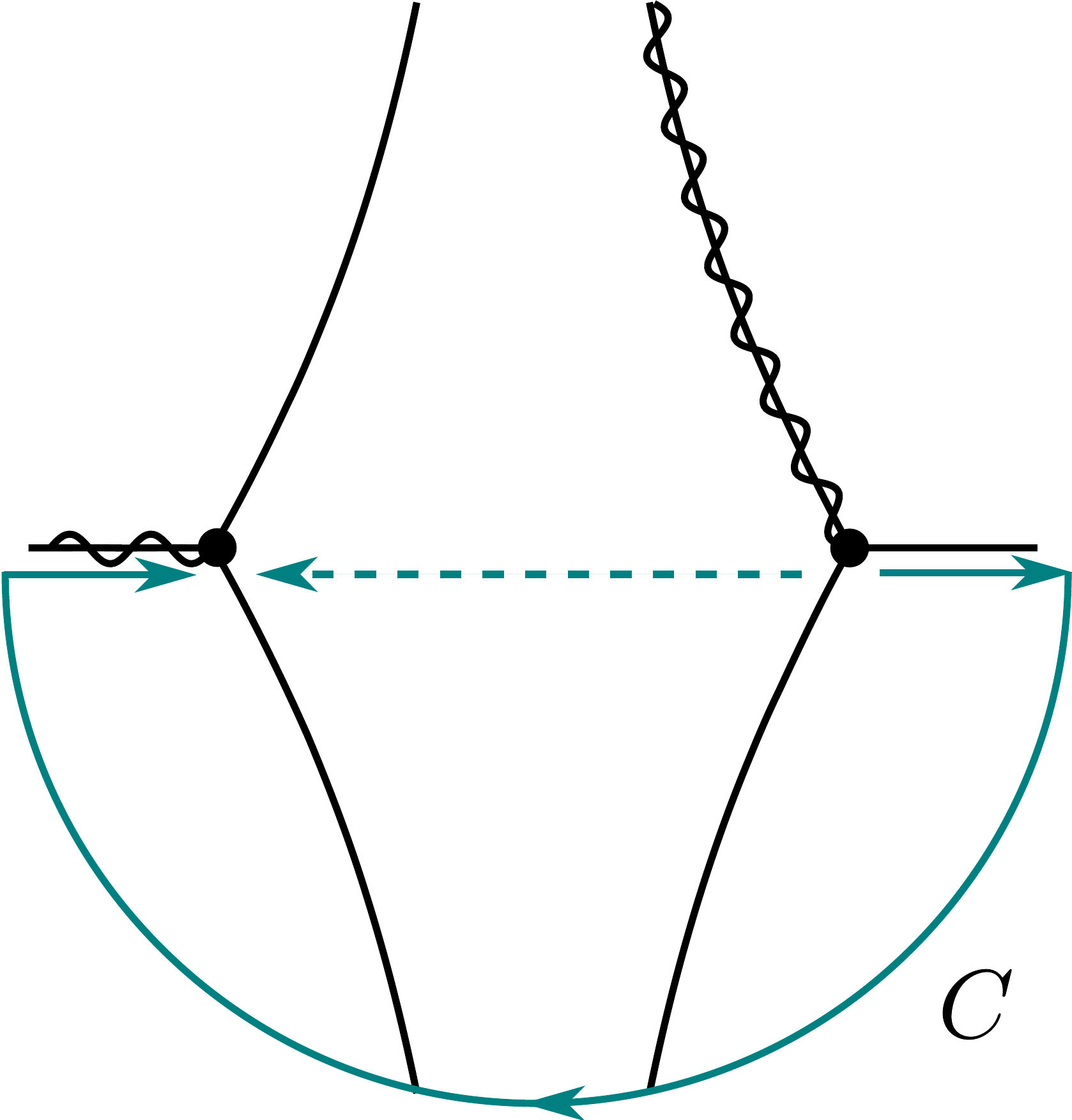}
    \caption{The Stokes diagram for an $n$-$p$ junction. The contour $C$ shows the path along which the integration should be performed in (\ref{eq:tc}). The dashed line shows the equivalent path.}
    \label{fig:pn-uni}
  \end{center}
\end{figure}

Following the reasoning in~\cite{Kemble35,Landau77}, we note that equality~(\ref{eq:t}) holds for any distance between the two turning points. To show this, we note that instead of the contour shown in figure~\ref{fig:Landaudiagrams}\,a, one can take an arbitrarily large half-circle in the lower complex plane. As a result, the integration in equation~(\ref{eq:eta1in}) should be performed along the contour $C$ shown in figure~\ref{fig:pn-uni}, giving rise to the transmission coefficient
\begin{equation}
 t=\exp\left(-\frac{i}{h} \int_{C} p_x(z)dz\right),
\label{eq:tc}
\end{equation}
Once we have obtained equation~(\ref{eq:tc}), we can deform the contour in the complex plane to reduce this result to equations~(\ref{eq:t}) and (\ref{eq:def-K}), see figure \ref{fig:pn-uni}.

In accordance with the theory from the previous section, we have to take the pink contour in figure~\ref{fig:Landaudiagrams}\,a) to reconstruct the reflected wave. Since we now go around the left turning point in the counterclockwise direction, we find that on the \emph{upper} lip of the cut
\begin{equation}
  (v^2-p_y^2)^{1/2} = \sqrt{v^2(x)-p_y^2} , \quad x < x_- .
\end{equation}
Studying the analytic continuation of $G(x)$, we see that it turns into $-G(x)$ on this lip of the cut. Therefore the reflected wave is given by
\begin{equation}
 \eta_1(x)= t \exp\left(K/h\right) \left(-\frac{G(x)}{p_x(x)}\right)^{1/2} \exp\left(-\frac{i}{h}\int_{x_-}^x p_x(x)dx\right) . \label{eq:eta1r}
\end{equation}
Since we now have a negative number under the square root, we have to determine its phase. To this end we consider how the incoming wave is transformed into the reflected wave along the circle circumventing the left turning point. Since the direction of this path is counterclockwise, $p_x(x)$ on the lower lip of the cut turns into $e^{i\pi} p_x(x)$ on the upper lip of the cut. Similarly, $1/G(x)$ is transformed into $G(x)$ on the upper lip of the cut. Therefore the reflected wave becomes
\begin{equation}
 \eta_1(x)= -i t \exp\left(K/h\right) \sqrt{\frac{G(x)}{p_x(x)}} \exp\left(-\frac{i}{h}\int_{x_-}^x p_x(x)dx\right) , \label{eq:eta1r2}
\end{equation}
and with the help of equation~(\ref{eq:t}) we conclude that the reflection coefficient equals
\begin{equation}
r=-i.
\label{eq:refl-i}
\end{equation}
The modulus of the reflection coefficient can be refined from (\ref{eq:conservation}). We obtain
\begin{equation}
|r|=\sqrt{1-|t|^2}=\sqrt{1-e^{-2K/h}}, \label{eq:refl-i-sqrt-notheta}
\end{equation}
which gives the uniform approximation for the modulus of the reflection coefficient. Combining these two equations, $r$ can be written as
\begin{equation}
 r=e^{-i\pi/2+i\theta}\sqrt{1-e^{-2K/h}},
\label{eq:refl-i-sqrt}
\end{equation}
where $\theta$ is an additional phase factor. From our previous considerations, we conclude that $\theta$ is small whenever both turning points are substantially separated, i.e. when $|p_y|$ is not too small. The method we used until now fails to give a uniform approximation for the phase $\theta$. This is related to the fact that any contour that transforms the outgoing wave into the reflected wave passes through the region between the turning points, if we do not want it to cross the cut. When both turning points become nearly degenerate, $\eta_1(z)$ in this region ceases to be semiclassical. In the next subsection we will use the method of comparison equations to obtain a uniform approximation for the phase $\theta$.

With the results~(\ref{eq:t}) and~(\ref{eq:refl-i-sqrt}), we can write down the full transfer matrix for an $n$-$p$ junction using equation~(\ref{eq:transfer-np}). The determinant of the transfer matrix equals
\begin{equation}
  \det T_{np} = \frac{|r|^2-1}{|t|^2} .
\end{equation}
The approximation~(\ref{eq:refl-i}) gives $\det T_{np} = 0$, so that $T_{np}$ cannot be inverted. Therefore the transfer matrix obtained by using $r=-i$ is \emph{unidirectional}. The unidirectionality closely relates to the fact that we neglected the subdominant solution between the two turning points. Using expression~(\ref{eq:refl-i-sqrt}) for the reflection coefficient, one obtains $\det T_{np}=-1$. Thus, the transfer matrix obtained with the use of equality~(\ref{eq:refl-i-sqrt}) can be inverted. It is therefore~\emph{bidirectional}.

Using equation~(\ref{eq:transfer-pn}), one immediately finds the transfer matrix for a $p$-$n$ junction. Alternatively, one can derive the same result using the contour shown in figure~\ref{fig:Landaudiagrams}\,b). In~\ref{app:WKB}, we show how to derive these matrices using a different formulation of the complex WKB method.

\subsection{An $n$-$p$-$n$ junction and the method of comparison equations}

To obtain the transmission coefficient for a full $n$-$p$-$n$ junction, we now use equation~(\ref{eq:FP}). Inserting the WKB result~(\ref{eq:refl-i}), we obtain a transmission coefficient that diverges at the transversal momenta satisfying the semiclassical quantization condition
\begin{equation}
  \frac{1}{h} \int_{x_{1+}}^{x_{2-}} d x' \, \sqrt{v^2(x') -p_y^2} = \pi\left(n+\frac{1}{2}\right) ,
\end{equation}
where $x_{1+}$ and $x_{2-}$ are the left and right turning points at the border of the hole region, respectively. This divergence is due to the one-directional nature of the transfer matrix, as explained in the previous subsection. If we use (\ref{eq:refl-i-sqrt}) instead of (\ref{eq:refl-i}) for the reflection coefficient and (\ref{eq:t}) for the transmission coefficient, we obtain 
\begin{equation}
  t_{npn} = \frac{e^{-K_{np}/h}e^{-K_{pn}/h} e^{-i L/h}}{1 - \sqrt{1 - e^{-2 K_{np}/h}}\sqrt{1 - e^{-2 K_{pn}/h}} e^{-2 i L/h + i\pi-i\theta_{np}-i\theta_{pn}}} ,
  \label{eq:transmission-npn-WKB-FP}
\end{equation}
where the quantity $L$ is the classical action in the hole region, given by equation~(\ref{eq:def-L}), $K_{np}$ and $K_{pn}$ are the action integrals in the classically forbidden regions for the $n$-$p$ and $p$-$n$ junction respectively, and are given by equation~(\ref{eq:def-K}). Finally, $\theta_{np}$ and $\theta_{pn}$ are the phases of the reflection coefficients (\ref{eq:refl-i-sqrt}). For angular scattering, formula (\ref{eq:transmission-npn-WKB-FP}) gives a rather good result for the transmission coefficient even if we put $\theta=0$, as we show numerically in section~\ref{sec:numerics}, see also~\cite{Tudorovskiy12}. However for nearly normal incidence this result is no longer accurate. Thus the final step in the construction of the uniform approximation is to find phase $\theta$ in (\ref{eq:refl-i-sqrt}).

In order to obtain a uniform approximation for the reflection coefficient, we use the method of comparison equations. In~\ref{subsec:comp-eq-nppn}, we explain how to map the potential of a general $n$-$p$ or $p$-$n$ junction to a quadratic potential. The latter can be solved explicitly, and with the help of the mapping we can then construct an approximate solution of the original equation.
We find that $\theta$ is given by
\begin{equation}
  \theta = \mathrm{Arg} \left[ \Gamma \left( 1 + i \frac{K}{\pi h}  \right) \right] - \frac{\pi}{4} + \frac{K}{\pi h} - \frac{K}{\pi h} \ln\left(\frac{K}{\pi h}\right) . \label{eq:theta-uniform}
\end{equation}
Using the asymptotic expansion of the $\Gamma$-function~\cite{Abramowitz65,WolframFunctions}, one easily finds that $\theta\to 0$ when $K/h$ is large, which agrees with the result of the previous section. Equation~(\ref{eq:theta-uniform}) was already anticipated in~\cite{Tudorovskiy12}, where it was obtained by explicitly solving the case of a linear potential, and then replacing the action between the two turning points by $K$. In~\ref{subsec:comp-eq-nppn} we now give a rigorous proof for this result. Further discussion of equation~(\ref{eq:transmission-npn-WKB-FP}) is postponed until section~\ref{sec:numerics}.

\section{Semiclassical treatment of above-barrier transmission} \label{sec:above-barr}

In this section we consider the second regime from section~\ref{sec:regimes}, namely above-barrier scattering. This regime can be split in two cases, \textit{i)} scattering above a potential hump, and \textit{ii)} scattering above a monotonous finite range potential.

The first case describes for instance finite-range gating in graphene. The Stokes diagram for this potential is shown in figure~\ref{fig:Stokes_diagrams}\,b). One sees that there are four turning points, two in the upper half-plane and two in the lower half-plane. In what follows we will assume that the potential tends to a constant at $|x|\to\infty$. An example of such a potential is $u(x)=\exp(-x^2)$. The Stokes diagram corresponding to this potential has infinitely many turning points. The approximation we made in section~\ref{sec:WKB} is that only the four turning points closest to the real axis should be taken into account, while the others can be neglected.

The second case is a monotonous finite range potential, that is, a finite increase of the potential. It corresponds to a single $n$-$n$ junction, which can be used to model a transition between two macroscopically wide areas with different gates applied. The $n$-$n$ junction can be simulated by the potential $u(x)=\tanh(x)$. It is important to note that a finite increase cannot be captured by a finite polynomial, or a rational polynomial function. Therefore, the Stokes diagram corresponding to this potential differs substantially from that of case i), and should be considered separately. In section~\ref{sec:tanh}, we will construct the exact solution for this case, and also give a semiclassical treatment.

Let us return to the Stokes diagram from figure~\ref{fig:Stokes_diagrams}\,b). In the first approximation, we can neglect the two outermost turning points $z_{2\pm}$, that correspond to $v(z_{\pm})=|p_y|$, and consider only those in the middle, namely $z_{1\pm}$, that correspond to $v(z_{1\pm})=-|p_y|$. The Stokes diagram we obtain in this way coincides with the one for a usual Schr\"odinger equation. The reflection coefficient for this case was first obtained in~\cite{Pokrovskii58a,Pokrovskii58b}, using a perturbation expansion. It was shown that the reflection coeffficient is exponentially small, in agreement with classical mechanics, where the above-barrier transmission is always equal to unity. We will use a different approach, that can be found in~\cite{Pokrovskii61,Heading62,Froeman65,Landau77,Pokrovskiiweb}. In our derivation we will implicitly assume that the energy is comparable to the height of the potential hump and will not consider the transition to the Born approximation, i.e. to the case $E\to \infty$, which was studied in~\cite{Pokrovskii61}.

\begin{figure}[ht]
\begin{center}
\includegraphics[height=4cm]{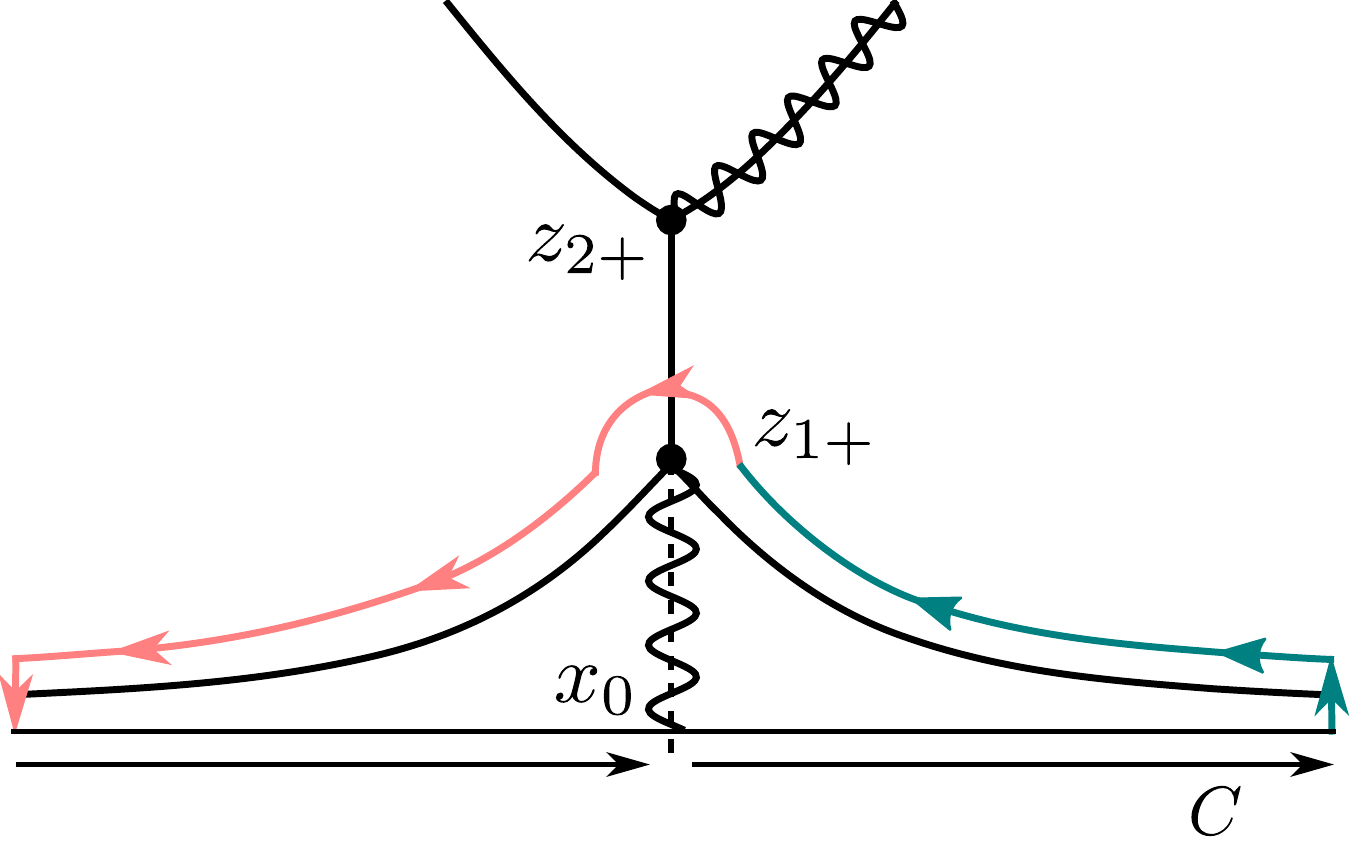}
\end{center}
\caption{The upper half of the Stokes diagram for scattering above a short-range potential. The solid lines depict anti-Stokes lines, while the dashed line depicts a Stokes line, and the wavy line depicts a cut. The path in the complex plane is shown by the green and pink contour, and transforms the transmitted wave into the reflected wave. The integration contour $C$ connects $x_0$ on the right lip of the cut with the same point on the left lip of the cut and consists of the two black lines, the green line and the pink line.}
\label{fig:Stokes-above-hump}
\end{figure}

Let us first turn to the definition of the scattering states. When calculating the current~(\ref{eq:el-current}), we assumed that the action $s(z_0,x)$ was purely real. However, in the case of above-barrier scattering the turning point $z_0$ is complex, which means that we cannot take it as the lower limit of integration in the action, since it adds a complex part. Therefore, we take this lower limit to be a point on the real axis. More specifically, we introduce so-called Stokes lines by the requirement that $\textrm{Re}[s(z_0,z)]$ is zero, which implies that the action is purely imaginary. It turns out that the middle two turning points are connected by such a Stokes line, which therefore crosses the real axis, see figure~\ref{fig:Stokes-above-hump}. We call the point where this line crosses the real axis $x_0$, and take it as the reference point for the action.

Let us now choose a convenient path in the complex plane. Starting at positive infinity, we first make a transition to the anti-Stokes line above or below the real axis. If the potential tends to a constant at plus (and minus) infinity, this transition does not change the expansion coefficients~\cite{Pokrovskiiweb}, since the asymptotic solutions become exact and the distance between the real axis and the anti-Stokes line remains finite.

To decide whether we should take a path in the upper or in the lower half plane, we start from the expected result. With respect to the reference point $x_0$, we have an incoming wave with unit amplitude, and we expect an exponentially small reflection. When changing the reference point to the upper turning point~$z_{1+}$, the coefficient in front of the incoming wave is multiplied by an exponentially small factor, and the coefficient in front of the reflected wave is multiplied by an exponentially large factor, in such a way that they are of the same order of magnitude at the upper anti-Stokes line. This condition is necessary to pass between two anti-Stokes lines that emanate from the same turning point, see section~\ref{sec:WKB}. It turns out that this condition is not satisfied in the lower half plane.

After these preliminaries, we can follow the procedure that was used for Schr\"odinger equations~\cite{Pokrovskii61,Heading62,Froeman65,Landau77,Pokrovskiiweb}.
We start with a transmitted wave at positive infinity,
\begin{equation}
  \eta_1(x)= t \sqrt{\frac{1}{p_x(x) G(x)}} \exp\left(\frac{i}{h}\int_{x_0^+}^x \sqrt{v(x')^2-p_y^2} dx'\right) ,
  \label{eq:out-above}
\end{equation}
where $p_x(x) = \sqrt{v^2(x)-p_y^2}$ and by $x_0^\pm$ we denote the point $x_0\pm\varepsilon$, where $\varepsilon\to 0$.\footnote{Note that in the terminology of section~\ref{sec:WKB}, we would say that the point $x_0^+$ is on the left lip of the cut, which is somewhat counterintuitive.} We choose the analytic continuation of the square root as
\begin{align}
  (v^2(x)-p_y^2)^{1/2} &= \sqrt{v^2(x)-p_y^2} , & x>x_0 , \\
  (v^2(x)-p_y^2)^{1/2} &= e^{i\pi}\sqrt{v^2(x)-p_y^2} , & x<x_0 .
\end{align}
which implies that $\eta_1(x)$ coincides with $\eta_1^+(z)$ along the positive real axis. Let us consider $\eta_1^+(z)$ along the path shown in figure~\ref{fig:Stokes-above-hump}. It becomes subdominant above the anti-Stokes line on the right. Therefore on the anti-Stokes line between $z_{1+}$ and $z_{2+}$, the function $\eta_1^+(z)$ correctly reproduces the behavior of $\eta(z)$, see section~\ref{sec:WKB}. When we continue along the contour, the transmitted wave is transformed into the reflected wave. Following the complex continuation, we see that $p_x(x)$ is transformed into $e^{i\pi} p_x(x)$, and that $1/G(x)$ is transformed into $G(x)$. Therefore, at negative infinity, we end up with
\begin{multline}
  \eta_1(x)= -i t \sqrt{\frac{G(x)}{p_x(x)}} \exp\left(\frac{i}{h}\oint_C (v^2(z)-p_y^2)^{1/2} dz \right) \\ \times \exp\left( -\frac{i}{h} \int_{x_0^-}^x \sqrt{v(x')^2-p_y^2} dx'\right) , \label{eq:eta1-above-rt-rel}
\end{multline}
Note that the integral in the second exponent is to be performed from $x_0^-$. The first integral is associated with the change of reference point; it goes along a contour $C$ that connects $x_0^+$ with $x_0^-$. Since the square root has opposite signs on opposite lips of the cut, it can be rewritten as
\begin{equation}
  \exp\left(\frac{i}{h}\oint_C (v^2(z)-p_y^2)^{1/2} dz \right)=\exp\left(\frac{2i}{h}\int_{x_0^+}^{z_{1+}} (v^2(z)-p_y^2)^{1/2} dz \right) .
\end{equation}
Approximating the transmission coefficient by one, we find that the reflection coefficient is given by
\begin{equation}
  r = -i e^{K/h} , \quad K = 2i \int_{x_0^+}^{z_{1+}} (v(z)^2-p_y^2)^{1/2} dz < 0 . \label{eq:r-above-hump-WKB}
\end{equation}
The fact that $K$ is a negative real number can be seen by performing the calculation for the prototype potential $v^2-p_y^2 = z^2+a^2$, see also~\cite{Heading62,Froeman65}.

Approximation~(\ref{eq:r-above-hump-WKB}) does not hold when $|p_y|$ is close to $E-u_0$, since in this case the middle two turning points are close together. We can get a more accurate prediction for the modulus of the reflection coefficient by considering the current conservation~(\ref{eq:conservation}). In fact, equation~(\ref{eq:eta1-above-rt-rel}) is equivalent to
\begin{equation}
  r = -i t e^{K/h} . \label{eq:r-t-Schrod}
\end{equation}
Combining this with $|r|^2+|t|^2=1$, we find that 
\begin{equation}
  |t| = \frac{1}{\sqrt{1+e^{2K/h}}} . \label{eq:t-above-WKB-cont}
\end{equation}
Note that we can now deform the contour into a large semi-circle, and as long as the upper-most turning point $z_{2+}$ does not come into play, the above derivation still holds. Therefore we conclude that equation~(\ref{eq:t-above-WKB-cont}) holds regardless of the distance between the middle two turning points. The same result is obtained by applying the method of comparison equations, see~\ref{subsec:comp-eq-conv}, from which we also find the correct phases,
\begin{equation}
  t = \frac{e^{i\phi}}{\sqrt{1+e^{2K/h}}}, \quad r=-i \frac{e^{i \phi}e^{K/h}}{\sqrt{1+e^{2K/h}}} , \label{eq:rt-above-middle}
\end{equation}
where $K$ was defined in equation~(\ref{eq:r-above-hump-WKB}), and
\begin{equation}
  \phi = \mathrm{Arg}\left[ \Gamma\left( \frac{1}{2} + \frac{i K}{\pi h} \right) \right] + \frac{K}{\pi h} - \frac{K}{\pi h} \ln\left( \frac{|K|}{\pi h} \right) . \label{eq:def-phi}
\end{equation}
All the above results coincide with the results for an ordinary Schr\"odinger equation~\cite{Froeman02}.

\begin{figure}[tb]
\begin{center}
\includegraphics[height=4cm]{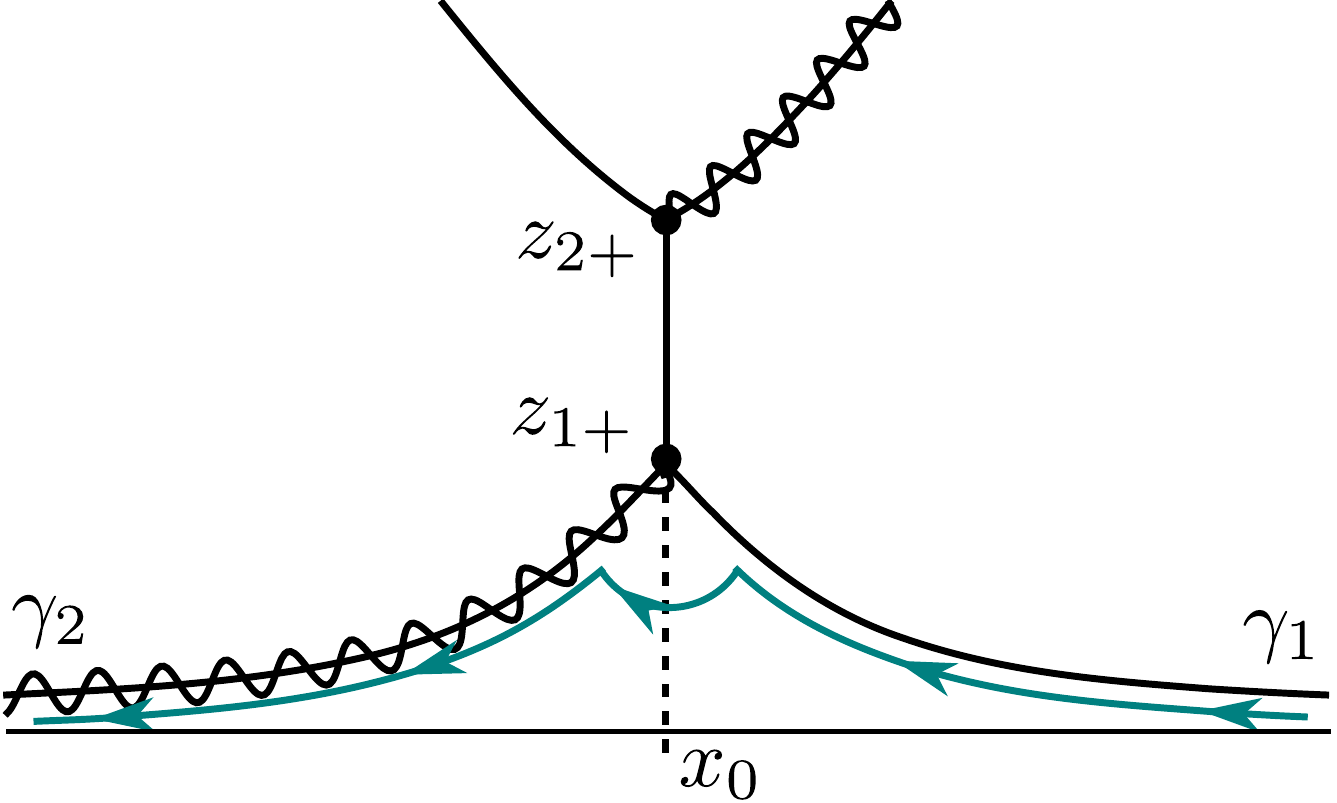}
\end{center}
\caption{The upper half of the Stokes diagram for scattering above a short-range potential. The solid lines depict anti-Stokes lines, two of which are labelled by $\gamma_{1,2}$, while the dashed line depicts a Stokes line, and the wavy line depicts a cut. The green line shows the path taken in the complex plane.}
\label{fig:Stokes-above-hump-contour2}
\end{figure}

Since the above results are identical to those for an ordinary Schr\"odinger equation, they fail to explain the total transmission at normal incidence. Indeed, when the transversal momentum $|p_y|$ becomes small, that is, for near-normal incidence, the upper two turning points come close together (as do the lower two) and therefore they have to be treated as a cluster. As long as $E-u_0$ is sufficiently large, the two clusters can be treated separately. To derive the reflection coefficient, we now position the branch cut differently, and use the contour shown in figure~\ref{fig:Stokes-above-hump-contour2}.

To find the reflection coefficient in this approximation, we start with the outgoing wave~(\ref{eq:out-above}) with $t$ equal to one. Then we change the reference point, i.e. the lower limit of the integral, to $z_{1+}$ and obtain
\begin{equation}
  \eta_1(z) = e^{K/2h} \frac{(-g)^{-1/2}(z)}{p_x^{1/2}(z)}e^{i s(z_{1+},z)/h} . \label{eq:above-2-tra}
\end{equation}
In~\ref{subsec:comp-eq-aboveup} we use the method of comparison equations, to show that upon passing from the anti-Stokes line $\gamma_1$ to $\gamma_2$, the accurate representation of the exact solution becomes
\begin{equation}
  \eta_1(z) = e^{K/2h} \frac{(-g)^{-1/2}(z)}{p_x^{1/2}(z)}e^{i s(z_{1+},z)/h} - i a e^{K/2h} \frac{(-g)^{1/2}(z)}{p_x^{1/2}(z)}e^{-i s(z_{1+},z)/h}, \label{eq:above-2-ref-in}
\end{equation}
where
\begin{equation}
  a = \frac{1}{\Gamma(-S/\pi h)} \sqrt{\frac{2\pi}{-S/\pi h}} e^{-(S/\pi h) \ln(-S/\pi h) - (-S/\pi h)} , \label{eq:a-const-asymp/gamma}
\end{equation}
and 
\begin{equation}
  S = \int_{z_{1+}}^{z_{2+}} (v^2(z)-p_y^2)^{1/2} d z < 0 . \label{eq:def-S}
\end{equation}
Changing the reference point back to $x_0$, one sees that the first term in equation~(\ref{eq:above-2-ref-in}) is the incoming wave normalized by one. The second term is the reflected wave, and the reflection coefficient equals
\begin{equation}
  r = -i a e^{K/h} . \label{eq:r-t-upper}
\end{equation}
It is readily seen from the Stirling approximation for the $\Gamma$-function~\cite{Abramowitz65,WolframFunctions} that $a$ approaches one when $-S/\pi h$ becomes large. Upon normal incidence, the upper two turning points merge and $S$ vanishes. One can check that $a$ vanishes in this case, and therefore equation~(\ref{eq:r-t-upper}) correctly predicts total transmission at normal incidence. We see that similar to the case of conventional Klein tunneling, or scattering in the presence of hole states, total transmission comes from the merging of two turning points. However, this time they do not merge on the real axis, but in the complex plane.

Equation~(\ref{eq:r-t-upper}) was derived under the assumption that the upper two turning points are close together, whereas equation~(\ref{eq:rt-above-middle}) was derived under the assumption that the middle two turning points are close together. In the intermediate regime, we can combine the two expressions into
\begin{equation}
  |r| = \frac{a e^{K/h}}{\sqrt{1+e^{2K/h}}} ,
  \label{eq:r-above-guess}
\end{equation}
which shows the correct behavior for normal incidence. The transmission coefficient can be derived from this from the current conservation $|r|^2+|t|^2=1$. In section~\ref{sec:numerics} we show that, surprisingly, equation~(\ref{eq:r-above-guess}) is in better agreement with our numerical result than equations~(\ref{eq:rt-above-middle}) and~(\ref{eq:r-t-upper}).

\section{Tunneling through a barrier without hole states} \label{sec:barr-noholes}

Now let us consider the third regime from section~\ref{sec:regimes}, the conventional tunneling regime. We consider a short-range potential, for which the Stokes diagram is shown in figure~\ref{fig:regimes}\,c). Two of the four turning points are real, and the other two are imaginary. In the previous section, we saw that imaginary turning points give rise to exponentially small reflections, so we will start by neglecting their influence.

To relate the transmission coefficient to the reflection coefficient, we use the contour shown in figure~\ref{fig:Landaudiagrams}\,c. Following the green and pink contours in a way similar to section~\ref{sec:overdense}, we find
\begin{equation}
  r = -i , \quad t = e^{-K/h} , \quad K = \frac{1}{h} \int_{x_-}^{x_+} d x \, \sqrt{p_y^2-v^2(x)} > 0 .  \label{eq:transmission-nn-WKB}
\end{equation}
Note that in this case the analytical continuation of $g(z)$, see equation~(\ref{eq:eta1-sol-cmplx-g}), does not give rise to an additional phase factor. A second way to see this is by realizing that both turning points $x_-$ and $x_+$ correspond to $v(x_\pm) = -|p_y|$. Considering the additional amplitude factor in equation~(\ref{eq:eta1-sol-cmplx}), and integrating it both turning points, we find
\begin{equation}
  \exp\left[ - i \int_{x_-}^{x_+} \frac{v'(x) d x }{\sqrt{p_y^2-v(x)^2}}  \right] = 1 .
\end{equation}
Therefore, the result is essentially the same as that for an ordinary Schr\"odinger equation, and we call this regime the conventional tunneling regime.

We note that equation~(\ref{eq:transmission-nn-WKB}) does not hold when the two turning points come close together. However, by similar reasoning as in section~\ref{sec:overdense}, see also~\cite{Landau77}, we can conclude that the relation
\begin{equation}
  r = -i t e^{K/h} , \label{eq:r-t-rel-Schrod}
\end{equation}
does hold regardless of the distance between the two turning points, as long as there are no other turning points close to the contour in the complex plane. Combining equation~(\ref{eq:r-t-rel-Schrod}) with the current conservation $|r|^2+|t|^2=1$, we can obtain the modulus of the transmission coefficient. In~\ref{subsec:comp-eq-conv}, we outline how we can use the method of comparison equations to reconstruct the correct phases, with the result
\begin{equation}
  t = \frac{e^{i\phi}}{\sqrt{e^{2K/h}+1}}, \quad r=-i \frac{e^{i \phi} e^{K/h}}{\sqrt{e^{2K/h}+1}} , \label{eq:rt-conventional-compeq}
\end{equation}
where $\phi$ is defined by equation~(\ref{eq:def-phi}). This result coincides with the one derived for above-barrier scattering, equation~(\ref{eq:rt-above-middle}), except for the fact that $K$ is now positive instead of negative. The same result is found for an ordinary Schr\"odinger equation~\cite{Froeman02}.

In section~\ref{sec:numerics}, we compare our result to numerical calculations, and see that the agreement is reasonable. The discrepancy is due to the influence of the other two turning points, that were neglected in the above treatment. Finally, we note that equation~(\ref{eq:transmission-nn-WKB}) also holds for cases when there are more than two complex turning points, as long as these are not too close to the real axis, and the turning points on the real axis are not too close together.

\section{The exactly solvable model of the monotonous finite range potential} \label{sec:tanh}

Up to now we have considered scattering of massless Dirac fermions by a potential hump.
Implicitly, we assumed that we only have to keep four turning points in the Stokes diagram to find the major contribution to the scattering. This assumption naturally led to Stokes diagrams topologically equivalent to those for a parabolic potential, $u(x)=-x^2$. 
In contrast to a potential hump, a monotonous finite range potential (finite increase) cannot be modeled by a polynomial function. This leads to a topologically different Stokes diagram. As an example let us consider an exactly solvable model for a finite range potential, provided by the function
\begin{equation}
  u(x) = \frac{u_0}{2} (1+\tanh(x)). \label{eq:tanh-pot}
\end{equation}
The exact solution for this potential was constructed in~\cite{Miserev12}. Here we present a slightly different approach, following the general method outlined in~\cite{Landau77,Morse53}. Similar techniques was employed in~\cite{Hartmann10}, where the eigenvalue problem for a potential well $u(x) = -1/\cosh(x)$ was solved.

Inserting the potential~(\ref{eq:tanh-pot}) into equation~(\ref{eq:reduction}), we find the differential equation
\begin{equation}
  h^2 \frac{d^2\eta_1}{d x^2} + \left[ q_2 \tanh^2(x) + q_1 \tanh(x) + q_0 \right] \eta_1 = 0 ,
  \label{eq:tanh-reduction}
\end{equation}
where
\begin{equation}
  q_2 = \frac{u_0}{2}\left(\frac{u_0}{2} - i h \right), \quad
  q_1 = u_0\left(\frac{u_0}{2}-E\right), \quad
  q_0 = \left(\frac{u_0}{2}-E\right)^2 - p_y^2 + i h \frac{u_0}{2}. \label{eq:tanh-pars}
\end{equation}
To solve this equation, we first perform the substitution
\begin{equation}
\xi = (1-\tanh(x))/2,
\label{eq:tanh-substitution}
\end{equation}
leading to
\begin{multline}
  4\xi^2(1-\xi)^2\frac{d^2\eta_1}{d \xi^2} + 4\xi(\xi-1)(2\xi-1) \frac{d \eta_1}{d \xi} \\
   + h^{-2}\left[ q_2(1-2 \xi)^2 + q_1(1-2\xi) + q_0 \right]\eta_1 = 0 .
  \label{eq:tanh-varchange}
\end{multline}
Substitution (\ref{eq:tanh-substitution}) maps the real axis to the interval $0\leq \xi \leq1$ in such a way the limit $x\to\infty$ corresponds to $\xi\to 0$ and  $x\to-\infty$ corresponds to $\xi\to 1$. When $\xi\to 0$ the part in square brackets in equation~(\ref{eq:tanh-reduction}) tends to $p_1^2$, where
\begin{equation}
  p_1 = \sqrt{(u_0-E)^2-p_y^2}, \label{eq:tanh-asymp1}
\end{equation}
and when $\xi\to 1$, it becomes $p_2^2$, where
\begin{equation}
p_2 = \sqrt{E^2-p_y^2}.
\end{equation}
Therefore, let us make the substitution
\begin{equation}
  \eta_1=\xi^{i p_1/2h}(1-\xi)^{i p_2/2h} w. \label{eq:tanh-ansatz}
\end{equation}
After some algebraic calculations, one finds that $w$ satisfies the hypergeometric differential equation,
\begin{equation}
  (1-\xi)\xi \frac{d^2 w}{d \xi^2} + \bigl( c - ( a + b + 1)\xi \bigr) \frac{dw}{d \xi} - a b w = 0 , \label{eq:hypergeom}
\end{equation}
with the parameters
\begin{equation}
  a = 1 + \frac{i p_1}{2h} + \frac{i p_2}{2h} + \frac{i u_0}{2 h} , \quad
  b = \frac{i p_1}{2 h} + \frac{i p_2}{2 h} - \frac{i u_0}{2 h} , \quad
  c = 1 + \frac{i p_1}{h}. \label{eq:hypergeom-pars}
\end{equation}
Two linearly independent solutions of (\ref{eq:hypergeom}) can be taken as formulas (15.5.3) and (15.5.4) in \cite{Abramowitz65},
\begin{equation}
 w_1={}_2F_1(a,b,c;\xi), \quad w_2=\xi^{1-c}(1-\xi)^{c-a-b}{}_2F_1(1-a,1-b,2-c;\xi).
 \label{eq:hypersol}
\end{equation}
Hence $\eta_1$ can be written as
\begin{multline}
  \eta_1 = c_1 \xi^{i p_1/2h} (1 - \xi)^{i p_2/2h} {}_2F_1(a,b,c;\xi)\\
   + c_2 \xi^{-i p_1/2h} (1 - \xi)^{-i p_2/2h} {}_2F_1(1-a,1-b,2-c;\xi) . \label{eq:tanh-sol}
\end{multline}
When $x\to \infty$ we can use the approximate relation $\xi\simeq e^{-2x}$. From the equality ${}_2F_1(a,b,c;\xi=0)=1$, we find the asymptotic behavior of $\eta_1$
\begin{equation}
   \eta_1\to c_1 e^{-i p_1 x/h}+c_2 e^{i p_1 x/h}.
   \label{eq:tanh-asympt}
\end{equation}
Therefore we conclude that the function
\begin{equation}
 \eta_1^{(t)}=\xi^{i p_1/2h} (1 - \xi)^{i p_2/2h} {}_2F_1(a,b,c;\xi) ,
\end{equation}
where the superscript `$t$' stands for tunneling, gives the solution for the scattering problem in the regime of Klein tunneling ($E+|p_y|<u_0$). On the other hand, the function
\begin{equation}
 \eta_1^{(a)}=\xi^{-i p_1/2h} (1 - \xi)^{-i p_2/2h} {}_2F_1(1-a,1-b,2-c;\xi),
\end{equation}
where the superscript `$a$' stands for above, gives the solution for the scattering problem in the regime of above-barrier scattering ($E-|p_y|>u_0$). Now we use formula (15.3.6) from~\cite{Abramowitz65},
\begin{eqnarray}
 F(a,b,c;\xi)=\frac{\Gamma(c)\Gamma(c-a-b)}{\Gamma(c-a)\Gamma(c-b)}{}_2F_1(a,b,a+b+1-c;1-\xi)\nonumber\\
 +\frac{\Gamma(c)\Gamma(a+b-c)}{\Gamma(a)\Gamma(b)}(1-\xi)^{c-a-b}{}_2F_1(c-a,c-b,c+1-a-b;1-\xi) ,
\end{eqnarray}
that relates the values of the hypergeometric function at the singular points $\xi=0$ and $\xi=1$. Using that for $x\to-\infty$, $\xi\simeq 1-e^{2x}$, we find that
\begin{align}
 &\eta_1^{(a)}\to \frac{\Gamma(2-c)\Gamma(c-a-b)}{\Gamma(1-a)\Gamma(1-b)}e^{i k_2 x/h}
 +\frac{\Gamma(2-c)\Gamma(a+b-c)}{\Gamma(1+a-c)\Gamma(1+b-c)} e^{-i k_2 x/h},\nonumber\\
 &\eta_1^{(t)}\to \frac{\Gamma(c)\Gamma(c-a-b)}{\Gamma(c-a)\Gamma(c-b)} e^{i k_2 x/h}
 +\frac{\Gamma(c)\Gamma(a+b-c)}{\Gamma(a)\Gamma(b)}e^{-i k_2 x/h}.
\end{align}
To find the proper reflection and transmission coefficients, we remember that the semiclassical scattering states are defined as~(\ref{eq:scat-sol}).
From these equations, we conclude that the ratio of the coefficients in front of $e^{-ik_2x/h}$ and $e^{ik_2x/h}$ at $x\to-\infty$ is equal to
\begin{equation}
 r^{(a,t)}\frac{E-p_2}{|p_y|} .
\end{equation}
This gives
\begin{align}
 r^{(a)}&=\frac{|p_y|}{E-p_2}\frac{\Gamma(a+b-c)\Gamma(1-a)\Gamma(1-b)}{\Gamma(1+a-c)\Gamma(1+b-c)\Gamma(c-a-b)},\\
 r^{(t)}&=\frac{|p_y|}{E-p_2}\frac{\Gamma(a+b-c)\Gamma(c-a)\Gamma(c-b)}{\Gamma(a)\Gamma(b)\Gamma(c-a-b)}.
\end{align}
Analogous arguments give the transmission coefficients as
\begin{align}
  t^{(a)} &= \sqrt{\frac{p_1}{p_2}} \sqrt{\frac{E-U_0-p_1}{E-p_2}}
  \frac{\Gamma(1-a) \Gamma(1-b)}{\Gamma(2-c) \Gamma(c-a-b)} , \label{eq:t-tanh-nn} \\
  t^{(t)} &= \sqrt{\frac{p_1}{p_2}} \frac{|p_y|}{\sqrt{(U_0-E+p_1)(E-p_2)}}
  \frac{\Gamma(c-a) \Gamma(c-b)}{\Gamma(c) \Gamma(c-a-b)}. \label{eq:t-tanh-np}
\end{align}
Of course these transmission coefficients should equal one at normal incidence. To see this, one has to take the limit $|p_y| \to 0$, upon which $p_1 \to |U_0-E|$, $p_2 \to E$ and $E-p_2 \approx p_y^2/(2E)$. This means that for both the Klein tunneling regime and for above-barrier scattering, the factor in front of the $\Gamma$-functions becomes one upon normal incidence. Furthermore, we note that at $p_y=0$, we have $b=0$ for the regime of Klein tunneling and $c=1+b$ for above-barrier scattering. Therefore in both cases the quotient of $\Gamma$-functions equals one and the exact solution shows total transmission at normal incidence. By the same methods, one can construct the solution for the decreasing potential $u(x) = u_0[1-\tanh(x)]/2$.

\begin{figure}[htbp]
\begin{center}
\includegraphics[height=6cm]{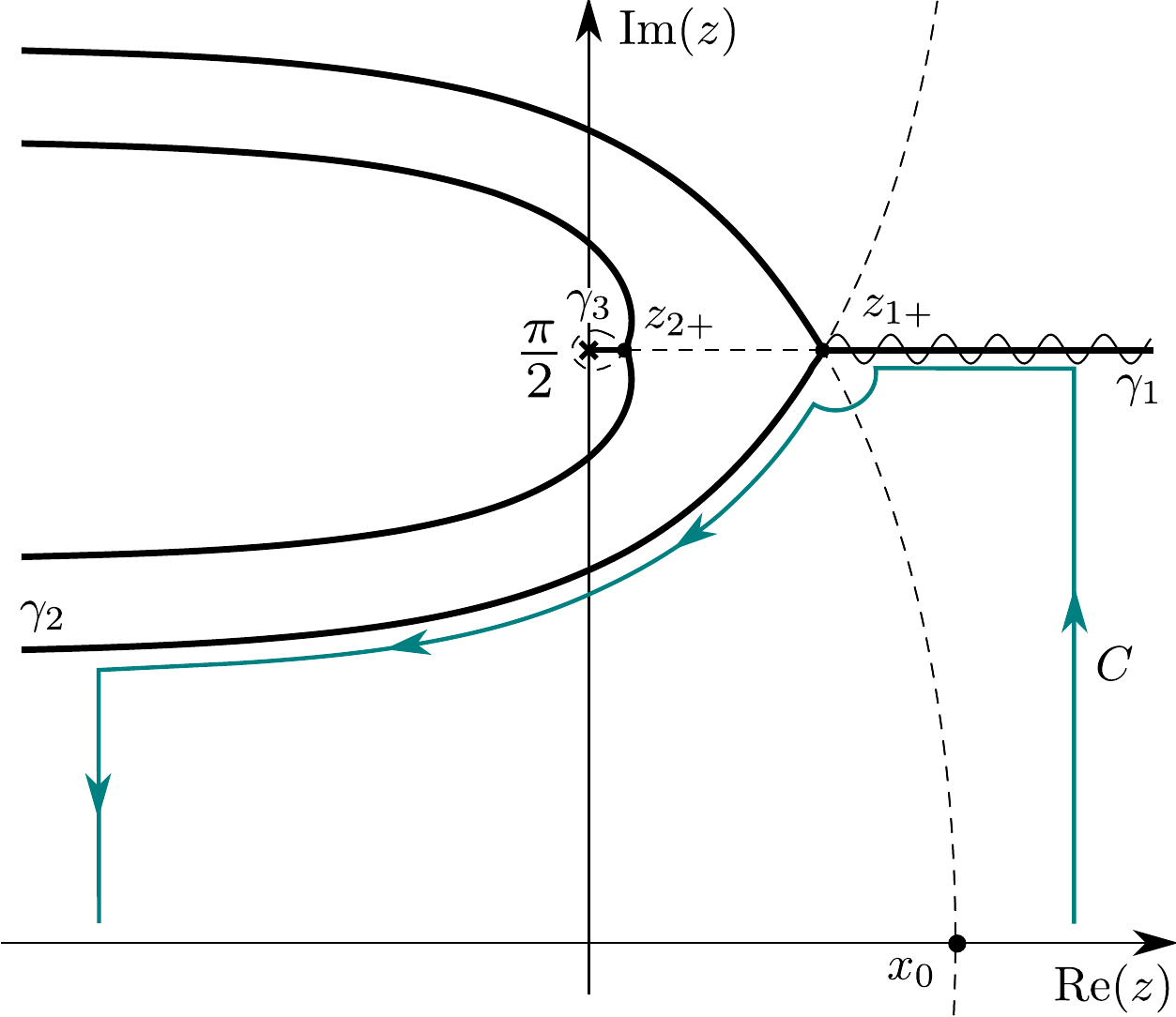}
\end{center}
\caption{Part of the Stokes diagram for scattering by a tangent hyperbolic potential. The cross at $z=i\pi/2$ depicts the pole of the tangent, and $\gamma_3$ denotes the anti-Stokes line that starts at $z_{2+}$ and ends at the pole. Going along the contour $C$, and making use of the method of comparison equations, one finds the transmission coefficient~(\ref{eq:refl-above-tanh-compeq}).}
\label{fig:tanh}
\end{figure}

To get more generic insight in the process of above-barrier scattering by a finite range monotonous potential, let us now treat scattering by a tangent hyperbolic potential semiclassically. In figure~\ref{fig:tanh} one sees the corresponding Stokes diagram. The main peculiarity of this diagram is the existence of a pole at $z=i\pi/2$, and the existence of a finite anti-Stokes line, that ends at the pole.
To obtain an approximation for the transmission coefficient in the case when $z_{1+}$ and $z_{2+}$ are far apart, we can apply the WKB approximation in the same way as in section~\ref{sec:above-barr}. This takes only the right-most turning point into account, and we once again find the transmission coefficient~(\ref{eq:t-above-WKB-cont}). This answer however fails to explain total transmission at normal incidence. The reason for this is that at normal incidence: turning points $z_{1+}$ and $z_{2+}$ merge, and the approximation considered above is no longer valid.

To obtain a prediction that is also valid for near-normal incidence, one has to treat the two turning points in the upper half plane as a cluster. This problem can be solved by the method of comparison equations, see~\ref{subsec:comp-eq-tanh}, and gives the reflection coefficient as
\begin{equation}
  r = -i e^{K/h} \sqrt{1-e^{-2 S/h}} e^{-i \theta} , \label{eq:refl-above-tanh-compeq}
\end{equation}
where $K$ is given by equation~(\ref{eq:r-above-hump-WKB}), and
\begin{equation}
  S = i \int_{z_{2+}}^{z_{1+}} d z (v^2(z)-p_y^2)^{1/2} > 0 . \label{eq:S-tanh-above-compeq}
\end{equation}
In turn, $\theta$ is given by
\begin{equation}
  \theta = \textrm{Arg}\left[ \Gamma\left( 1 + \frac{i S}{\pi h} \right) \right] + \frac{S}{\pi h} - \frac{S}{\pi h} \ln\left( \frac{S}{\pi h} \right) - \frac{\pi}{4} . \label{eq:theta-tanh-above-compeq}
\end{equation}
Note that this result does give total transmission at normal incidence, since $S$ tends to zero when $|p_y|$ tends to zero.

The same answer can also be derived in a simpler way, by making effective use of our previous results for the Klein tunneling regime. Indeed, let us consider the half axis $z=x+i\pi/2$, $x>0$. Along this axis we have $\tanh(x+i\pi/2)=\coth(x)$. When $x\to 0$, the potential $u(x)$ is proportional to $1+\coth(x)\to\infty$, which means that we are dealing with a $p$-$n$ junction along this line!
Along this line, the equation for $\eta_1(x+i\pi/2)$ reads
\begin{equation}
  \left(h^2\frac{d^2}{dx^2}+V^2(x)-p_y^2+ihV'(x)\right)\eta_1=0 ,
  \label{eq:eta1-V}
\end{equation}
where $V(x) = U_0(1+\coth(x))/2-E$. Sufficiently far from the pole at $x=0$, we can use the transfer matrix for a $p$-$n$ junction, equation (\ref{eq:transfer-pn}), to establish the connection between the wave function on the anti-Stokes line $\gamma_1$, on the the right of the classically forbidden region, and on anti-Stokes line $\gamma_3$, to the left of the classically forbidden region. This gives the relation
\begin{equation}
  T_{pn,11} \frac{\sqrt{G(x)}}{\sqrt{p_x(x)}} e^{-i S(x_{2+},x)/h} + 
  T_{pn,21} \frac{e^{i S(x_{2+},x)/h}}{\sqrt{p_x(x)}\sqrt{G(x)}} \leftrightarrow
  \frac{e^{i S(x_{1+},x)/h}}{\sqrt{p_x(x)}\sqrt{G(x)}}, \label{eq:p-n-tanh}
\end{equation}
where all quantities relate to equation~(\ref{eq:eta1-V}), and $x_{1,2+} = z_{1,2+}-i\pi/2$. One can now establish the relation between $G(x)$ in equation~(\ref{eq:p-n-tanh}) and $g(x+i \pi/2)$, to find $\eta_1(z)$ in the complex plane. Since the term with coefficient $T_{pn,21}$ turns out to be dominant between $\gamma_3$ and $\gamma_2$, it provides one of the two terms of the asymptotic expansion along $\gamma_2$. When continues to the real axis, this term becomes the reflected wave. Changing the reference point of the action from $z_{2+}$ to $x_0$, one finally arrives at the previous result~(\ref{eq:refl-above-tanh-compeq}) for the reflection coefficient. 

Upon normal incidence, the turning points $z_{1+}$ and $z_{2+}$ merge. Along the line $x+i \pi/2$ we have conventional Klein tunneling. Therefore $T_{pn,21}$ vanishes, i.e. there is no reflected wave. From the previous arguments, we then conclude that the reflection coefficient for above-barrier scattering vanishes as well. Thus total transmission for a particle that is normally incident on an monotonous increasing potential is related to conventional Klein tunneling in the complex plane. This effect can therefore be referred to as ``virtual Klein tunneling''.

Of course the same calculations can be done for a finite decrease of the potential. In this case one finds that the reflection coefficient is given by
\begin{equation}
  r = -i e^{K/h} \sqrt{1-e^{-2 S/h}} e^{i \theta} , \label{eq:refl-above-tanh-compeq-decr}
\end{equation}
where $S>0$ is the action between the two complex turning points in the upper half of the complex plane, similar to equation~(\ref{eq:S-tanh-above-compeq}). Given this definition of $S$, the phase $\theta$ is defined by equation~(\ref{eq:theta-tanh-above-compeq}).

Glueing the increasing and decreasing potentials together, we obtain a finite step potential of an arbitrary width. In contrast to the potential hump, considered in section~\ref{sec:above-barr}, the transmission through this structure reveals Fabry-P\'erot oscillations,
\begin{equation}
  |t_{nnn}| = \frac{|t_1| |t_2|}{\left|1 - r_1 r_2 e^{2 i L/h}\right|} , \label{eq:t-nnn}
\end{equation}
where $r_1$ and $r_2$ are given by equation~(\ref{eq:refl-above-tanh-compeq-decr}) with parameters corresponding to the left and right junction respectively. The transmission coefficients are given by $|t_{1,2}| = \sqrt{1-|r_{1,2}|^2}$. Finally, $L$ is the action between the two reference points for the separate junctions, $x_{0-}$ and $x_{0+}$,
\begin{equation}
  L = \int_{x_{0,-}}^{x_{0,+}} \sqrt{v^2(x)-p_y^2} d x .
\end{equation}

\section{Comparison with numerical results} \label{sec:numerics}

In this section we compare our semiclassical predictions from the previous sections to numerical results. These are obtained by approximating the potential by a series of small steps. Since the potential is constant between each of them, one can use the exact solution for a constant potential~\cite{KatsnelsonNovGeim2006}. Matching the coefficients at each interface with the help of a computer, we obtain the reflection and transmission coefficients.

Let us start by considering a finite increase of the potential, which corresponds to an $n$-$p$ junction for $E<u_0$. We model it by
\begin{equation}
  u(x/l_1) = 0.5\,u_0\left[1 + \tanh(10x/l_1 - 5)\right] . \label{eq:numericpotential}
\end{equation}
When $x$ changes from $0$ to $l_1$, the potential saturates with an accuracy of $0.01\%$. Therefore, the junction can be cut at these points, without any substantial numerical error. An $n$-$p$-$n$ junction is modeled as an $n$-$p$ junction with length~$l_1$, a $p$-$n$ junction with length $l_3$ and a constant potential of length $l_2$ in between.

\begin{figure}[h]
\begin{minipage}{0.65\textwidth}
\begin{center}
\includegraphics[height=7cm]{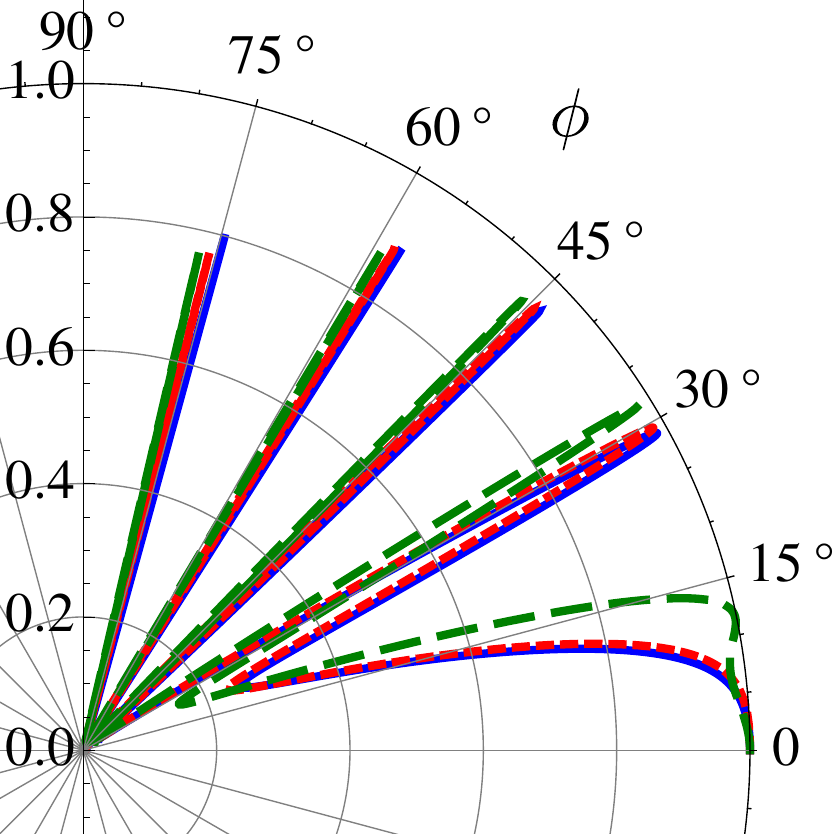}
\end{center}
\end{minipage}
\begin{minipage}{0.32\textwidth}
\includegraphics[width=\textwidth]{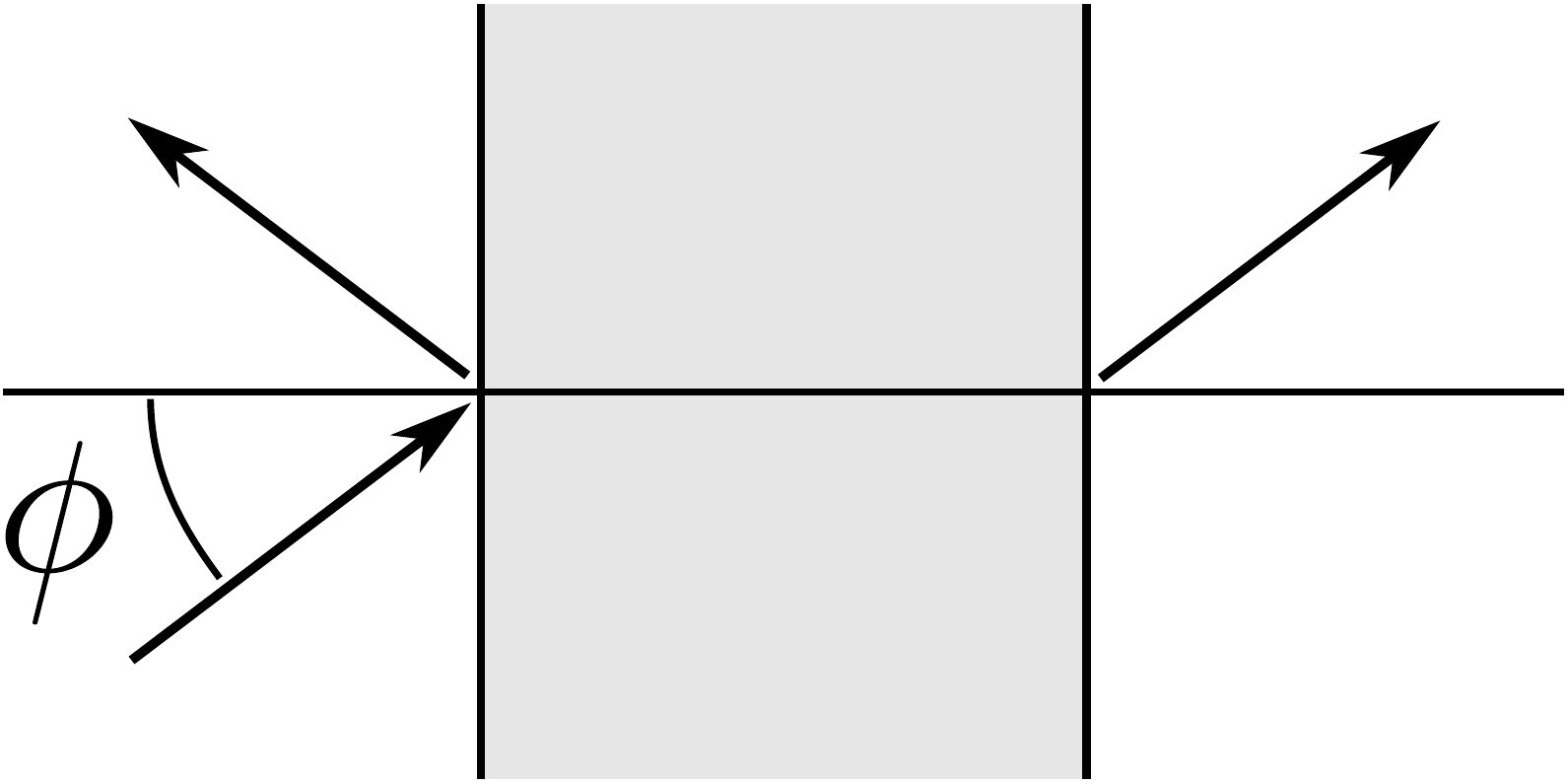}

\bigskip
\bigskip

\includegraphics[width=\textwidth]{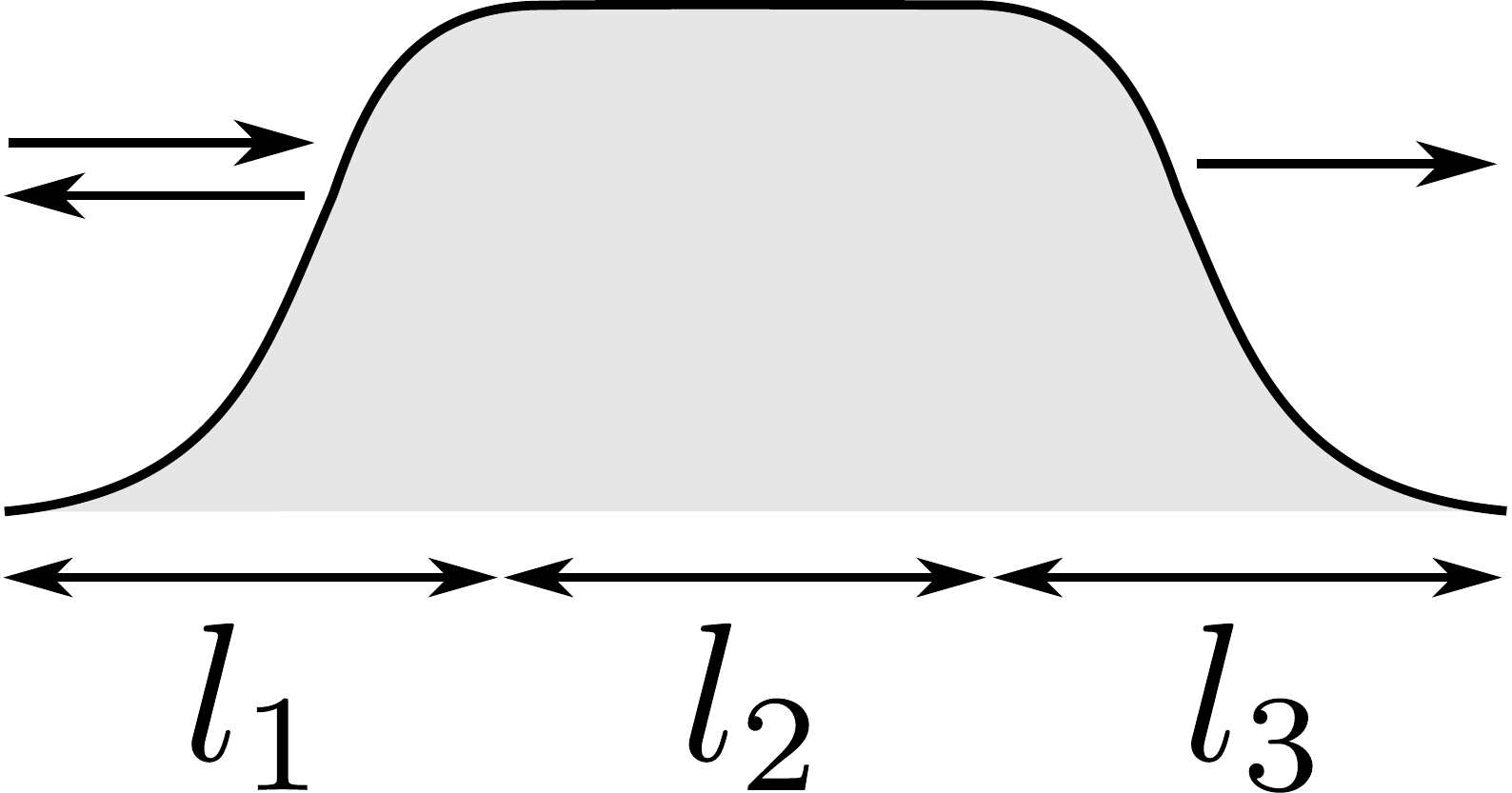}
\end{minipage}
\caption{Left panel: The transmission for an electron incident on an \mbox{$n$-$p$-$n$} junction as a function of the angle of incidence $\phi$. The dimensionless parameters are $h=0.08$ and $\widetilde E=0.4$. The barrier width $l_2/l = 4.3$ and the $n$-$p$ and $p$-$n$ regions have lengths $l_1/l = 2$ and $l_3/l = 2.6$, respectively. The blue line (solid) shows the numerical results for 99 steps, the green line (large dashes) shows the WKB approximation, equation~(\ref{eq:transmission-npn-WKB-FP}) where $\theta=0$, and the red line (small dashes) shows the uniform approximation, where $\theta$ is given by equation~(\ref{eq:theta-uniform}). Right panel: Top view of the potential, showing the angle of incidence $\phi$, and side view, showing the length scales $l_1$-$l_3$.}
\label{fig:double-hump}
\end{figure}

Figure~\ref{fig:double-hump} shows the transmission $|t|^2$ depending on the angle of incidence~$\phi$, which is related to the transversal momentum $p_y$ by $p_y = E \sin \phi$. We compare the numerical transmission for an asymmetric $n$-$p$-$n$ junction with the semiclassical result, equation~(\ref{eq:transmission-npn-WKB-FP}) with $\theta=0$, and the uniform result, where $\theta$ is given by equation~(\ref{eq:theta-uniform}). In the computation of the dimensionless parameters we take the maximum of the potential $u_0$ and the length $l_1/2$ as the typical values $v p_0$ and $l$ introduced in section~\ref{sec:regimes}. Then figure~\ref{fig:double-hump} corresponds to  $h=0.08$ and $\widetilde E=0.4$. In the case of graphene, this would correspond to an electron energy of 100 meV, a barrier height of 250 meV, and length scales  $l_1=70$ nm, $l_2=150$ nm and $l_3=90$ nm.

We see that the agreement between the numerical result and equation~(\ref{eq:transmission-npn-WKB-FP}) with $\theta=0$ becomes better as the angle of incidence increases, that is, deep in the semiclassical regime. Indeed, when we use equation~(\ref{eq:theta-uniform}) for $\theta$, we uniformly approximate the numerical data over the entire range of incidence angles. Concerning the validity of the semiclassical approximation, we note that the agreement improves when the potential is smoother, i.e. when $l_1$ and $l_3$ are large. The exact solution obtained in section~\ref{sec:tanh} perfectly coincides with the numerical results.

One sees that apart from total transmission at normal incidence, there are also additional side resonances. This is a well-known phenomenon for transmission through a metastable hole state. For a more detailed consideration, it is convenient to rewrite equation~(\ref{eq:transmission-npn-WKB-FP}) in the form
\begin{equation}
  |t_{npn}|^2 = \frac{e^{-2(K_{np} + K_{pn})/h}}{\left[1-\sqrt{(1 - e^{-2 K_{np}/h})(1 - e^{-2 K_{pn}/h})}\right]^2 + 4 \sqrt{(1 - e^{-2 K_{np}/h})(1 - e^{-2 K_{pn}/h})} \sin^2 \vartheta} ,
\end{equation}
where $\vartheta = L/h + \theta_{np}/2 + \theta_{pn}/2 - \pi/2$. From this expression one immediately sees that the transmission coefficient is maximal at $\vartheta = n \pi$, and that its modulus equals 
\begin{equation}
  |t_{npn}|_{\mathrm{res}} = \frac{e^{-(K_{np} + K_{pn})/h}}{1-\sqrt{(1 - e^{-2 K_{np}/h})(1 - e^{-2 K_{pn}/h})}}
\end{equation}
For a perfectly symmetric junction, $K_{np} = K_{pn}$, the amplitude is unity, and the corresponding angles were called `magic'~\cite{KatsnelsonNovGeim2006,Tudorovskiy12}. For a generic asymmetric junction the height of the resonances \emph{decays}. When we consider the truly semiclassical regime $K_{np}/h \gg 1$, $K_{pn}/h \gg 1$, we find that 
\begin{equation}
  |t_{npn}|_{\mathrm{res}} \approx \frac{1}{\cosh(K_{np}/h-K_{pn}/h)} .
\end{equation}
The possible consequences of such behavior for graphene-based electronics were mentioned in~\cite{Tudorovskiy12}.

\begin{figure}[htbp]
\begin{center}
\includegraphics[width=\textwidth]{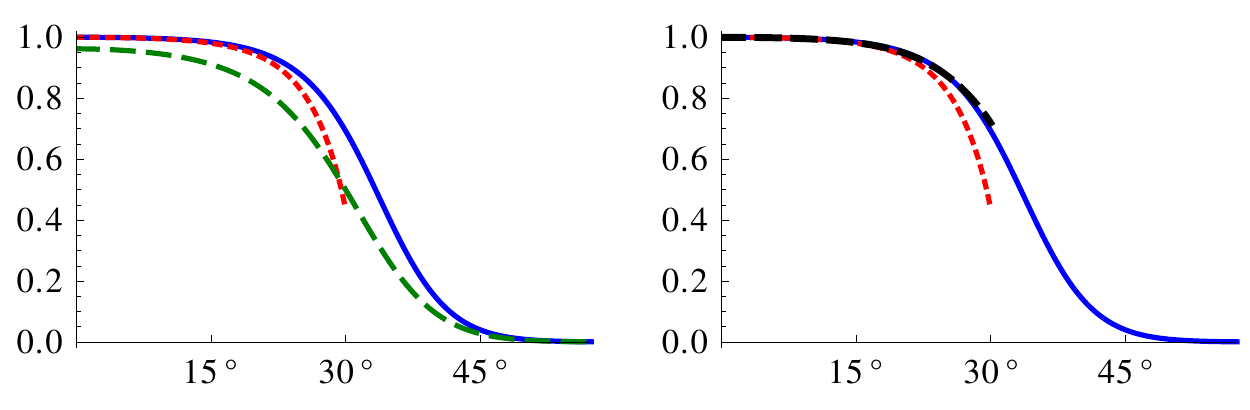}
\end{center}
\caption{The angular dependence of the transmission for an electron incident on an short-range potential profile~(\ref{eq:Ushortrange}). The dimensionless parameters are $h=0.2$ and $\widetilde E=2$. The length scale in the potential is given by $l_1/l =2$. In both panels the blue line (solid) shows the numerical result with 49 steps, and the red line (small dashes) shows the semiclassical prediction for the transmission in the regime of above-barrier scattering derived from equation~(\ref{eq:r-t-upper}). In the left panel, the green line (large dashes) shows the semiclassical result~(\ref{eq:rt-above-middle}), that smoothly goes over in the result~(\ref{eq:rt-conventional-compeq}) for the conventional tunneling regime. In the right panel, the black line (large dashes) shows the transmission derived from the reflection coefficient~(\ref{eq:r-above-guess}) for the regime of above-barrier scattering.}
\label{fig:nn-above}
\end{figure}

To test our semiclassical results for scattering above a short-range potential, we use the model potential
\begin{equation}
   u(x/l_1) = \frac{u_0}{\cosh(10 x/l_1-10)} . \label{eq:Ushortrange}
\end{equation}
Without substantial numerical error, we can cut this junction at $x=0$ and $x=2 l_1$. In figure~\ref{fig:nn-above}, we show the transmission for an electron incident on such a short-range potential. The dimensionless parameters are given by $h = 0.2$ and $\widetilde E=2$. For the case of graphene, this corresponds to a particle of energy 200 meV, a potential height of $100$ meV and a length $l_1 = 70$ nm.

In the left panel of figure~\ref{fig:nn-above}, we compare the numerical transmission with the semiclassical result derived from~(\ref{eq:r-t-upper}) and the equality $|t|^2=1-|r|^2$, and the semiclassical result~(\ref{eq:rt-above-middle}). Note that we use cartesian plots instead of angular plots from now on, in order to make the difference between the different approximations more pronounced. As anticipated in section~\ref{sec:above-barr}, the semiclassical prediction~(\ref{eq:r-t-upper}), that takes the upper two turning points into account, works well at normal incidence. However, once we get closer to the point where the above-barrier scattering regime turns into the conventional tunneling regime (around 30 degrees in the figure), the discrepancy becomes larger. Equation~(\ref{eq:rt-above-middle}), which was derived by considering the middle two turning points as a cluster, gives a slightly better result at this point. Note that this result for the transmission smoothly goes over in~(\ref{eq:rt-conventional-compeq}), which is seen to give a reasonable prediction for the conventional tunneling regime. The discrepancy is due to the influence of the two complex turning points, which were not taken into account in the derivation.

In the right panel of figure~\ref{fig:nn-above}, we compare the numerical result with the results obtained from the semiclassical prediction~(\ref{eq:r-t-upper}) and the equality $|t|^2=1-|r|^2$, and our heuristic formula~(\ref{eq:r-above-guess}). We see that for above-barrier scattering, the heuristic expression~(\ref{eq:r-above-guess}) is more accurate than equation~(\ref{eq:r-t-upper}). Overall, the agreement improves when the potential gets smoother, that is, when $l_1$ increases.

\begin{figure}[htbp]
\begin{center}
\includegraphics[width=\textwidth]{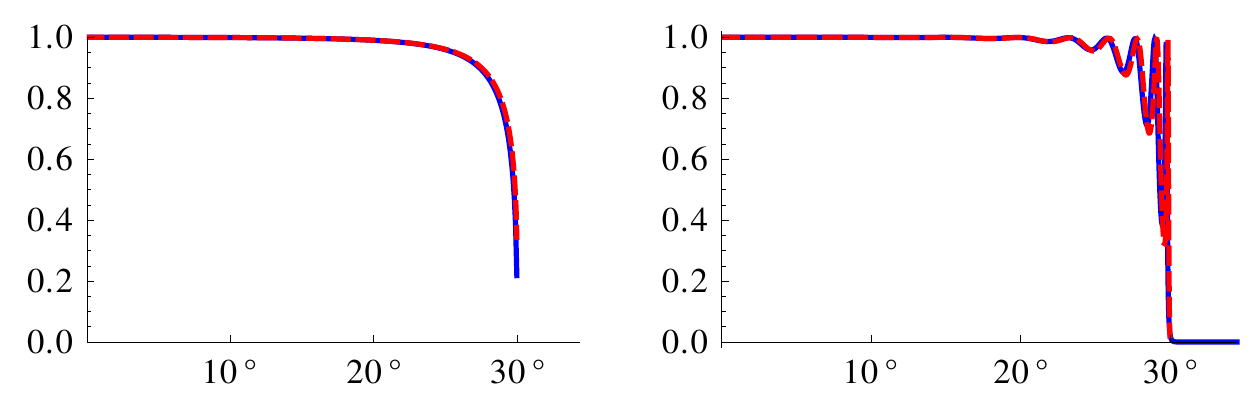}
\end{center}
\caption{The angular dependence of the transmission for an electron. In the left panel we consider a finite range increase of the potential, with $h=0.3$, $\widetilde E=2$ and a length $l_1/l=2$. The blue (solid) line shows the numerical result with $49$ steps, and the red (dashed) line shows the semiclasical result~(\ref{eq:refl-above-tanh-compeq}). In the right panel, we consider a broad potential barrier with $h=0.2$ and $\widetilde E=2$. The width of the constant part $l_2/l = 2.9$ and  the potential increase and decrease have lengths $l_1/l=2$ and $l_3=2.6$, respectively. The blue (solid) line shows the numerical result with $99$ steps and the red (dashed) line shows the semiclassical result, equations~(\ref{eq:t-nnn}) and~(\ref{eq:refl-above-tanh-compeq-decr}).}
\label{fig:nn-tanh}
\end{figure}

To test our results for scattering above a finite increase of the potential, we go back to the potential~(\ref{eq:numericpotential}).
In the left panel of figure~\ref{fig:nn-tanh}, we compare our semiclassical result~(\ref{eq:refl-above-tanh-compeq}) for this case, with the numerical result and the exact solution~(\ref{eq:t-tanh-nn}). The dimensionless parameters are $h=0.3$, $\widetilde E=2$ and $l_1/l = 2$. In the case of graphene, this corresponds to an energy of 200 meV, a potential of height $100$ meV and a length $l_1=50$ nm. The exact solution and the numerical transmission coincide exactly, and therefore only the numerical result is shown. We see that there is good agreement between the numerical result and the semiclassical prediction. Once again, this agreement improves as the potential gets smoother. 

In the right panel, we consider above-barrier scattering for a broad potential hump that is modeled as a potential increase of length $l_1$, a constant part of length $l_2$, and a decrease of length $l_3$. The dimensionless parameters are $h=0.2$, $\widetilde E=2$ and $l_1/l=2$. In the case of graphene, this corresponds to an energy of 200 meV and a potential of height $100$ meV. The lengths are $l_1=70$ nm, $l_2=100$ nm and $l_3=90$ nm. The semiclassical result is given by equation~(\ref{eq:t-nnn}), with reflection coefficient~(\ref{eq:refl-above-tanh-compeq-decr}), and the modulus of the transmission coefficient is constructed from $|r|^2=1-|t|^2$. Since the exact solution once again coincides with the numerical result, it is not shown. The agreement between the semiclassical prediction and the numerical result is quite good, and improves as the potential gets smoother. However, the positions of the maxima in the oscillations are not perfectly reproduced. This is due to the fact that the turning points above and below the real line are quite close in this case, that is, $K/h$ is of order one, and our approximation is not strictly valid.

\section{Conclusion}

In this paper we have studied potential scattering of massless Dirac fermions in semiclassical approximation. We have shown that, depending on the energy of the incoming particle and its angle of incidence, there are three different regimes. These are \textit{i)} the regime of the Klein tunneling, i.e. the regime when the scattering of electrons is mediated by hole states supported by the barrier, \textit{ii)} the above-barrier scattering regime, and \textit{iii)} the conventional tunneling regime. For each of these regimes we found easy-to-use analytic expressions for the transmission and reflection coefficients. We have shown that the conventional WKB method does not allow to study near-normal incidence, due to the degeneracy of turning points at normal incidence. Therefore we cannot obtain expressions for reflection and transmission coefficients uniformly valid for any incidence angle. For near-degenerate turning points, the initial problem has to be reduced to a certain comparison equation with a well-known analytic solution. Using the solution of this comparison equation, we were able to obtain reflection and transmission coefficients for near-normal incidence, which is clearly crucial for physical applications. We completed the analytical part of the paper with the consideration of an exactly solvable model that simulates a monotonous $n$-$n$ junction. This case is somewhat peculiar due to the pole of the potential in the complex plane. The unconventional Stokes diagram for this case is closely related to the one for the Klein tunneling regime. Therefore above-barrier scattering for such a junction can be treated as ``virtual Klein tunneling'' in the complex plane. The predictions provided by our analytic expressions show good agreement with numerical calculations.

\section*{Acknowledgments}

We are grateful to Sergey Dobrokhotov and Andrey Shytov for
helpful discussions.

We acknowledge financial support from the Stichting voor Fundamenteel Onderzoek der Materie (FOM),
which is financially supported by the Nederlandse Organisatie voor Wetenschappelijk Onderzoek (NWO).
This work is supported by the Dutch Science Foundation NWO/FOM and the EU-India FP-7 collaboration under MONAMI.

\appendix

\section{The complex WKB method} \label{app:WKB}

In this appendix we summarize the WKB approximation in the complex plane, as it was developed in~\cite{Zwaan29,Kemble35,Heading62,Fedoruk66,Froeman65,Froeman02,Berry72}. We start with the explanation of the general method, and then discuss its application to potential scattering for massless Dirac fermions.

\subsection{General formulation}
We start with the equation
\begin{equation}
  h^2 \frac{d^2 \psi}{d z^2} + q(z) \psi(z) = 0 , \label{eq:general-order2DE}
\end{equation}
where $h \ll 1$ is a small parameter and $q(z)$ is an analytic function of the complex variable $z$, that may also depend on $h$. It has two approximate solutions,
\begin{equation}
\begin{split}
  f_1(z_0,z) &= q^{-1/4} \exp\left( \frac{i}{h} \int_{z_0}^z d z' \, q^{1/2}(z') \right) ,  \\
  f_2(z_0,z) &= q^{-1/4} \exp\left( -\frac{i}{h} \int_{z_0}^z d z' \, q^{1/2}(z') \right) ,  \label{eq:WKB-fund-sol}
\end{split}
\end{equation}
that will be referred to as basis functions from now on. Similar to the main text, we introduce anti-Stokes lines by the condition that
\begin{equation}
  s(z_0,z) = \int_{z_0}^z q^{1/2}(z') d z'  \label{eq:def-antiStokes}
\end{equation}
is a real function, Here $z_0$ is a turning point, defined by the requirement that $q(z_0) = 0$. On each anti-Stokes line $\gamma$ the exact solution can then be represented as
\begin{equation}
  \psi(z) = C_1^{\gamma} f_1(z_0,z) + C_2^{\gamma} f_2(z_0,z) . \label{eq:exp-antiStokes}
\end{equation}
The main problem when approximating the exact solution in this way is given by the Stokes phenomenon~\cite{Stokes57,Heading62,Froeman65,Fedoruk66,Froeman02}, that we briefly touched in section~\ref{sec:WKB}: the exact solution $\psi(z)$ has different representations~(\ref{eq:exp-antiStokes}) in different sectors of the complex plane. This naturally leads to the connection problem~\cite{Heading62,Froeman65,Fedoruk66,Froeman02}; given certain constants $C_1^{\gamma}$, $C_2^{\gamma}$ on the anti-Stokes line $\gamma$, which constants $C_1^{\gamma_1}$, $C_2^{\gamma_1}$ are needed to represent the exact solution on the anti-Stokes line $\gamma_1$? To connect the coefficients along the anti-Stokes lines $\gamma$ and $\gamma_1$, we introduce the matrix $M$,
\begin{equation}
  \left( \begin{array}{c} C_1^{\gamma_1} \\ C_2^{\gamma_1}  \end{array} \right) = M \left( \begin{array}{c} C_1^{\gamma} \\ C_2^{\gamma}  \end{array} \right) . \label{eq:M-connect-coeff}
\end{equation}
In the rest of this appendix we will determine the matrix $M$ for the various transitions described in section~\ref{sec:WKB}. We will not give a precise estimate of the errors that are involved. Instead, we mention that precise estimates for the error terms are derived in~\cite{Froeman65,Froeman02}.

\begin{figure}[htbp]
\begin{center}
  \includegraphics[height=4.5cm]{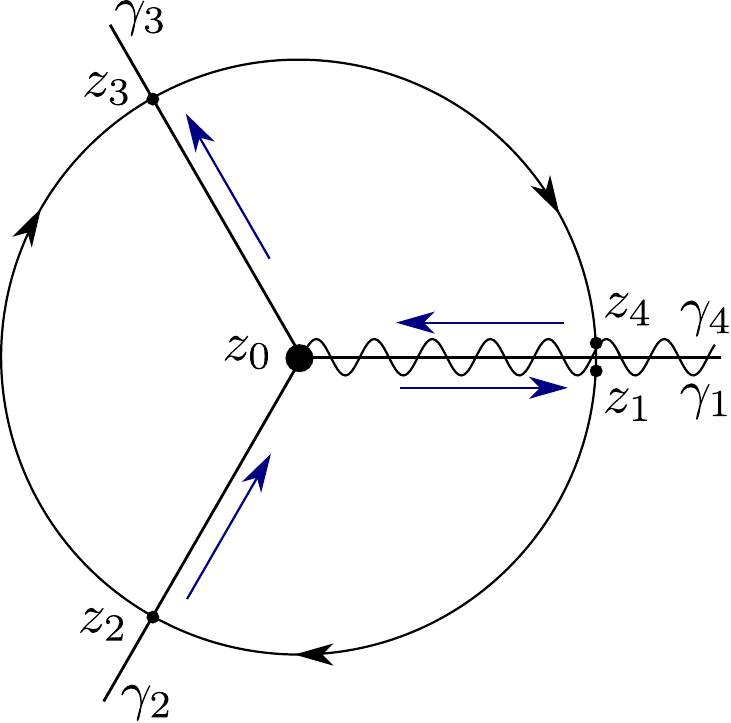}
  \caption{From a simple turning point (large circle) three anti-Stokes lines (solid) emanate. The wavy line indicates the branch cut, and the blue arrows indicate the direction in which the quantity $s(z_0,z)$ increases. The points $z_1$ and $z_4$ lie on the left and the right lip of the cut, respectively. Both $\gamma_1$ and $\gamma_4$ indicate the same anti-Stokes line, and are only used to distinguish different lips of the cut. The points $z_2$ and $z_3$ lie on the other two anti-Stokes lines $\gamma_2$ and $\gamma_3$ respectively.}
  \label{fig:around_tp}
\end{center}
\end{figure}

We start by considering the transition between two anti-Stokes lines that emanate from the same turning point. Let us consider the situation of figure~\ref{fig:around_tp}, with a simple turning point and a branch cut along the positive axis. Remember that in section~\ref{sec:WKB} we defined the left and right lip of the cut with respect to an observer standing on the cut with the turning point behind him. When we are on the right lip of the branch cut, we denote the anti-Stokes linealong the positive $x$-axis by $\gamma_1$. On the left lip of the branch cut, we use the notation $\gamma_4$. When proceeding from $\gamma_1$ in the clockwise direction, we first arrive at $\gamma_2$ and then at $\gamma_3$. Now assume that the expansion coefficients on $\gamma_1$ are given, and that $s(z_0,z)$ increases along the anti-Stokes line, as indicated by the blue arrow in figure~\ref{fig:around_tp}. This implies that when we move away from $\gamma_1$ in the clockwise direction, the action $s(z_0,z)$ obtains a negative complex part, as can be derived from the Cauchy-Riemann relations~\cite{Heading62}. Therefore the basis function $f_1(z_0,z)$ attains exponentially large values (\emph{dominant} term), whereas $f_2(z_0,z)$ becomes exponentially small (\emph{subdominant} term). At a certain distance from $\gamma_1$, the subdominant term will be much smaller than the error in the dominant term, and we cannot keep the subdominant term within the accuracy of the method. Closer to the anti-Stokes line $\gamma_2$, both terms become comparable again. However, the information about the coefficient in front of the subdominant term has been lost. The coefficient in front of the dominant term does not change, so we have $C_1^{\gamma_1} = C_1^{\gamma_2}$. In the most general form, the relation between the constants $C_{1,2}^{\gamma_2}$ and $C_{1,2}^{\gamma_1}$ reads
\begin{equation}
\left(\begin{array}{c}C_1^{\gamma_2} \\ C_2^{\gamma_2} \end{array}\right)=
\left(\begin{array}{cc}1 & 0 \\ \alpha & \beta \end{array}\right)
\left(\begin{array}{c}C^{\gamma_1}_1 \\ C^{\gamma_1}_2 \end{array}\right) .
\end{equation}
Let us consider two linearly independent solutions $\psi(z)$ and $\widetilde \psi(z)$, with coefficients $C^{\gamma_1}_{1,2}$ and $\widetilde C_{1,2}^{\gamma_1}$ on the anti-Stokes line $\gamma_1$, respectively. This can be written as
\begin{equation}
  \left( \begin{array}{cc} \psi & \widetilde\psi \\ \psi' & \widetilde\psi' \end{array} \right) = 
  \left( \begin{array}{cc} f_1 & f_2 \\ f_1' & f_2' \end{array} \right)
  \left( \begin{array}{cc} C^{\gamma_1}_1 & \widetilde C_{1}^{\gamma_1} \\ C^{\gamma_1}_2 & \widetilde C_{2}^{\gamma_1} \end{array} \right) .
\end{equation}
Taking the determinant on both sides, we see that the left-hand side is just the Wronskian, which is constant due to the current conservation for second order ordinary differential equations. On the right-hand side, the first determinant is also constant, which can be verified by inserting the definitions~(\ref{eq:WKB-fund-sol}). Hence the determinant of the second matrix is constant, and this constant does not depend on the anti-Stokes line $\gamma_1$. We can then consider the transition from the anti-Stokes line~$\gamma_1$ to $\gamma_2$, and write
\begin{equation}
\left( \begin{array}{cc} C^{\gamma_2}_1 & \widetilde C_{1}^{\gamma_2} \\ C^{\gamma_2}_2 & \widetilde C_{2}^{\gamma_2} \end{array} \right) =
\left(\begin{array}{cc}1 & 0 \\ \alpha & \beta \end{array}\right)
\left( \begin{array}{cc} C^{\gamma_1}_1 & \widetilde C_{1}^{\gamma_1} \\ C^{\gamma_1}_2 & \widetilde C_{2}^{\gamma_1} \end{array} \right) .
\end{equation}
Taking the determinant on both sides and using the fact that the determinants of the matrices with coefficients are equal, we find that the determinant of the first matrix on the left should equal one, and therefore $\beta=1$.

The change in the subdominant coefficient is therefore given by the so-called \emph{Stokes constant} times the dominant coefficient, see also~\cite{Heading62,Froeman65,Fedoruk66,Froeman02}, that is,
\begin{equation}
\left(\begin{array}{c}C_1^{\gamma_2} \\ C_2^{\gamma_2} \end{array}\right)=
\left(\begin{array}{cc}1 & 0 \\ \alpha & 1 \end{array}\right)
\left(\begin{array}{c}C^{\gamma_1}_1 \\ C^{\gamma_1}_2 \end{array}\right) , \label{eq:Stokescst-mat}
\end{equation}
where $\alpha$ is the Stokes constant. The fact that only the coefficient in front of the subdominant term changes was called the `principle of exponential dominance' in~\cite{Berry72}. Note that when we start with a subdominant term only, its coefficient is unchanged, as can also be seen from equation~(\ref{eq:Stokescst-mat}).

Now let us compute the actual value of the Stokes constant. Following~\cite{Furry47}, we start by noting that the exact solution should be single-valued when one makes a full turn around the turning point. However, the basis functions~(\ref{eq:WKB-fund-sol}) contain the square root of $z$, which has a branch cut in the complex plane. Let $z_4$ be a point on $\gamma_4$, on the left lip of the cut, and let $z_1$ be the same point, but this time on the right lip of the cut, on $\gamma_1$, see figure~\ref{fig:around_tp}. When we assume that we are dealing with a simple turning point, so that in the vicinity of the turning point, $q(z) = \alpha(z-z_0) = r e^{i\phi}$, we can write
\begin{equation}
  q(z_4) = r e^{i\delta} , \quad q(z_1) = r e^{i\delta + 2\pi i} , \quad q^{1/2}(z_1) = e^{i\pi} q^{1/2}(z_4) ,
\end{equation}
where $\delta$ is the angle at which the branch cut emanates from the turning point and equals zero in figure~\ref{fig:around_tp}. We find that
\begin{equation}
  f_1(z_0,z_1) = -i f_2(z_0,z_4) , \quad f_2(z_0,z_1) = -i f_1(z_0,z_4) . \label{eq:rel-sqrt-cuts}
\end{equation}
Since the exact solution is single-valued,
\begin{equation}
  \psi(z) = C_1^{\gamma_1} f_1(z_0,z_1) + C_2^{\gamma_1} f_2(z_0,z_1) = C_1^{\gamma_4} f_1(z_0,z_4) + C_2^{\gamma_4} f_2(z_0,z_4) .
\end{equation}
From equation~(\ref{eq:rel-sqrt-cuts}), we then find that
\begin{equation}
  C_1^{\gamma_4} = -i C_2^{\gamma_1}, \quad C_2^{\gamma_4} = -i C_1^{\gamma_1} .
\end{equation}
Hence, the Stokes constants have to be chosen in such away that when we go from $z_1$ to $z_4$ in the clockwise direction, see figure~\ref{fig:around_tp}, the matrix $M$ from equation~(\ref{eq:M-connect-coeff}) reads
\begin{equation}
  M = \left(
  \begin{array}{cc}
    0 & -i \\
    -i & 0
  \end{array}
  \right) .
\end{equation}
From the above logic we know that $f_1(z_0,z)$ is dominant in the region between $\gamma_1$ and $\gamma_2$, and therefore the matrix $A$ that connects the coefficients can be found from equation~(\ref{eq:Stokescst-mat}),
\begin{equation}
  A = \left(
  \begin{array}{cc}
    1 & 0 \\
    a & 1
  \end{array}
  \right) .
\end{equation}
Since $f_1(z_0,z)$ is subdominant between $\gamma_2$ and $\gamma_3$, the matrix $B$ that connects the coefficients on these anti-Stokes lines, is the transpose of the matrix in equation~(\ref{eq:Stokescst-mat}). Between $\gamma_3$ and $\gamma_4$, $f_1(z_0,z)$ is once again dominant, and we find the matrix $C$,
\begin{equation}
  B = \left(
  \begin{array}{cc}
    1 & b \\
    0 & 1
  \end{array}
  \right) ,
  \quad
  C = \left(
  \begin{array}{cc}
    1 & 0 \\
    c & 1
  \end{array}
  \right) .
\end{equation}
From the identity $M = A B C$, one then finds that
\begin{equation}
  \left(
  \begin{array}{cc}
    1+b c & b \\
    a(1+b c) + c & 1  + a b
  \end{array}
  \right) = \left(
  \begin{array}{cc}
    0 & -i \\
    -i & 0
  \end{array}
  \right) .
\end{equation}
One of these equations turns out to be redundant, and solving the remaining three we find that all three Stokes constants are equal;
\begin{equation}
  a = b = c =  -i . \label{eq:Stokescst-rot}
\end{equation}
When one goes in the counterclockwise direction, it turns out that all Stokes constants equal $i$. Since our basis functions~(\ref{eq:WKB-fund-sol}) are only accurate up to order $h$, we emphasize that the Stokes constant we derived here has the same accuracy. For a very precise error estimate, see~\cite{Froeman65, Froeman02}. Finally, we mention that for a cluster of two turning points, from which four anti-Stokes lines emerge, a similar procedure one only gives a relation between the different Stokes constants, but not their actual value~\cite{Froeman02}.

The second case, when two turning points are connected by a finite anti-Stokes line, was already completely discussed in the main text. We therefore turn to the third case, when we have a transition between two anti-Stokes lines $\gamma_1$ and $\gamma_2$, that emanate from different turning points $z_0$ and $z_1$. As discussed in the main text, one of the basis functions, say $f_1(z_0,z)$, for definiteness, will be dominant, and the other one, say $f_2(z_0,z)$ will be subdominant. Under the assumption that $C^{\gamma_1}_1$ and $C^{\gamma_1}_2$ are comparable, it is shown in~\cite{Froeman02} that sufficiently far from the turning point $z_0$, the solution is accurately represented by the dominant term only, and the subdominant term should be neglected. Furthermore, an estimate of the error is derived. In terms of the matrices, this statement can be cast in the form
\begin{equation}
  M = \left( \begin{array}{cc}
    e^{i s(z_0,z_1)/h} & 0 \\
    0 & 0
  \end{array} \right) ,\quad s(z_0,z_1) = \int_{z_0}^{z_1} d z' \, q^{1/2}(z') . \label{eq:Tmat-move-S-inc}
\end{equation}
The above reasoning leads to the simple statement that~\cite{Froeman02} ``one cannot proceed with an approximate solution, or an exact solution with approximately known initial conditions, in a classically forbidden region from the initial point in the direction in which the wave function decreases.'' This leads to the so-called `one-directional nature of the connection formulae'~\cite{Heading62,Froeman65,Froeman02,Berry72}; a connection formula between a classically forbidden region and a classically allowed region can only be used in one direction. In equation~(\ref{eq:Tmat-move-S-inc}) the one-directionality is manifest, since the matrix $M$ has zero determinant.

One may ask what happens when one does keep the subdominant solution between $\gamma_1$ and $\gamma_2$, assuming that the coefficient in front of it does not change. In~\cite{Froeman65,Froeman02}, the result of such a procedure was compared with the exact solution for a parabolic potential, and it was shown that such a naive procedure gives wrong results for the exponentially small corrections. We therefore stress that the WKB-method applied to simple turning points can only give results in the leading-order approximation. When exponentially small corrections are required, one needs to resort to either unitarity arguments or one needs to make use of an exact solution, in the way that is explained in the main text and in~\ref{app:comp-eq}.

\subsection{Application to potential scattering for massless Dirac fermions}
\label{subsec:WKB-applic}

We now want to make the connection between the abtract theory from the previous subsection, and the particular case of potential scattering for massless Dirac fermions which is considered in the main text. From equation~(\ref{eq:reduction}), we see that for the case of graphene
\begin{equation}
  q(z) = v^2(z)-p_y^2 + ihv'(z) \label{eq:q-def-sl} .
\end{equation}
This makes the basis functions~(\ref{eq:WKB-fund-sol}) slightly different from the asymptotic solutions~(\ref{eq:eta1-sol-cmplx}) introduced in the main text. However, we can recover the latter by expanding the square root of $q(z)$,
\begin{equation}
  \bigl( v^2(z)-p_y^2 + ihv'(z) \bigr)^{1/2} = \bigl(v^2(z)-p_y^2\bigr)^{1/2} + \frac{i h v'(z)}{2 (v^2(z)-p_y^2)^{1/2}} .
  \label{eq:sqrt-exp}
\end{equation}
Since we only considered the leading order when constructing the functions~(\ref{eq:WKB-fund-sol}), corrections of order $h^2$ do not play a role. Therefore we can also neglect the correction in the amplitude factor $q^{-1/4}$ and we recover the approximate solutions~(\ref{eq:eta1-sol-cmplx}). Also note that the `quantum' part in $q(z)$ also slightly shifts the turning points, as compared to the way they were introduced in the main text. However, one can show that this shift is of order $h$ and that it also does not change the results to leading order.

At this point we have to choose how we look at the term
\begin{equation}
  \exp\left( \frac{1}{2} \int_{z_0}^z \frac{v'(z)}{(v^2(z)-p_y^2)^{1/2}} \right) 
\end{equation}
in the solutions~(\ref{eq:eta1-sol-cmplx}). The first option is to look at this term as being part of the action. In that case, we have to change the lower limit in the integral when making a transition from one turning point to another. We see from the previous subsection that the Stokes constant for a clockwise rotation equals $-i$ in this case. Note that to leading order this result is not changed by the correction in equation~(\ref{eq:q-def-sl}). Using the method of comparison equations, we give an independent proof of this fact in~\ref{subsec:compeq-singletp}. This first option naturally arises for the method of comparison equations, see~\ref{app:comp-eq}.

The second option is to calculate the integral once, and to regard it as an amplitude factor. This was done in the main text, and leads to the solutions~(\ref{eq:eta1-sol-cmplx-g}) that include the additional amplitude factor $g(z)$. However, in this case equation~(\ref{eq:rel-sqrt-cuts}) is no longer valid. When $v(z_0)$ is positive, i.e. we are dealing with a turning point that separates a hole region from a classically forbidden region, one still has $g^{1/2}(z_1) = g^{-1/2}(z_4)$. However, when $v(z_0)$ is negative, i.e. we are dealing with a turning point that limits an electron region, one has $g^{1/2}(z_1) = -g^{-1/2}(z_4)$. These two statements can be combined as
\begin{equation}
  f_1(z_0,z_1) = -i\nu f_2(z_0,z_4) , \quad f_2(z_0,z_1) = -i\nu f_1(z_0,z_4) , \label{eq:rel-sqrt-cuts-g}
\end{equation}
where $\nu = \mathrm{sgn} v(x_0)$. Repeating the derivation presented in the previous subsection, one finds that the Stokes constant for a clockwise rotation equals $-i \nu$, cf. equation~(\ref{eq:eta1Stokes}). To comply with the main text, we will choose this second option in this appendix.

\begin{figure}[t]
  \begin{center}
    \includegraphics[width=0.4\textwidth]{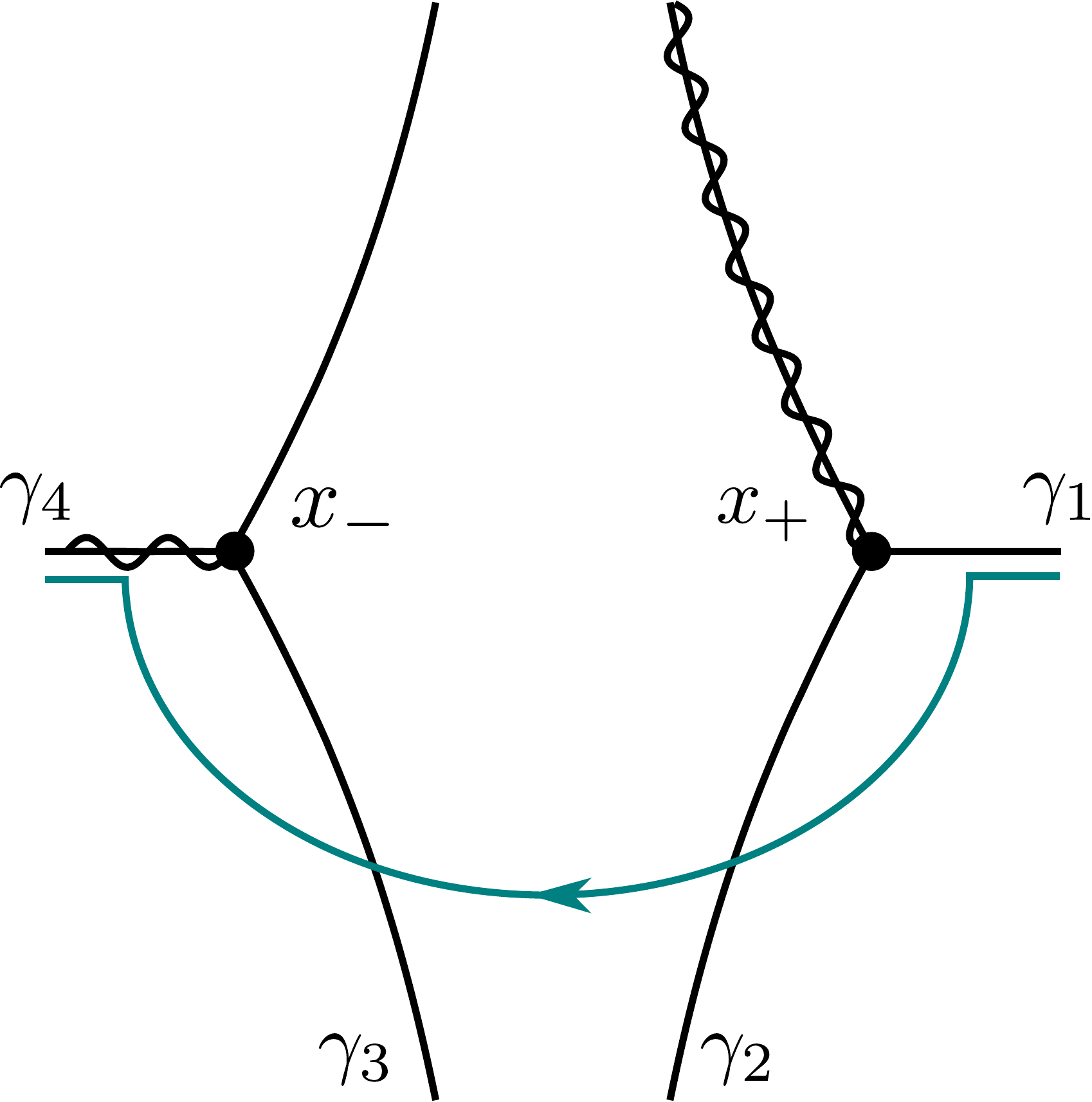}
    \caption{The Stokes diagram for an $n$-$p$ junction, together with the path we take in the complex plane. The solid circles represent the turning points $x_{\pm}$, with $v(x_{\pm})=\pm|p_y|$, and the wavy lines represent the cuts. The relevant anti-Stokes lines (solid lines) are labelled $\gamma_{1-4}$.}
    \label{fig:Stokes-np}
  \end{center}
\end{figure}

We can now rederive the results from section~\ref{subsec:contours} using the matrix approach that we just explained. The Stokes diagram for an $n$-$p$ junction is shown once more in figure~\ref{fig:Stokes-np}. In section~\ref{subsec:contours}, we had to carefully select the path that we take in the complex plane, but now we no longer have to be so careful. We therefore start on the anti-Stokes line $\gamma_1$ with two coefficients, $C^{\gamma_1}_1$ and $C^{\gamma_1}_2$. Taking the path indicated in the figure, the analytic continuation of the square root is defined as
\begin{equation}
\begin{array}{rll}
  (v^2(x)-p_y^2)^{1/2} &= \sqrt{v^2(x)-p_y^2}, \quad &x>x_+  \\
  (v^2(x)-p_y^2)^{1/2} &= e^{-i\pi/2} \sqrt{p_y^2-v^2(x)} , \quad &x_- < x < x_+ ,  \\
  (v^2(x)-p_y^2)^{1/2} &= e^{-i\pi}\sqrt{v^2(x)-p_y^2} , \quad &x < x_- .
\end{array}
\label{eq:sqrtphase}
\end{equation}
With this definition $f_1(x_+,z)$ is dominant between $\gamma_1$ and $\gamma_2$, and since \mbox{$v(x_+)>0$}, we use the matrix~(\ref{eq:Stokescst-mat}) with $\alpha=-i$, that is,
\begin{equation}
  \left( \begin{array}{c} C_1^{\gamma_2} \\ C_2^{\gamma_2} \end{array} \right) =
  \left( \begin{array}{cc} 1 & 0 \\ -i & 1 \end{array} \right)
  \left( \begin{array}{c} C_1^{\gamma_1} \\ C_2^{\gamma_1} \end{array} \right) .
\end{equation}
Upon going from the anti-Stokes line $\gamma_2$ to $\gamma_3$, we see that $\eta_1^+$ is subdominant, whereas $\eta_1^-$ is dominant. Therefore we use the matrix
\begin{equation}
  \left( \begin{array}{c} C_1^{\gamma_3} \\ C_2^{\gamma_3} \end{array} \right) =
  \left( \begin{array}{cc} 0 & 0 \\ 0 & e^{-i s(x_+,x_-)/h} \end{array} \right)
  \left( \begin{array}{c} C_1^{\gamma_2} \\ C_2^{\gamma_2} \end{array} \right)
\end{equation}
In the region between $\gamma_3$ and $\gamma_4$ the function $\eta_1^-(z)$ is dominant and since $v(x_-)<0$, our matrix becomes
\begin{equation}
  \left( \begin{array}{c} C_1^{\gamma_4} \\ C_2^{\gamma_4} \end{array} \right) =
  \left( \begin{array}{cc} 1 & i \\ 0 & 1 \end{array} \right)
  \left( \begin{array}{c} C_1^{\gamma_3} \\ C_1^{\gamma_3} \end{array} \right) .
\end{equation}
The final step is to match the coefficients of the semiclassical solutions~(\ref{eq:eta1-sol-cmplx-g}) to the coefficients $a_{r,l}^{el}$ and $a_{r,l}^{h}$ of the right- and left-moving electron and hole waves, see section~\ref{sec:overdense}. Using that $C_1^{\gamma_1} = a_l^h, C_2^{\gamma_1} = a_r^h, C_1^{\gamma_4} = -a_l^{el}$ and $C_2^{\gamma_4} = -a_r^{el}$ and multiplying the three matrices above, we find that for an $n$-$p$ junction
\begin{equation}
  \left( \begin{array}{c} a_r^{el} \\ a_l^{el} \end{array} \right) =
  \left( \begin{array}{cc}
    e^{K/h}    & -i e^{K/h} \\
    -i e^{K/h} & - e^{K/h}
  \end{array} \right)
  \left( \begin{array}{c} a_r^h \\ a_l^h \end{array} \right) , \label{eq:a-WKB}
\end{equation}                                                                                                                                                                                                                                                                                          
where $K$ is given by equation~(\ref{eq:def-K}). The result for a $p$-$n$ junction can be derived from this by using the relation between the transfer matrices~(\ref{eq:transfer-np}) and~(\ref{eq:transfer-pn}). The reflection and transmission coefficients that can be obtained from equation~(\ref{eq:a-WKB}), coincide with equations~(\ref{eq:t}) and~(\ref{eq:refl-i}). We once again note that this matrix has zero determinant. This is a direct consequence of the fact that we neglected the exponentially small solution within the barrier, and implies that we can only use these matrices in one direction, ``from the right to the left.''

\section{The method of comparison equations}  \label{app:comp-eq}

In this appendix we consider the second way to solve the connection problem, the method of comparison equations~\cite{Fock34,Langer49,Cherry50,Lynn70,Zauderer72}. The basic idea of this method is to express the solutions of the original differential equation in terms of the solutions of a `related' equation that is exactly solvable. This is the rigorous formulation of the approach usually taken in text books on quantum mechanics, see e.g.~\cite{Griffiths05}, where one approximates the potential by a linear function, and writes down its solution in terms of Airy functions. Generally speaking, the Airy equation is the prototype equation if a simple turning point is taken into account. If more turning points are considered, one can reduce the initial problem to more complicated prototype equations with the same number of turning points. In this appendix we use the formulation given in~\cite{Zauderer72}.

\subsection{Explanation of the method}
We consider the second order differential equation
\begin{equation}
  h^2 \frac{d^2 \psi}{d z^2} + R(z,h) \psi(z) = 0 ,
  \label{eq:order2DE-comp}
\end{equation}
where $h \ll 1$ is a small parameter, and $z$ lies in a (possibly complex) domain~$D$. The function $R(z,h)$ is supposed to be analytic, with asymptotic expansion
\begin{equation}
  R(z,h) = \sum_{n=0}^\infty R_n(z) h^n 
\end{equation}
with respect to $h$. A point $z_j$ at which $R_0(z)$ has a root of order $m_j$ is called a turning point of order $m_j$.\footnote{Note that this definition is different from the one employed in~\ref{app:WKB}, where we called a zero of $q(z) = R(z,h)$ a turning point, but that it coincides with the definition in the main text.} The total number of turning points is denoted by $N+1$ and we set $\mu=\sum_{j=0}^N m_j$.

We will reduce~(\ref{eq:order2DE-comp}) to the related equation 
\begin{equation}
  h^2 \frac{d^2 V}{d \phi^2} + Q(\phi,h) V(\phi) = 0 .
  \label{eq:related}
\end{equation}
The exact choice of $Q(\phi,h)$ will be specified below. Following~\cite{Zauderer72}, we write
\begin{equation}
  \psi(z,h) = (\phi'(z))^{-1/2} V(\phi(z)) .
  \label{eq:comp-solution}
\end{equation}
Substituting this into equation~(\ref{eq:order2DE-comp}), we find that it is satisfied if
\begin{equation}
  h^2 \left( \frac{3}{4} \frac{(\phi'')^2}{(\phi')^2} - \frac{\phi'''}{2\phi'} \right) - Q(\phi,h) (\phi')^2 + R(z,h) = 0
\end{equation}
We will solve this equation order by order in $h$, assuming that
\begin{align}
  Q(\phi,h) &= \sum_{n=0}^\infty Q_n(\phi) h^n \\
  \phi(z,h) &= \sum_{n=0}^\infty \phi_n(z) h^n . \label{eq:xi-exp}
\end{align}
Collecting all terms of order $h^0$, we find that
\begin{equation}
  Q_0(\phi_0) (\phi_0')^2 = R_0(z) . \label{eq:terms-h0}
\end{equation}
Gathering terms of order $h^1$, we obtain
\begin{equation}
  Q_1(\phi_0) (\phi_0')^2 + Q_0'(\phi_0) \phi_1 (\phi_0')^2 + 2 Q_0(\phi_0) \phi_0' \phi_1' = R_1(z) . \label{eq:terms-h1}
\end{equation}
Differentiating~(\ref{eq:terms-h0}) and substituting the result into~(\ref{eq:terms-h1}), we obtain
\begin{equation}
  \phi_1(z) = \frac{1}{2} \phi_0' R_0^{-1/2} \int_{z_0}^z d z' \, R_0^{-1/2} \left( R_1 - (\phi_0')^2 Q_1(\phi_0) \right) . \label{eq:phi1-comp-prev}  
\end{equation}
We assume that the mapping $\phi(z)$ is non-singular, i.e. $\phi'$ does not vanish within~$D$. According to equation~(\ref{eq:terms-h0}), this means that $Q_0(\phi_0)$ should vanish whenever $R_0(z)$ vanishes. By differentiating equation~(\ref{eq:terms-h0}), one sees that $Q_0(\phi_0)$ should have a root $\phi_0(z_j)$ of order $m_j$ at every turning point $z_j$. Thus we conclude that $Q_0(\phi_0)$ and $R_0(z)$ have the same number of turning points within~$D$, and that the order of their degeneracy coincides.

In the simplest cases, we can choose $Q_0(\phi)$ to be a polynomial function. Since we just concluded that $Q_0(\phi_0)$ should have the same number of turning points within $D$, we can write this polynomial as
\begin{equation}
  Q_0(\phi) = \gamma_{\mu 0} \prod_{j=0}^N (\phi - \phi_0(z_j))^{m_j} .
\end{equation} 
Taking the square root of~(\ref{eq:terms-h0}), substituting the above expression and integrating from a turning point to an arbitrary point $z$, we find
\begin{equation}
  \int_{\phi_0(z_0)}^{\phi_0(z)} d s \, \prod_{j=0}^N [s - \phi_0(z_j)]^{m_j/2} = \int_{z_0}^z d z' \, [\gamma_{\mu 0}^{-1} R_0(z') ]^{1/2} .
  \label{eq:phi0-implicit}
\end{equation}
If we take $z=z_j$ in~(\ref{eq:phi0-implicit}) to be one of the turning points, then this equation allows us to determine the constants $\phi_0(z_j)$. However, this gives only $N$ equations for the $N+1$ unknowns $\phi_0(z_j)$, and also leaves the constant $\gamma_{\mu0}$ undetermined. On a heuristic level, this means that in constructing the mapping one is free to choose both the origin and the scale. However, the sign of $\gamma_{\mu0}$ is important, since it determines whether we are dealing with a barrier-type, or a well-type problem.
Equation~(\ref{eq:phi0-implicit}) also determines the mapping $\phi_0(z)$ implicitly.

We can also choose $Q_1(\phi,h)$ to be a polynomial function,
\begin{equation}
  Q_1(\phi) = \sum_{k=0}^\infty \gamma_{k1} \phi^k .
\end{equation}
Then equation~(\ref{eq:terms-h1}) becomes
\begin{equation}
  \phi_1(z) = \frac{1}{2} \phi_0' R_0^{-1/2} \int_{z_0}^z d z' \, R_0^{-1/2} \left( R_1 - (\phi_0')^2 \sum_{k=0}^\mu \phi_0^k \gamma_{k1} \right) . \label{eq:phi1-comp}
\end{equation}
Now we require that this expression is non-singular at the turning points. This implies that~\cite{Lynn70, Zauderer72}
\begin{align}
  &\int_{z_0}^{z_j} d z' \, R_0^{-1/2} \left( R_1 - (\phi_0')^2 \sum_{k=0}^\mu \phi_0^k \gamma_{k 1} \right) = 0, \label{eq:cond-phi1-1} \\
  & \frac{d^q}{d z^q} \left( R_1 - (\phi_0')^2 \sum_{k=0}^\mu \phi_0^k \gamma_{k 1} \right) = 0 \; \mathrm{at} \; z=z_j ,
\end{align}
where $j$ runs from $0$ to $N$ and $q$ runs from $0$ and $m_j-2$. These conditions determine the constants $\gamma_{k1}$. One should note that there are only $\mu-1$ equations, while there are $\mu+1$ constants. Therefore, some of them can be set to zero, and this considerably simplifies the expression for $Q_1(\phi)$. When these constants are determined, equation~(\ref{eq:phi1-comp}) determines  $\phi_1(z)$.

In a similar way, one can show that $\phi_j$, with $j\geq 2$ exist. Therefore, the mapping~(\ref{eq:xi-exp}) is well-defined. Here we do not consider higher order corrections, for the purpose of the current paper it is enough to know that they exist~\cite{Lynn70,Zauderer72}.

\subsection{Application to a first-order turning point} \label{subsec:compeq-singletp}

In this subsection we illustrate the method explained in the previous section by the case of a single turning point on the real axis. We will use the method of comparison equations to solve the connection problem. We consider the case where the turning point separates a classically forbidden region (on the left) and a classically allowed region (on the right). From equation~(\ref{eq:reduction}) for massless Dirac fermions, we see that
\begin{equation}
  R_0(x) = v^2(x) - p_y^2, \quad R_1(x) = i v'(x) .  \label{eq:R0R1-graphene}
\end{equation}
Since we consider a first-order turning point $x_0$ on the real axis, we have $m_0=1$ and $\mu=1$, and for convenience we can set $\phi_0(x_0) = 0$. Since we assume that the classically forbidden region is on the left, we also have $\gamma_{10}=1$. Defining the square root as
\begin{align}
   x^{1/2} &= \sqrt{x} , & x>0 , \\
   x^{1/2} &= -i\sqrt{|x|} , & x<0 , \label{eq:sqrt-single}
\end{align}
from equation~(\ref{eq:phi0-implicit}) we obtain
\begin{equation}
  \tfrac{2}{3} \phi_0^{3/2}(x) = \int_{x_0}^x dx' R_0^{1/2}(x') . \label{eq:mapping-1tp-phi0}
\end{equation}
Since we have only one turning point, condition~(\ref{eq:cond-phi1-1}) does not give us any information, and we can set $\gamma_{11} = \gamma_{01} = 0$. Inserting these constants into equation~(\ref{eq:phi1-comp}), we obtain $\phi_1(x)$ as
\begin{equation}
  \phi_1(x) = \frac{1}{2} \phi_0' R_0^{-1/2} \int_{x_0}^x \frac{R_1(x')}{R_0^{1/2}(x')} d x' .  \label{eq:mapping-1tp-phi1}
\end{equation}
Equations~(\ref{eq:mapping-1tp-phi0}) and~(\ref{eq:mapping-1tp-phi1}) together determine the mapping $\phi(x)$.

Having established the mapping, we can now solve our related equation,
\begin{equation}
  h^2 \frac{d^2 V}{d \phi^2} + \phi V(\phi) = 0 .
\end{equation}
This is the well-known Airy (or Stokes) equation, and its solutions are given in terms of the Airy functions~\cite{WolframFunctions,Abramowitz65};
\begin{equation}
  V(\phi) = c_1 \mathrm{Ai}(-h^{-2/3} \phi) + c_2 \mathrm{Bi}(-h^{-2/3} \phi) .
\end{equation}
Assuming that we are sufficiently far from the turning point, we can use the asymptotic expansions of the Airy functions. For $\xi\to\infty$, they read
\begin{align}
 &\mathrm{Ai}(\xi) = \frac{e^{-\frac{2}{3}\xi^{3/2}}}{2\sqrt\pi\,\xi^{1/4}} , &
 &\mathrm{Bi}(\xi) = \frac{e^{\frac{2}{3}\xi^{3/2}}}{\sqrt\pi\,\xi^{1/4}} , \label{eq:Airy-infty} \\
 &\mathrm{Ai}(-\xi) = \frac{\sin \left(\frac{2}{3}\xi^{3/2}+\frac{1}{4}\pi \right)}{\sqrt\pi\,\xi^{1/4}}, &
 &\mathrm{Bi}(-\xi) = \frac{\cos \left(\frac{2}{3}\xi^{3/2}+\frac{1}{4}\pi \right)}{\sqrt\pi\,\xi^{1/4}}. \label{eq:Airy-mininfty}
\end{align}
We can then find the solutions to the original equation~(\ref{eq:order2DE-comp}) from equation~(\ref{eq:comp-solution}) and the mapping. From equation~(\ref{eq:xi-exp}), we find that
\begin{equation}
(h^{-2/3} \phi)^{3/2} = \frac{1}{h} \phi_0^{3/2} + \frac{3}{2} \phi_0^{1/2} \phi_1 + \mathcal{O}(h) ,
\end{equation}
and from equation~(\ref{eq:mapping-1tp-phi0}) that
\begin{equation}
  \phi_0^{1/2} \phi_0' = R_0^{1/2} .
\end{equation}
We then construct the solution to the original equation using~(\ref{eq:comp-solution}) and insert the mapping~(\ref{eq:mapping-1tp-phi0}) and~(\ref{eq:mapping-1tp-phi1}). Let us define
\begin{equation}
  \widetilde \eta_{1}^{\,\pm}(x) = \frac{1}{R_0^{1/4}} \exp\left(\pm \frac{i}{2} \int_{0}^x \frac{R_1}{R_0^{1/2}} dx' \right) \exp\left(\pm \frac{i}{h} \int_{0}^x R_0^{1/2} d x' \right) . \label{eq:def-eta1pm-tilde}
\end{equation}
Note that these waves are equal to the ones defined in equation~(\ref{eq:eta1-sol-cmplx}). Furthermore, they coincide with the basis functions~(\ref{eq:WKB-fund-sol}) defined in~\ref{app:WKB} when we do not regard the first exponent as an amplitude, but instead as part of the action, see the discussion in~\ref{subsec:WKB-applic}.
After a short computation, one obtains that for $x\to\infty$,
\begin{equation}
  \psi(x) = \frac{e^{i\pi/4}}{2\pi^{1/2}} (-i c_1+c_2) \widetilde \eta_1^{\,+}(x) + \frac{e^{-i\pi/4}}{2\pi^{1/2}} (i c_1+c_2) \widetilde \eta_1^{\,-}(x) .
  \label{eq:single-tp-allowed}
\end{equation}
On the other hand, we find that for $x\to -\infty$,
\begin{equation}
  \psi(x) = \frac{e^{-i\pi/4}}{2\pi^{1/2}} c_1 \widetilde \eta_1^{\,+}(x) + \frac{e^{-i\pi/4}}{\pi^{1/2}} c_2 \widetilde\eta_1^{\,-}(x) , \label{eq:single-tp-forbidden}
\end{equation}
where $R_0(x)$ in equation~(\ref{eq:def-eta1pm-tilde}) is now negative and we have used the definition~(\ref{eq:sqrt-single}) of the square root.

In~\ref{app:WKB} we stated that inside the classically forbidden region, only the term that increases along a given path should be kept~\cite{Froeman02}, and we put the other coefficient to zero by hand. Therefore one has to be careful when interpreting the results~(\ref{eq:single-tp-allowed}) and~(\ref{eq:single-tp-forbidden}). When going from the classically allowed into the classically forbidden region, these equalities give~\cite{Heading62,Froeman65,Froeman02}
\begin{equation}
  c_{r,\infty} \widetilde \eta_1^{\,+} + c_{l,\infty} \widetilde \eta_1^{\,-} \to (-i c_{r,\infty} + c_{l,\infty}) \widetilde \eta_1^{\,-}
\end{equation}
where $c_{r,\infty} = e^{i\pi/4}\pi^{-1/2} (-i c_1+c_2)/2$ and $c_{l,\infty} = e^{-i\pi/4}\pi^{-1/2} (i c_1+c_2)/2$. Comparing this result with the ones obtained in~\ref{app:WKB}, we conclude that the factor $-i$ in front of $c_{r,\infty}$ on the right is nothing but the Stokes constant.

Let us  also see what happens when we go from the classically forbidden region into the classically allowed region. Then the connection formula reads
\begin{equation}
  \widetilde \eta_1^{\,+}(x) \to \widetilde \eta_1^{\,+}(x) + i \widetilde \eta_1^{\,-}(x) .
\end{equation}
This concludes our discussion of the connection formulae for a single turning point and their one-directional nature.

\subsection{Application to $n$-$p$ and $p$-$n$ junctions} \label{subsec:comp-eq-nppn}

When we want to construct a uniform approximation for the transmission coefficient through an $n$-$p$ junction, we should take into account two turning points $x_-<x_+$. We assume that they are nondegenerate, i.e. $m_0=m_1=1$ and $\mu=2$. Since we consider a barrier type problem, we have $\gamma_{20}=1$, and we set $\phi_0(x_-) = -\phi_0(x_+) = -a$.
Defining the square root as in equation~(\ref{eq:sqrtphase}), one finds from equation~(\ref{eq:phi0-implicit}) that
\begin{equation}
  \frac{\pi a^2}{2} = \int_{x_-}^{x_+} \sqrt{p_y^2-v^2(x)} \, dx = K , \label{eq:maptpK-single-np}
\end{equation}
where the last equality is implied by equation~(\ref{eq:def-K}). This determines the constant $a$ in the related equation in terms of the parameters of our initial problem. We use the same equation~(\ref{eq:phi0-implicit}) to determine the mapping $\phi_0(z)$. Then we formally use the expansion $\phi_0 \gg a$, though the mapping may not be defined for such $\phi_0$. However, this does not influence the coefficients in front of the asymptotic solutions. One finds that the mapping $\phi_0(z)$ is given by
\begin{equation}
\begin{aligned}
  \int_{x_-}^x \sqrt{v^2(x')-p_y^2} \, d x' &\cong -\frac{1}{2}\phi_0^2 + \frac{a^2}{4} + \frac{a^2}{2}\ln\left(-2\frac{\phi_0}{a}\right), && x<x_-, \\
\int_{x_+}^x \sqrt{v^2(x')-p_y^2} \, d x' &\cong \frac{1}{2}\phi_0^2 - \frac{a^2}{4} - \frac{a^2}{2}\ln\left(2\frac{\phi_0}{a}\right), && x>x_+.
\end{aligned}   \label{eq:mapxl0-single-np}
\end{equation}
For the case of an ordinary Schr\"odinger equation, one has $R_1 = 0$ and hence the first correction $\phi_1$ is zero. This leads to the scattering matrices stated in~\cite{Froeman02}. However, for massless Dirac fermions $\phi_1$ does not vanish, since $R_1 = i v'(x)$. Inserting this into equation~(\ref{eq:cond-phi1-1}), and setting the upper limit to $x_+$, we find that
\begin{equation}
  \gamma_{01} = i .
\end{equation}
From equation~(\ref{eq:phi1-comp}), we then obtain $\phi_1(x)$ as
\begin{equation}
\begin{aligned}
  \phi_1(x) &\cong \frac{1}{2\phi_0} \left[ - \int_{x_-}^x \frac{i v'}{\sqrt{v^2-p_y^2}}\,d x' + i \ln\left(-\frac{a}{2 \phi_0}\right) \right], && x<x_-, \\
  \phi_1(x) &\cong \frac{1}{2\phi_0} \left[ \int_{x_+}^x  \frac{i v'}{\sqrt{v^2-p_y^2}}\,d x' - i \ln\left(\frac{2 \phi_0}{a}\right) \right], && x>x_+,
\end{aligned}   \label{eq:phi1xl0-single-np}
\end{equation}
where we have once more made an expansion for large $\phi_0$.

We find that the related equation~(\ref{eq:related}) reduces to
\begin{equation}
  h^2 \frac{d^2 V}{d \phi^2} + \left( \phi^2 - a^2 + i h \right) V(\phi) = 0 , \label{eq:related-single-np}
\end{equation}
which is exactly the equation for an $n$-$p$ junction in graphene with a linear potential~\cite{Cheianov06,Tudorovskiy12,Sonin09}. Its solution is given by
\begin{equation}
  V(\xi) = c_1 D_\nu(\sqrt{2} e^{i\pi/4} h^{-1/2}\phi) + c_2 D_{-\nu-1}(\sqrt{2} e^{3 i\pi/4} h^{-1/2} \phi) , \quad \nu = \frac{i a^2}{2 h} ,
\end{equation}
where $D_\nu(x)$ are the parabolic cylinder functions~\cite{Whittaker50,WolframFunctions}.
Its asymptotic expansions are given by
\begin{equation}
D_\nu(z) = \left\{
\begin{aligned}
&z^\nu e^{-z^2/4}, && -\pi/2<\mathrm{arg}\,(z)\leq\pi/2 \\
&z^\nu e^{-z^2/4}-z^{-\nu-1}e^{z^2/4}e^{-i\pi\nu}\frac{\sqrt{2\pi}}{\Gamma(-\nu)}, && \mathrm{arg}\,(z)\leq-\pi/2\\
&z^\nu e^{-z^2/4}-z^{-\nu-1}e^{z^2/4}e^{i\pi\nu}\frac{\sqrt{2\pi}}{\Gamma(-\nu)}, && \mathrm{arg}\,(z)>\pi/2
\end{aligned}
\right. , \label{eq:parcyl-as}
\end{equation}
Just as in the previous subsection, the solution of the original differential equation is now given by
\begin{equation}
  \psi(x) = (\phi'(x))^{-1/2} V(\phi(x)) ,
\end{equation}
in which one has to insert the mapping~(\ref{eq:mapxl0-single-np}), (\ref{eq:phi1xl0-single-np}). Then we match the resulting expressions to the scattering states~(\ref{eq:scat-sol}). After some calculations, one finds that the transfer matrix~(\ref{eq:transfer-np}) connecting the hole states on the right and the electron states on the left is given by
\begin{equation}
  T_{np} =  \left( \begin{array}{cc} e^{K/h} & \sqrt{e^{2K/h}-1} \, e^{-i\theta-i\pi/2} \\ \sqrt{e^{2K/h}-1}\,e^{i\theta-i\pi/2} & -e^{K/h} \end{array} \right) , \label{eq:T-np-uniform}
\end{equation}
where $K$ is given by equation~(\ref{eq:def-K}) and $\theta$ by equation~(\ref{eq:theta-uniform}). To find the transfer matrix for a $p$-$n$ junction, one can either do a similar calculation, or use the connection between the transfer matrices~(\ref{eq:transfer-np}) and~(\ref{eq:transfer-pn}). Either way, one obtains
\begin{equation}
  T_{pn} =  \left( \begin{array}{cc} e^{K/h} & \sqrt{e^{2K/h}-1}\, e^{i\theta+i\pi/2} \\ \sqrt{e^{2K/h}-1} \, e^{-i\theta+i\pi/2} & -e^{K/h} \end{array} \right) . \label{eq:T-pn-uniform}
\end{equation}
From these matrices, one can easily derive the transmission coefficient~(\ref{eq:t}) and the reflection coefficient~(\ref{eq:refl-i-sqrt}).
Finally, note that in the limit $K/h\to\infty$, the matrix~(\ref{eq:T-np-uniform}) reduces to the matrices obtained from the WKB approximation, equation~(\ref{eq:a-WKB}).

\subsection{Application to Schr\"odinger-like cases} \label{subsec:comp-eq-conv}

In the previous subsection, one of the turning points corresponded to $v(z)=-|p_y|$, and the other one to $v(z)=|p_y|$. In this subsection we consider the situation where the two turning points correspond to $v(z)=-|p_y|$. This includes both the conventional tunneling regime from section~\ref{sec:barr-noholes}, where we took only the two real turning points into account, as well as the regime of above-barrier scattering from section~\ref{sec:above-barr} when we consider only the middle two turning points. The final answers for the reflection and transmission coefficients in this case are similar to those for an ordinary Schr\"odinger equation. Therefore we speak of Schr\"odinger-like cases. 

The computations for both cases are similar~\cite{Froeman02}, but the one for above-barrier scattering is slightly more complicated. Therefore we will focus on it below, leaving the other one to the reader. For above-barrier scattering the turning points are complex. Hence we have to consider the method of comparison equations in a complex domain $D$, containing these turning points as well as the real axis.

Let us consider two simple complex turning points, $z_{1-}$ and $z_{1+}$. Since the problem is of barrier-type, $\gamma_{20}=1$, and we set $\phi_0(z_{1-}) = -\phi(z_{1+}) = -i b$. We assume that the branch cut is placed between the two turning points, and define
\begin{align}
  x^{1/2} &= \sqrt{x} , & x>x_0 , \\
   x^{1/2} &= e^{-i\pi} \sqrt{x} , & x<x_0 , \label{eq:sqrt-above}
\end{align}
where $x_0$ is the point where the Stokes line from $z_{1+}$ to $z_{1-}$ crosses the real axis, see section~\ref{sec:above-barr}. Performing the integration on the positive side of the cut, we find that
\begin{equation}
  K = 2 i \int_{x_0^+}^{z_{1+}} (v^2(z) -p_y^2)^{1/2} d z = i \int_{z_{1-}}^{z_{1+}} (v^2(z) -p_y^2)^{1/2} d z = - \frac{\pi b^2}{2} . \label{eq:maptpK-single-conv}
\end{equation}
We remind the reader that $x_0$ is the point where the Stokes line from $z_{1+}$ to $z_{1-}$ crosses the real axis, see section~\ref{sec:above-barr}, and that with $x_0^\pm$, we mean the point $x_0\pm\varepsilon$ when $\varepsilon\to 0$. The first equality above follows from the definition in equation~(\ref{eq:r-above-hump-WKB}), and the second one from the fact that $v(x)$ is a real function. The third equality, which is the most important for this subsection, follows from equation~(\ref{eq:phi0-implicit}). 

With the help of equation~(\ref{eq:phi0-implicit}), one then determines the mapping $\phi_0(z)$. We set the lower limit to $z_{1-}$, and then split the integrals into two parts. The first part goes along the cut, and connects the turning point to the real axis. Using equation~(\ref{eq:maptpK-single-conv}) and expanding for large $\phi_0$, we find that
\begin{equation}
\begin{aligned}
  \int_{x_0^-}^x \sqrt{v^2(x')-p_y^2}\, d x' & \cong -\frac{1}{2}\phi_0^2 - \frac{b^2}{4} - \frac{b^2}{2}\ln\left(-2\frac{\phi_0}{b}\right) , && x<x_0 \\
  \int_{x_0^+}^x \sqrt{v^2(x')-p_y^2} \, d x' & \cong \frac{1}{2}\phi_0^2 + \frac{b^2}{4} + \frac{b^2}{2}\ln\left(2\frac{\phi_0}{b}\right) , && x>x_0 .
\end{aligned}
  \label{eq:mapxl0-single-conv}
\end{equation}
Contrary to the two previous examples, equation~(\ref{eq:cond-phi1-1}) shows that this time
\begin{equation}
  \gamma_{01} = 0 ,
\end{equation}
due to the fact that both turning points correspond to $v(z_{1\pm})=-|p_y|$. This implies that the comparison equation will be identical with the one for an ordinary Schr\"odinger equation, and that we can therefore expect similar results. Using equation~(\ref{eq:phi1-comp}), we find that $\phi_1(x)$ equals
\begin{equation}
\begin{aligned}
  \phi_1(x) &\cong - \frac{1}{2} \phi_0^{-1} \int_{-|p_y|}^{v(x)} \frac{i d v}{\sqrt{v^2-p_y^2}} , && x<x_0, \\
  \phi_1(x) &\cong \frac{1}{2} \phi_0^{-1} \int_{-|p_y|}^{v(x)} \frac{i d v}{\sqrt{v^2-p_y^2}} , && x>x_0 .
\end{aligned}
\label{eq:phi1xl0-single-conv}
\end{equation}

The related equation~(\ref{eq:related}) reduces to
\begin{equation}
  h^2 \frac{d^2 V}{d \phi^2} + \left( \phi^2 + b^2 \right) V(\phi) = 0 , \label{eq:related-single-conv}
\end{equation}
which is indeed the same as for an ordinary Schr\"odinger equation. Its solution is given by
\begin{equation}
  V(\phi) = c_1 D_\nu(\sqrt{2} e^{i\pi/4} h^{-1/2} \phi) + c_2 D_{-\nu-1}(\sqrt{2} e^{3 i\pi/4} h^{-1/2} \phi) , \quad \nu = -\frac{1}{2} - \frac{i b^2}{2 h} .
\end{equation}
As in the previous subsection, we now construct the exact solution using equation~(\ref{eq:comp-solution}). Then we make asymptotic expansions of the parabolic cylinder functions and apply the mapping~(\ref{eq:mapxl0-single-conv}), (\ref{eq:phi1xl0-single-conv}). Matching the result to the scattering states~(\ref{eq:scat-sol}), we obtain the reflection and transmission coefficients~(\ref{eq:rt-above-middle}), which are similar to those for an ordinary Schr\"odinger equation~\cite{Froeman02}.

The computation for tunneling through the barrier runs entirely similar, the main difference being that this time the turning points are real and that $K$ is positive instead of negative. However, as indicated in section~\ref{sec:barr-noholes} and shown in~\cite{Froeman02}, up to these differences the final answer is exactly the same.

\subsection{Application to above-barrier scattering} \label{subsec:comp-eq-aboveup}

In the previous subsection we applied the method of comparison equations to above-barrier scattering for two turning points that both correspond to \mbox{$v(z) = -|p_y|$}. However, in section~\ref{sec:above-barr}, we saw that upon near-normal incidence on a short-range potential, the two complex turning points $z_{1+}$ and $z_{2+}$, that correspond to $v(z_{1+})=-|p_y|$ and $v(z_{2+})=|p_y|$ respectively, merge. So if we want to derive an expression for the reflection coefficient that is valid at near-normal incidence, we should apply the method of comparison equations to these two turning points. Since they are connected by an anti-Stokes line, we are dealing with a well-type problem, and hence $\gamma_{20}=-1$. The turning points $z_{1+}$ and $z_{2+}$ are mapped to $-i b$ and $i b$ respectively, and the branch cut is placed between these two points. Applying the mapping~(\ref{eq:phi0-implicit}), we find that
\begin{equation}
  S = \int_{z_{1+}}^{z_{2+}} (v^2(z)-p_y^2)^{1/2} d z = - \frac{\pi b^2}{2} , \label{eq:map-aboveupper-S}
\end{equation}
where $S$ was defined earlier in equation~(\ref{eq:def-S}). From figure~\ref{fig:Stokes_diagrams}\,b), we see that four anti-Stokes lines emerge from the cluster. Since we are interested in the wave function along the lower two lines, we consider $\phi_0$ in the lower half-plane. Applying equation~(\ref{eq:phi0-implicit}) once more, we find
\begin{equation}
  \int_{z_{1+}}^z (v^2-p_y^2)^{1/2} d z' = -\frac{\pi b^2}{4} + \frac{i}{2} \phi_0 (\phi_0^2+b^2)^{1/2} + \frac{i b^2}{2} \ln\left[\frac{\phi_0}{b}+\left( \frac{\phi_0^2}{b^2} +1 \right)^{1/2} \right] . \label{eq:phi0-aboveupper}
\end{equation}
We proceed by determining $\gamma_{01}$ with equation~(\ref{eq:cond-phi1-1}), and obtain $\gamma_{01} = 1$. This allows us to calculate $\phi_1(z)$. We find that
\begin{equation}
  \phi_1(z) = \tfrac{1}{2} \phi_0' R_0^{-1/2} \left( \int_{z_{1+}}^z \frac{i v'(z')}{(v^2-p_y^2)^{1/2}} d z' + i \ln\left[\frac{\phi_0}{b}+\left( \frac{\phi_0^2}{b^2} +1 \right)^{1/2} \right] -\frac{\pi}{2} \right) . \label{eq:phi1-aboveupper}
\end{equation}

The related equation~(\ref{eq:related}) reduces to
\begin{equation}
  h^2 \frac{d^2 V}{d \phi^2} + \left( -\phi^2 - b^2 + h \right) V(\xi) = 0,
\end{equation}
and has the solution
\begin{equation}
  V(\phi) = c_1 D_\nu(\sqrt{2}h^{-1/2} \phi) + c_2 D_{-\nu-1}(i\sqrt{2}h^{-1/2} \phi), \quad \nu = -\frac{b^2}{2 h} ,
\end{equation}
in terms of the aforementioned parabolic cylinder functions. Looking at the asymptotic expansions~(\ref{eq:parcyl-as}), we see the solutions represent traveling waves along the lines $\mathrm{Arg}(\phi) = -\pi/4$ and $\mathrm{Arg}(\phi) = -3\pi/4$ in the lower half plane. Therefore we make an expansion along these lines, and then make use of the mapping~(\ref{eq:phi0-aboveupper}) and~(\ref{eq:phi1-aboveupper}). We then obtain the asymptotic expansion of the solution of the original equation on the anti-Stokes lines $\gamma_1$ and $\gamma_2$ in figure~\ref{fig:Stokes-above-hump-contour2}. In terms of the functions $\widetilde \eta_1^{\,\pm}(z)$, defined in~(\ref{eq:eta1-sol-cmplx}), let us write
\begin{equation}
  \psi(z) = c_1^{\gamma_1} \widetilde \eta_1^{\,+}(z) ,
\end{equation}
on $\gamma_1$ and
\begin{equation}
  \psi(z) = c_1^{\gamma_2} \widetilde \eta_1^{\,+}(z) + c_2^{\gamma_2} \widetilde \eta_1^{\,-}(z) ,
\end{equation}
on $\gamma_2$. Then we find that 
\begin{equation}
  \frac{c_1^{\gamma_2}}{c_1^{\gamma_1}} = 1, \quad \frac{c_2^{\gamma_2}}{c_1^{\gamma_1}} = -i a ,
\end{equation}
where $a$ was defined in equation~(\ref{eq:a-const-asymp/gamma}). The first exponent in~(\ref{eq:eta1-sol-cmplx}) becomes $(-g)^{\mp 1/2}$, and we immediately see that upon passing from $\gamma_2$ to $\gamma_1$, equation~(\ref{eq:above-2-tra}) goes over in equation~(\ref{eq:above-2-ref-in}).

\subsection{Application to the exactly solvable potential increase} \label{subsec:comp-eq-tanh}

In this final subsection, we apply the method of comparison equations to the situation considered in section~\ref{sec:tanh}. As before, we consider a cluster of two turning points, namely the turning points $z_{2+}$ and $z_{1+}$ that lie above the real axis, with $\mathrm{Re}(z_{2+}) < \mathrm{Re}(z_{1+})$, and $v(z_{2+})=|p_y|$, $v(z_{1+})=-|p_y|$. These turning points are mapped to $-a$ and $a$ respectively, and we see that the situation exactly coincides with the $p$-$n$ junction that was considered in~\ref{subsec:comp-eq-nppn}. However, this time we want to consider the asymptotic representation of the exact solutions along different anti-Stokes lines. To find the transmitted wave, we need the solution along the anti-Stokes line $\gamma_1$ in figure~\ref{fig:tanh}, which corresponds to $\mathrm{Arg}(\phi) = 0$ in the comparison equation. On the other hand, we find the incoming and reflected waves by considering the solution along the anti-Stokes line $\gamma_2$, that corresponds to the line $\mathrm{Arg}(\phi) = -\pi/2$ in the comparison equation. A second difference is that we want to have all wavefunctions defined with respect to the reference point $z_{1+}$.

Applying the method of comparison equations as in~\ref{subsec:comp-eq-nppn}, one can show that the comparison equation for this case equals
\begin{equation}
  h^2 \frac{d^2 V}{d \phi^2} + \left( \phi^2 - a^2 - i h \right) V(\phi) = 0 ,
\end{equation}
which has the solution
\begin{equation}
  V(\xi) = c_1 D_{\mu-1}(\sqrt{2} e^{i\pi/4} h^{-1/2} \phi) + c_2 D_{-\mu}(\sqrt{2} e^{3 i\pi/4} h^{-1/2} \phi) , \quad \mu = \frac{i a^2}{2 h} ,
\end{equation}
in terms of the parabolic cylinder functions. We then use the asymptotic expansions of the parabolic cylinder functions and apply the mapping $\phi = \phi(z)$ to find the asymptotic expansion of the exact solution along the anti-Stokes lines $\gamma_1$ and $\gamma_2$ in figure~\ref{fig:tanh}. Introducing the coefficients $c_1^{\gamma_1}$ and $c^{\gamma_2}_{1,2}$ as in the previous subsection, one finds that
\begin{equation}
  \frac{c^{\gamma_2}_1}{c^{\gamma_1}_1}=1, \quad \frac{c^{\gamma_2}_2}{c^{\gamma_1}_1}= -i \left( 1 - e^{-2 S/h} \right)^{1/2} e^{-i \theta} , \label{eq:Stokescst-tanh-compeq}
\end{equation}
in terms of the quantities $S$ and $\theta$ defined by equations~(\ref{eq:S-tanh-above-compeq}) and~(\ref{eq:theta-tanh-above-compeq}), respectively.

The reflection coefficient~(\ref{eq:refl-above-tanh-compeq}) is then obtained by a procedure similar to that in section~\ref{sec:above-barr}. One starts with a transmitted wave, that is defined with respect to the point $x_0$ on the real axis. Then one changes the reference point to $z_{1+}$, and uses the result~(\ref{eq:Stokescst-tanh-compeq}) to make the transition from the anti-Stokes line $\gamma_1$ to the anti-Stokes line $\gamma_2$. Finally, one changes the reference point back to $x_0$ and finds the reflection coefficient~(\ref{eq:refl-above-tanh-compeq}).

\section{Relations between asymptotic scattering states} \label{app:scatstates}

In a previous paper~\cite{Tudorovskiy12} the authors placed particular emphasis on the geometric interpretation of the amplitude factor of the function $\Psi(x)$ in (\ref{eq:sigma-single}) in terms of the Berry phase. The scattering states defined there are different from those we have introduced in the current paper. Here we establish the relationship between different scattering states.

First we notice that using (\ref{eq:connection}) we can write relation (\ref{eq:psiexpeta}) as
\begin{eqnarray}
\Psi=\left(\begin{array}{c}1+i\,\textrm{sgn}(p_y)\hat g \\
1-i\,\textrm{sgn}(p_y)\hat g
\end{array}\right)\eta_1(x), \qquad
\hat g=\frac{1}{|p_y|}\left(-ih\frac{d}{dx}+v(x)\right).
\end{eqnarray}
Using equation (\ref{eq:scat-sol}) and~(\ref{eq:equal-G}) we find
\begin{eqnarray}
\Psi_\pm=\frac{e^{\pm iS(x_0,x)/h}}{\sqrt{p_x(x)}}\left(\begin{array}{c}
G^{\mp 1/2}(x)+i\,\nu\,\textrm{sgn}(p_y)\,G^{\pm 1/2}(x) \\
G^{\mp 1/2}(x)-i\,\nu\,\textrm{sgn}(p_y)\,G^{\pm 1/2}(x)
\label{eq:PsiPM}
\end{array}\right).
\end{eqnarray}
In the electron region ($\nu=-1$) the latter can be written as
\begin{eqnarray}
\Psi_\pm=\frac{e^{\pm iS(x_0,x)/h}}{\sqrt{p_x(x)}}\sqrt{\frac{2|v(x)|}{|p_y|}}
\left(\begin{array}{c}
e^{-i\phi^\pm_p/2} \\
e^{i\phi^\pm_p/2}
\end{array}\right), \label{eq:PsiPMel}
\end{eqnarray}
where $\phi^+_p+\phi^-_p=\pi\,\textrm{sgn}(p_y)$,
\begin{equation}
\cos(\phi^+_p)=\frac{p_x}{|v(x)|}, \qquad
\sin(\phi^+_p)=\frac{p_y}{|v(x)|}.
\end{equation}
For the hole region ($\nu=1$) we can write (\ref{eq:PsiPM}) as
\begin{eqnarray}
\Psi_\pm=\frac{e^{\pm iS(x_0,x)/h}}{\sqrt{p_x(x)}}
e^{i\pi\,\textrm{sgn}(p_y)/2}\left(\begin{array}{c}
G^{\pm 1/2}(x)-i\,\textrm{sgn}(p_y)\,G^{\mp 1/2}(x) \\
-G^{\pm 1/2}(x)-i\,\textrm{sgn}(p_y)\,G^{\mp 1/2}(x)
\end{array}\right),
\end{eqnarray}
or
\begin{equation}
\Psi_\pm=e^{i\pi\,\textrm{sgn}(p_y)/2}
\frac{e^{\pm iS(x,x_0)/h}}{\sqrt{p_x(x)}}\sqrt{\frac{2|v(x)|}{|p_y|}}
\left(\begin{array}{c}
e^{-i\phi^\pm_p/2} \\
-e^{i\phi^\pm_p/2}
\end{array}\right).
\label{eq:PsiPMh}
\end{equation}
Comparing (\ref{eq:PsiPMel}), (\ref{eq:PsiPMh}) with asymptotic scattering states $\widetilde\Psi_\pm$ defined in \cite{Tudorovskiy12} we find that in the electron region
\begin{equation}
  \Psi_\pm = \sqrt{2} |p_y|^{-1/2} \widetilde\Psi_\pm,
\end{equation}
while in the hole region
\begin{equation}
  \Psi_\pm = \sqrt{2}e^{i\pi\,\textrm{sgn}(p_y)/2}|p_y|^{-1/2} \widetilde\Psi_\pm.
\end{equation}
The change in the definition of the asymptotic scattering states leads to a corresponding change of the phase of the transmission coefficient for tunneling from an electron region to a hole region, cf. equation~(\ref{eq:t}) of the current paper and equation~(125) in~\cite{Tudorovskiy12}.

\end{document}